\RequirePackage{fix-cm}
\documentclass[twocolumn]{svjour3}          
\smartqed  
\usepackage{theorem}
\usepackage{color}
\usepackage{graphicx}
\usepackage{amsfonts,amsmath,amssymb}
\usepackage{leftidx}
\usepackage{float}
\usepackage[round]{natbib}

\DeclareMathOperator{\FRni}{\mathsf{F^R_{ni}}}
\DeclareMathOperator{\MRni}{\mathsf{M^R_{ni}}}
\DeclareMathOperator{\MRrni}{\mathsf{M^{R\to r}_{ni}}}
\DeclareMathOperator{\Frnj}{\mathsf{F^r_{nj}}}

\DeclareMathOperator{\M}{\mathsf{M_n}}
\DeclareMathOperator{\F}{\mathsf{F_n}}
\DeclareMathOperator{\FR}{\mathsf{F^R_n}}
\DeclareMathOperator{\MR}{\mathsf{M^R_n}}
\DeclareMathOperator{\MRr}{\mathsf{M^{R\to r}_n}}
\DeclareMathOperator{\Fr}{\mathsf{F^r_n}}

\DeclareMathOperator{\FN}{\mathsf{F_N}}
\DeclareMathOperator{\MRN}{\mathsf{M^R_N}}
\DeclareMathOperator{\MrN}{\mathsf{M^r_N}}

\DeclareMathOperator{\MRNN}{\mathsf{M^R_{N-1}}}
\DeclareMathOperator{\MrNN}{\mathsf{M^r_{N-1}}}

\DeclareMathOperator{\MRrN}{\mathsf{M^{R\to r}_N}}
\DeclareMathOperator{\MrrN}{\mathsf{M^{r\to r}_N}}

\DeclareMathOperator{\Fn}{\mathsf{F_{n+1}}}
\DeclareMathOperator{\Mn}{\mathsf{M_{n+1}}}
\DeclareMathOperator{\MRn}{\mathsf{M^R_{n+1}}}
\DeclareMathOperator{\MRrn}{\mathsf{M^{R\to r}_{n+1}}}
\DeclareMathOperator{\Mrn}{\mathsf{M^r_{n+1}}}

\DeclareMathOperator{\Zr}{\mathsf{Z^r_n}}
\DeclareMathOperator{\ZR}{\mathsf{Z^R_n}}
\DeclareMathOperator{\R}{\mathsf{R}}

\newtheorem{obs}{Remark}

\begin{document}

\title{Bayesian inference in Y-linked two-sex branching processes with mutations: ABC approach}

\author{Miguel Gonz{\'a}lez \and Cristina Guti{\'e}rrez \and Rodrigo Mart{\'i}nez}


\institute{Gonz{\'a}lez, M. \and Guti{\'e}rrez, C. \and Mart{\'i}nez, R.\at
              Department of Mathematics and ICCAEx,  University of Extremadura,
              Avda. Elvas s/n, 06006-Badajoz, SPAIN.\\
              Tel.: +34-924-289-300 (Ext. 57571)\\
              Fax: +34-924-272-911\\
              \email{cgutierrez@unex.es}          
              \and
              Gonz{\'a}lez, M.\\
              \email{mvelasco@unex.es}\\ \\
              Mart{\'i}nez, R.\\
              \email{rmartinez@unex.es}
}

\date{Received: date / Accepted: date}

\maketitle

\begin{abstract}

A Y-linked two-sex branching process with mutations and blind
choice of males is a suitable model for analyzing the evolution of
the number of carriers of an allele and its mutations of a
Y-linked gene. Considering a two-sex monogamous population, in
this model each female chooses her partner from among the male
population without caring about his type (i.e., the allele he
carries).

In this work, we deal with the problem of estimating the main
parameters of such model developing the Bayesian inference in a
parametric framework. Firstly, we consider, as sample scheme, the
observation of the total number of females and males
up to some generation as well as the number of males of each genotype at last generation. Later, we introduce the information of the mutated males only in the last generation obtaining in this way a second sample scheme. For both samples, we apply the Approximate Bayesian Computation (ABC) methodology to approximate the posterior
distributions of the main parameters of this model. The accuracy
of the procedure based on these samples is illustrated and discussed
by way of simulated examples.

\end{abstract}

\section{Introduction}

In \cite{ggm2012a}, a stochastic model in the field of branching
processes was introduced with the aim of describing the evolution
of the number of carriers of a Y-linked gene and its mutations in a
two-sex monogamic population. This model allows to study the interesting and important problem of how mutations of Y-linked genes evolve in a population. In a general sense, we use the term mutation for any change in the genetic material
which gives rise to the transmission of a different trait. We consider a population where two types of alleles could coexist. We denote them as $\R$ and $\mathsf{r}$. The $\R-$allele is considered a marker allele or an allele which transmits a trait of interest (not expressed in the phenotype of the male) and the $\mathsf{r}-$allele is considered an allele which transmits any other trait different of that transmitted by $\R$. Moreover, we assume that
$\R-$allele could mutate transmitting a different trait of $\R$ and therefore, we also denote this mutated allele as $\mathsf{r}$. That is, in our context, $\mathsf{r}$-allele means all alleles which transmit a trait different of that transmitted by $\R$, stemming or not from mutations.
We also assume that backmutation is not allowed, i.e. the $\mathsf{r}-$allele never can return to the $\R-$form. Therefore, there could exist a flow from $\R$ to $\mathsf{r}$ but not vice versa.  Notice that, if in the population there would only be $\R$-alleles, it could appear later $\mathsf{r}-$alleles which would stem from mutations.

This model, called Y-linked two-sex branching process (Y-BBP) with
mutations, considers a population formed by females and males who mate with blind choi\-ce to produce offspring, i.e. each female chooses her partner from among the male
population without caring about his genotype (because the trait is not expressed in the phenotype of the male or it is not decisive at mating time). Applying the genetic inheritance rules, every couple
gives birth to females and males, with every male progeny
inheriting the genetic material corresponding to the Y-chromosome
from his father. But, during reproduction, there could occur a
mutation in the transmitted allele by a father with $\R$-allele, altering
the characteristic of the son with respect
to his progenitor. Hence, under these assumptions, a male with $\R-$allele could
give birth either a male offspring who is a clone of his genetic
material (the same allele) or a mutant with a new type of allele ($\mathsf{r}$).

As important example of such mutations, one could suppose that an
alteration in the allele might impair the individuals reproductive
capacity. In this way, the process could be applied to model
problems of fertility. In particular, it would allow one to study
the case of mutations which end in different levels of fertility
including total infertility (aspermia). A particular case of this
situation is presented in \cite{sun}, in which it is suggested
that a mutation in the USP9Y Y-chromosomal gene causes the absence
of sperm in semen. Another possibility is that the mutation may
represent the beginning of a new paternal lineage, as for example
the one that gave rise to the haplogroup I which is related to
risk of suffering coronary disease, see \cite{Charchar}.

The
aforementioned work \ \cite{ggm2012a} should be consulted for
further background motivation and information about conditions
guaranteeing a positive probability of survival of the alleles in the population. Such conditions depend on several parameters of the model:
the reproduction mean of each genotype, the probability of being
female and the probability of mutation. Therefore, from a practical point of view, it is necessary to develop estimation procedures for these parameters.

The aim of this paper is to develop the Bayesian inferential
theory for a Y-BBP with mutations considering an enough informative and realistic
sample scheme (in the sense of the minimum amount of information that it is necessary to be observed in order to obtain accurate estimates). The branching process theory has
usually assumed that the entire family tree is needed to be
observed in order to make accurate inferences. However, to observe
such quantity of information is hard in practice. In this sense, the authors published in a previous work a study about the inference of the parameters of a Y-BBP model without mutations (see \cite{ggm2013a}),
based on a more realistic sampling scheme where the
total number of females and males
up to some generation as well as the number of males of each genotype
in the last generation is observed.
Carrying on with these ideas, in this paper and for the Y-BBP with mutations, we consider firstly the same sample.
However, in contrast with the model without mutation, the son's genotype is not determined directly from the father's one.
As consequence, this sample could determine the global behavior of alleles in the population, but might not provide enough information in order to make inference on the parameters of the model with mutations.
Therefore, to overcome this lack of information, some knowledge about the number of
mutated males in the last generation should be added. This will be considered as the second sampling scheme.

Moreover, in the Bayesian framework, a Markov cha\-in Monte Carlo (\!MCMC) me\-thod was used for the model without mutations in \cite{ggm2013a}, with very good results. However, although in general MCMC me\-thod works well in many substantive problems, it can perform poorly when is applied to large data sets or complex models, as the model presented in this paper. In fact, the approximation to this problem using the MCMC methodology has provided poor results, failing to provide accurate posterior approximations in a reasonable computational time. Besides, at least in our context, such methodology often needs to make use
of the conjugate family theory representing a lack of generality.

Due to these limitations, we are interested in applying a
different statistical tool to solve this incomplete data problem,
the Approximate Bayesian Computation (ABC) methodology (see, for example, \cite{marin}, \cite{sunnaker} or \cite{lintusaari} for a recent survey). This
method is being developed during last decades as an alternative to
such more traditional MCMC methods. These likelihood-free
techniques are very well-sui\-ted to models for which the likelihood
of the data are either mathematically or computationally
intractable but it is easy to simulate from them, so that they
look very appropriate, a priori, for studying the inference of the Y-BBP with mutations.

Besides this Introduction, the paper is organized in 8 sections as follows. In Section 2, it is described
in detail the Y-BBP with mutations as well as the asymptotic
behavior of the different types of alleles in the population. Section 3 is devoted to introduce the Tolerance Rejection-ABC Algorithm. We apply it in Section 4 to a simulated example based on the sample des\-cribed in this Introduction. In Section 5 we set out a more informative sampling scheme.  We apply again the algorithm, but now with this new sample, in Section 6, developing a series of simulated examples which cover the different situations that can be observed in the sample. After that, in Section 7, we examine the robustness of the methodology, and in Section 8 we use the approximation of the posterior distributions of the parameters to infer the predictive posterior distribution of the size of future generations. Finally, in Section 9, we provide some concluding remarks.

\section{Description of the model}

The genetic frame we model is given by a Y-linked gene
which presents two allelic forms, denoted as $\R$ and $\mathsf{r}$, where $\R$ can
mutate giving rise to new (different) alleles, all denoted also as
$\mathsf{r}$. This allele represents the
transmission of any trait different from the characteristic
transmitted by the $\mathsf{R}$-allele (stemming or not from its mutations).

Since the Y-chromosome is specific to males, we deal with a
two-sex population formed by females, by males which carry the
$\mathsf{R}$-allele (called $\mathsf{R}$-males), and by $\mathsf{r}$-males which carry the $\mathsf{r}$-allele. It is assumed that each individual
mates with only one individual of the opposite sex if available
(perfect fidelity or monogamous mating), forming a couple.
Therefore, in the population one could find two types of couples,
denoted by $\mathsf{R}$- and $\mathsf{r}$-couples, depending on
whether its male is of type $\mathsf{R}$ or of type $\mathsf{r}$,
respectively.

According to the rules of genetic inheritance, and taking into
account the possibility of mutation, an $\mathsf{R}$-couple can
give birth to females, $\mathsf{R}$-males, and $\mathsf{r}$-males,
whereas, given the assumption of no backmutation and that mutations of $\mathsf{r}$-allele are also named as $\mathsf{r}$, an
$\mathsf{r}$-couple gives birth to females and $\mathsf{r}$-males.

Assuming non-overlapping generations and given the number of
$\mathsf{R}$- and $\mathsf{r}$-couples in generation $\mathsf{n}$,
denoted by $\ZR$ and $\Zr$, respectively, the number of females,
males, and couples of each genotype in the $\mathsf{(n+1)}$st
generation is determined by considering a two-stage structure,
reproduction and mating, similarly as it was described in \cite{ghmm06} and \cite{gmm09} for others Y-BBP without mutations.

In the reproduction phase, couples of the $\mathsf{n}$th
generation produce offspring independently of each other and
according to certain reproduction law which is the same for a
given genotype but may be different for different genotypes since
the mutation could affect the reproductive capacity. Moreover,
these reproduction laws are independent of the generation the
couples belong to. Mathematically, the number of females and males of each genotype
stemming from each type of couple is identified with the following
independent sequences of independent, identically distributed,
non-negative, and integer-valued random vectors:
$$\{(\FRni,\MRni,\MRrni), \mathsf{i=1,2,...;n=0,1,...}\}$$
and $$\{(\Frnj,\mathsf{M_{nj}^{r\to r}}),
\mathsf{j=1,2,...;n=0,1,...}\}.$$
Here, $\FRni$ and $\Frnj$ are,
respectively, the number of females stemming from the
$\mathsf{i}$th $\mathsf{R}$-couple and the $\mathsf{j}$th
$\mathsf{r}$-couple of generation $\mathsf{n}$; $\MRni$ is the
number of males stemming from the $\mathsf{i}$th
$\mathsf{R}$-couple of the $\mathsf{n}$th generation which have
preserved the $\mathsf{R}$-allele, and $\MRrni$ is the
number of males stemming from the $\mathsf{i}$th
$\mathsf{R}$-couple of the $\mathsf{n}$th generation, whose
alleles have mutated and now are of type $\mathsf{r}$; and
finally, $\mathsf{M_{nj}^{r\to r}}$ is the number of males
stemming from the $\mathsf{j}$th $\mathsf{r}$-couple of the
$\mathsf{n}$th generation, and which therefore carry also the
$\mathsf{r}$-allele.

We assume that the distributions of $\FRni\!+\!\MRni\!+\!\MRrni$ and
$\Frnj+\mathsf{M_{nj}^{r\to r}}$ have finite means,
$\mathsf{m_R}$ and $\mathsf{\mathsf{m_r}}$, respectively, and variances.

Moreover, the conditional distribution of the vector
$(\FRni,\MRni,\MRrni)$ given $\FRni+\MRni+\MRrni=k$ is multinomial
with parameters ($k$, $\alpha$,
 $(1-\alpha)(1-\beta)$, $(1-\alpha)\beta)$, for
$k\geq 0$, and $0<\alpha<1$, $0\leq \beta <1$ with
$\alpha$ representing the probability for an offspring to
be female and $\beta$ the probability of mutation. Then,
in accordance with this multinomial scheme, the average numbers of
females, $\R$-males, and $\mathsf{r}$-males generated by an
$\R$-couple are, respectively, $\mathsf{\alpha m_R}$,
$\mathsf{(1-\alpha)(1-\beta) m_R}$ and $\mathsf{(1-\alpha)\beta
m_R}$. Notice that, if $\beta=0$, then mutations do not happen a.s. so, if in the population both alleles coexist, $\mathsf{r}-$allele stems from the $\mathsf{r}$-couples in the initial generation and one has the Y-BBP without mutation studied in \cite{gmm09}. The case $\beta=1$ is not considered in this paper because in such case, from the first generation on, only the $\mathsf{r}-$allele would survive in the population a.s. and then one has the classical bisexual branching process introduced by \cite{daleyA} describing the evolution of this allele.

With respect to the mutant-allele, the conditional distribution of
$(\Frnj,\mathsf{M_{nj}^{r\to r}})$ given
$\Frnj+\mathsf{M_{nj}^{r\to r}}=l$ is also multinomial with
parameters ($l$,$\mathsf{\alpha}$, $\mathsf{(1-\alpha)}$), for
$l\geq 0$, and $0<\alpha<1$, with
$\alpha$ the same for both genotypes, i.e., the gene has
no influence on sex designation. Then, the average numbers of
females and $\mathsf{r}$-males are, respectively, $\mathsf{\alpha
m_r}$ and $\mathsf{(1-\alpha)m_r}$.

At the end of the reproduction phase, one has the total number of
females, $\mathsf{R}$-males, and $\mathsf{r}$-males, denoted by
$\Fn$, $\MRn$, and $\Mrn$, respectively, which together constitute
the $\mathsf{(n+1)}$th generation. Specifically, one obtains such
variables by means of the following expressions:
\begin{equation}\label{eqfem}
\Fn=\sum_{\mathsf{i}=1}^{\ZR}\FRni+\sum_{\mathsf{j}=1}^{\Zr}\Frnj,\end{equation}
\begin{equation}\label{eqmales}\MRn=\sum_{\mathsf{i}=1}^{\ZR}\MRni \ \mbox{
and }
 \Mrn=\MRrn+\mathsf{M^{r\to r}_{n+1}},\end{equation} where
$$\MRrn=\sum_{\mathsf{i}=1}^{\ZR}\MRrni \quad \mbox{
and } \quad \mathsf{M^{r\to
r}_{n+1}}=\sum_{\mathsf{j}=1}^{\Zr}\mathsf{M_{nj}^{r\to r}},$$
with the empty sum defined as 0, and $\MRrn$ and $\mathsf{M^{r\to
r}_{n+1}}$ denoting the total number of males with
$\mathsf{r}$-genotype in generation $\mathsf{n+1}$ which stemming
from $\mathsf{R}$- and $\mathsf{r}$-couples, respectively.

Given the total numbers of females, $\mathsf{R}$-males, and
$\mathsf{r}$-males in the $\mathsf{(n+1)}$st generation, the
number of couples of each type ($\R$ or $\mathsf{r}$) in this generation is
determined in the mating phase as follows: perfect fidelity mating is assumed, hence
if the total number of females is greater than or equal to the
total number of males then every male finds a mate in the female
population resulting in $\mathsf{Z^R_{n+1}}=\MRn$ couples of type
$\mathsf{R}$ and $\mathsf{Z^r_{n+1}}=\Mrn$ couples of type
$\mathsf{r}$. On the other hand, every female mate when  the total number of males
exceeds the total number of females. Moreover, since it is assumed that the genotype has no impact on the mating mechanism, females choose its mate in a blind way. Hence, the total number of $\mathsf{R}$-couples in the $\mathsf{(n+1)}$th generation, $\mathsf{Z^R_{n+1}}$, follows a hypergeometric distribution with parameters $\Fn$,
$\Mn=\MRn+\Mrn$, and $\MRn$, while the total number of
$\mathsf{r}$-couples in this generation equals the number of
remaining females, i.e., $\mathsf{Z^r_{n+1}}=\Fn-\mathsf{Z^R_{n+1}}$, whose distribution is also hypergeometric with parameters $\Fn$, $\Mn$, and $\Mrn$.

The bivariate sequence $\{(\ZR,\Zr)\}_{n\geq0}$, describing the
evolution of the number of couples of each type over
generations, is called Y-linked two-sex bran\-ching process with
mutations and blind choice of males. It is
shown in \cite{ggm2012a} that the process above is a homogeneous
multitype Mar\-kov chain and that each genotype shows the dual
behavior typical for bran\-ching processes known as the
extinction-explosion dicho\-tomy. However, the behavior of the $\mathsf{r}-$allele dependents on the behavior of the $\R$-allele. In concrete, if the $\mathsf{R}$-allele becomes extinct, the survival or not of the $\mathsf{r}$-allele depends on its own reproductive capacity. Whereas, considering $\beta>0$, if the $\mathsf{R}$-allele explodes, the $\mathsf{r}$-allele also explodes due to the mutations, independently of the $\mathsf{m_r}$ value, so that the coexistence set is a.s. $\{\mathsf{Z^R_n}\rightarrow \infty,\mathsf{Z^r_n}\rightarrow \infty\}=\{\mathsf{Z^R_n}\rightarrow \infty\}$. Moreover, this set has a positive probability if $\min\{\alpha,(1-\alpha)\}(1-\beta)\mathsf{m_R}>1$ (see \cite{ggm2012a} for details).

In \cite{cgutierrez}, a simulation-based study was developed
to determine the behavior of the different types of alleles
in the population on the coexistence set. So, we established that the asymptotic behavior of the $\mathsf{r}$-allele depends on the relation between $\mathsf{m_r}$, and $(1-\beta)\mathsf{m_R}$. In particular, when $\mathsf{m_r}\geq (1-\beta)\mathsf{m_R}$, the $\mathsf{r}-$genotype is the dominant one in the sense that,
a.s. on $\{\mathsf{Z^R_n}\rightarrow \infty\}$, the sequence
$\{\Zr/\ZR\}_{\mathsf{n\geq 0}},$ converges to infinity.
In the case $\mathsf{m_r}<(1-\beta)\mathsf{m_R}$, there is no
dominant genotype because the previous sequence converges, a.s. on $\{\mathsf{Z^R_n}\rightarrow \infty\}$, to a positive and finite value.

Specifically, when $\mathsf{m_r}>(1-\beta)\mathsf{m_R}$, for $\mathsf{n}$ large enough, it can be stated that
$$\frac{\Zr}{\ZR}\simeq \left(\frac{\mathsf{m_r}}{\mathsf{(1-\beta)m_R}}\right)^n \mathsf{W},$$ with $\mathsf{W}$ \!a certain non-degenerate random variable.
When $\mathsf{m_r}=(1-\beta)\mathsf{m_R}$, the sequence  $\{\Zr/\ZR\}_{\mathsf{n\geq 0}}$ also grows a.s. to infinity, however now it does so linearly, that is, for $\mathsf{n}$ large enough, it is satisfied that
$$\frac{\Zr}{\ZR}\simeq \mathsf{n\frac{\beta}{1-\beta}} + \mathsf{W^*},$$
where $\mathsf{W^*}$ is a non-degenerate random variable. Finally, for $\mathsf{n}$ large enough,
$$\frac{\Zr}{\ZR}\simeq \mathsf{\frac{\beta m_R}{(1-\beta)m_R-m_r}}$$
in the case $\mathsf{m_r}<(1-\beta)\mathsf{m_R}$, that is,  $\{\Zr/\ZR\}_{\mathsf{n\geq 0}}$ converges a.s. to the constant $\mathsf{\beta m_R((\!1-\beta)m_R\!-\!m_r)^{-1}}$ which had been determined empirically.

Moreover, we have determined computationally the asymptotic ratio of the quotient between the total number of $\mathsf{r}-$couples in consecutive generations,  $\mathsf{Z^r_{n+1}}/\Zr$, and we have concluded that such ratio is, a.s. on $\{\mathsf{Z^R_n}\rightarrow \infty\}$,  $\min\{\alpha,(1-\alpha)\}\max\{\mathsf{m_r},(1-\beta)\mathsf{m_R}\}$. Finally, it was proved in \cite{ggm2012a} that the asymptotic ratio of $\mathsf{Z^R_{n+1}}/\ZR$ is, a.s. on $\{\mathsf{Z^R_n}\rightarrow \infty\}$, $\min\{\alpha,(1-\alpha)\}(1-\beta)\mathsf{m_R}$.

Based on these previous results and a deeper study of the simulations, it is easy to deduce the rates of growth of every type of couple in every case, on the set where both genotypes survives. The knowledge of such ratios is important for the development of the results of this paper. So, when $\mathsf{m_r}>(1-\beta)\mathsf{m_R}$, the sequence $\{\ZR\}_{\mathsf{n\geq 0}}$ grows geometrically at a rate $\mathsf{\tau_R}=\min\{\alpha,(1-\alpha)\}(1-\beta)\mathsf{m_R}$ while
$\{\Zr\}_{\mathsf{n\geq 0}}$ grows, also geometrically, at a rate $\mathsf{\tau_r}=\min\{\alpha,(1-\alpha)\}\mathsf{m_r}$, i.e., each type of couple have a different rate of growth, being the $\mathsf{r}$-allele the dominant one.

On the other hand, when $\mathsf{m_r}<(1-\beta)\mathsf{m_R}$, $\{\ZR\}_{\mathsf{n\geq 0}}$ and $\{\Zr\}_{\mathsf{n\geq 0}}$ have the same rate of geometric growth given by $\mathsf{\tau_R}$, and moreover it is verified that, as $\mathsf{n}$ tends to infinity, the limit of $\Zr/\mathsf{\tau_R^n}$ is, a.s. on $\{\mathsf{Z^R_n}\rightarrow \infty\}$, proportional to the limit of $\ZR/\mathsf{\tau_R^n}$ with proportionality constant $\mathsf{\beta m_R((1-\beta)m_R-m_r)^{-1}}$.

Finally, when $\mathsf{m_r}=(1-\beta)\mathsf{m_R}$,  $\{\ZR\}_{\mathsf{n\geq 0}}$ grows at a geometric rate of $\mathsf{\tau_R}$ while the sequence that normalizes $\{\Zr\}_{\mathsf{n\geq 0}}$ is $\mathsf{\{n\tau_R^{n}\}_{n\geq 0}}$. Moreover, as $\mathsf{n}$ tends to infinity, the limit of $\Zr/\mathsf{n\tau_R^{n}}$ is, a.s. on $\{\mathsf{Z^R_n}\rightarrow \infty\}$, proportional to the limit of $\ZR/\mathsf{\tau_R^{n}}$ with proportionality constant $\beta/(1-\beta)$.

As we indicated at the Introduction, our aim in this paper is to apply the ABC methodology to obtain accurate approximations to the posterior distributions of the parameters of the model, that is, of $\alpha$, $\beta$, $\mathsf{m_R}$ and $\mathsf{m_r}$ and to verify that this methodology works adequately in all the possible situations given by the explained above relations between $\mathsf{m_r}$ and $(1-\beta)\mathsf{m_R}$, always on the coexistence set. To do that, previously, we must select the sample we are going to observe. We are interested in finding a sufficiently informative sampling scheme observing the minimum amount of information that leads us to obtain good estimates. Related to this question, as we also indicated at the Introduction, the authors published (see \cite{ggm2013a}) a study about the estimation of the main parameters of a Y-BBP (without considering mutations) based on a sample where only the total number of females and males (without knowing the genotype of the males) up to some generation $\mathsf{N}$ as well as
the different types of males only in the last generation $\mathsf{N}$ were assumed to be observed. Following these ideas, initially we set out in this paper the Bayesian estimation of the parameters of the Y-BBP with mutations based on that same sample.

\section{Approximate Bayesian Computation}\label{tres}

Let $\mathcal{FM}_\mathsf{N}$ denote the observed data until
generation $\mathsf{N}$ which is assumed that has been generated from a
model with parameter vector
$\theta=(\alpha,\beta,\mathsf{m_R},\mathsf{m_r})$. In particular
\begin{equation}\label{muestrainicial}
\mathcal{FM}_\mathsf{N}=\{\mathsf{FM}_0,\mathsf{FM}_1,...,\mathsf{FM}_{\mathsf{N-1}},\mathsf{FMRr}_{\mathsf{N}}\},\end{equation} where
$\mathsf{FM}_{\mathsf{n}}=(\F,\M)$, $\mathsf{n}=0,...,\mathsf{N-1}$, is the vector given by the total
number of females and males in generation
$\mathsf{n}$ and $\mathsf{FMRr}_{\mathsf{N}}=(\mathsf{F_{N}},\mathsf{M^R_N},\mathsf{M^r_N})$ is the vector given by the total
number of females and males of each genotype at last generation. Note that $\mathsf{FM}_0$ could be fixed -initial generation
at an experiment- or random -representing the first generation one
observes, non necessarily the initial fixed generation.
Henceforward, we shall focuss on the first interpretation. Moreover, we shall assume that $\mathsf{F_{N}}>0$, $\mathsf{M^R_N}>0$ and $\mathsf{M^r_N}>0$. Notice that this assumption implies that $\F>0$ and $\M>0$, for all $\mathsf{n}=1,...,\mathsf{N-1}$ and also implies that both genotypes have coexisted at least in the last generation.

The aim of Bayesian approach is to derive the posterior
distribution of the parameter vector,
$\theta|\mathcal{FM}_\mathsf{N}$. ABC methodology offers good approximations to the posterior
distributions of parameters for models which ha\-ve intractable
likelihoods but are easy to simulate.

The use of ABC ideas initially comes from the field of population
genetics (see \ \cite{beaumont}, \ \cite{pri} and \cite{tbgd1997}), although  these were quickly extended to a great variety
of scientific applications areas. The basic ideas are to simulate
a large number of data from a model depending on a parameter
vector that is drawn from a prior distribution and compare the
simulated data with the values from the observed sample. The aim
of the ABC methodology is to provide samples from a posterior-type
distribution (in the sense that it includes the sample
information) which is a good (enough) approximation of the
posterior distributions of the parameters of the model. Several
algorithms have been proposed in the literature to solve the
problem of how to choose this approximation, surveys on ABC
algorithms can be read in \cite{lintusaari}, \cite{marin} and \cite{sunnaker}.

These general ideas can be properly adapted to our model which is very easy to simulate given the parameter vector, some information about the initial
generation, as for example, the total number of females and males
of each type, and the family of probability distributions the reproduction laws belongs to. In our case, as we have a complete absence of knowledge on the reproduction laws of the model that has generated the observed data, we will assume, for simplicity, a parametric setting with Poisson distributions as reproduction laws. This distribution is frequently used as offspring distribution, see for example \cite{Bertoin08}, \cite{fg}, \cite{fkg}, \cite{modesleemam}, \cite{pakes} or \cite{Bl}. Another parametric reproduction law could also be considered without substantial changes in the estimates (see the sensitivity analysis showed in Section \ref{sensitivity}).

Moreover, in our case, it is not possible to calculate explicitly the
likelihood function, $f(\mathcal{FM}_\mathsf{N}|\theta)$, because the complete branching structure cannot be derived due to the fact that the total number of males of each genotype, the total number of $\mathsf{r}$-males stemming from $\mathsf{R}$-couples and the total number of each type of couple are not observed in each generation.

\subsection{Description of the algorithm}\label{sub3.1}

In our particular case, the proposed algorithm is the Tolerance
Rejection-ABC Algorithm which is an adaptation of that proposed in
\cite{pri} which works as follows. For a
Y-BBP with mutations, assuming observed the sample in
(\ref{muestrainicial}), it is easy to simulate for each specific
vector of parameters $\mathsf{\theta}$ (sampled from a prior
distribution $\pi(\theta)$) the entire
family tree up to the current $\mathsf{\mathsf{N}}$th generation
and to obtain the random vectors ($\FR$, $\MR$, $\MRr$, $\Fr$,
$\mathsf{M^{r\to r}_{n}}$, $\ZR$, $\Zr$), \
$\mathsf{n}=0,\ldots,\mathsf{N}$. Then,  using Equations
(\ref{eqfem}) and (\ref{eqmales}), can be obtained a simulated sample of
$(\F,\M)$,\ $\mathsf{n}=0,\ldots,\mathsf{N-1}$ and $(\mathsf{F_N},\mathsf{M^R_N},\mathsf{M^r_N})$,
renamed as
$$\mathcal{FM}_\mathsf{N}^{\mbox{\tiny{sim}}}=\{\mathsf{FM}_0^{\mbox{\tiny{sim}}},\mathsf{FM}_1^{\mbox{\tiny{sim}}}
,...,\mathsf{FM}_{\mathsf{N-1}}^{\mbox{\tiny{sim}}},\mathsf{FMRr}_{\mathsf{N}}^{\mbox{\tiny{sim}}}\}.$$

Notice, $\mathcal{FM}_\mathsf{N}^{\mbox{\tiny{sim}}}$ depends on  $\mathcal{FM}_\mathsf{N}$ only through $\mathsf{FM}_0$.
Actually, $\mathsf{F_0^{\mbox{\tiny{sim}}}}=\mathsf{F_0}$ and the vector $(\mathsf{M^R_0}^{\mbox{\tiny{sim}}},\mathsf{M^r_0}^{\mbox{\tiny{sim}}})$
is simulated from the uniform distribution, subject to the constraint $\mathsf{M^R_0}^{\mbox{\tiny{sim}}}+\mathsf{M^r_0}^{\mbox{\tiny{sim}}}=\mathsf{M_0}$.
Moreover, we consider only paths simulated by the algorithm where both alleles have coexisted in the last generation, i.e., where $\mathsf{F_N^{\mbox{\tiny{sim}}}}>0$,  $\mathsf{M^R_N}^{\mbox{\tiny{sim}}}>0$ and $\mathsf{M^r_N}^{\mbox{\tiny{sim}}}>0$, as it occurred in the observed sample $\mathcal{FM}_\mathsf{N}$.

Now, for a
given $\varepsilon>0$, known as tolerance level, and a distance,
$\rho(\cdot,\cdot)$, the algorithm compares (in terms of metric)
the simulated paths, $\mathcal{
FM}_\mathsf{N}^{\mbox{\tiny{sim}}}$, with the observed sample $\mathcal{
FM}_\mathsf{N}$.
This allows us to obtain an approximation of $\theta\mid
\mathcal{FM}_\mathsf{N}$ by the distribution
$$\theta|\rho(\mathcal{
FM}_\mathsf{N}^{\mbox{\tiny{sim}}},\mathcal {FM}_\mathsf{N})\leq
\varepsilon,$$ using a small enough $\varepsilon$. In our case, we shall use a small enough quantile of the sample of the distances as it is usual in ABC studies (see, for example, \cite{marin}).

To quantify the distance between $\mathcal{FM}_\mathsf{N}^{\mbox{\tiny{sim}}}$ and
$\mathcal{FM}_\mathsf{N}$ we use
\begin{eqnarray*}
\rho(\mathcal{FM}_\mathsf{N}^{\mbox{\tiny{sim}}},\mathcal{FM}_\mathsf{N})=&&\\
&&\hspace*{-2.6cm}\left(\displaystyle{\sum_{\mathsf{n}=1}^\mathsf{N}}\left(\frac{\F^{\mbox{\tiny{sim}}}}{\F}-
\frac{\F}{\F^{\mbox{\tiny{sim}}}}\right)^2+\displaystyle{\sum_{\mathsf{n}=1}^\mathsf{N-1}}\left(\frac{\M^{\mbox{\tiny{sim}}}}{\M}- \frac{\M}{\M^{\mbox{\tiny{sim}}}}\right)^2\right.\\
&&\hspace*{-3cm}+\left.\left(\frac{\MRN^{\mbox{\tiny{sim}}}}{\MRN}- \frac{\MRN}{\MRN^{\mbox{\tiny{sim}}}}\right)^2+\left(\frac{\MrN^{\mbox{\tiny{sim}}}}{\MrN}-
\frac{\MrN}{\MrN^{\mbox{\tiny{sim}}}}\right)^2\right)^{1/2}
\end{eqnarray*}

Notice that we have re-scaled each coordinate of the vectors since their magnitudes can be extremely different, depending on generation, sex and genotype (see \cite{lintusaari} and \cite{pri}).

Then, the Tolerance Rejection-ABC Algorithm is formulated as,

\begin{center}

\textsc{Tolerance Rejection-ABC Algorithm}

\vspace*{0.35cm}

\parbox{10cm}{
\tt \noindent For $i=1$ to $m$ do

\noindent repeat

\noindent generate $(\alpha^\mathsf{sim},\gamma^\mathsf{sim},\phi^\mathsf{sim})\!\sim \! U(0,1)\!\times \!U(0,1)\! \times \! U(0,1)$

\noindent generate $\beta^\mathsf{sim}=0$ with probability $\gamma^\mathsf{sim}$ and

\hspace*{1.5cm}    $\beta^\mathsf{sim}\sim \pi(\mathsf{\beta})$ with probability $1-\gamma^\mathsf{sim}$

\noindent generate $\mathsf{m_r^{sim}}=0$ with probability $\phi^\mathsf{sim}$ and

\hspace*{1.5cm}    $\mathsf{m_r^{sim}}\sim \pi(\mathsf{m_r})$ with probability $1-\phi^\mathsf{sim}$

\noindent generate $\mathsf{m_R^{sim}}\sim \pi(\mathsf{m_R})$

\noindent let
$\widetilde{\theta}=(\alpha^\mathsf{sim},\beta^\mathsf{sim},\mathsf{m_R^{sim}},\mathsf{m_r^{sim}})$

\noindent simulate $\mathcal{FM}_{\mathsf{N}}^{\mathsf{sim}}$ from the likelihood
$f(\mathcal{FM}_\mathsf{N}|\widetilde{\theta})$

\noindent until
$\rho(\mathcal{FM}_\mathsf{N}^{\mathsf{sim}},\mathcal{FM}_\mathsf{N}) \leq
\varepsilon$,

\noindent set $\theta^{(i)}=\widetilde{\theta}$

\noindent end for }
\end{center}

Note that, we generate the parameter $\alpha^\mathsf{sim}$ from a uniform distribution on $(0,1)$ and the parameter $\mathsf{m_R^{sim}}$ from a generic prior distribution $\pi(\mathsf{m_R})$ on $(0,\infty)$. This is consistent with the fact that $\FN>0$ and $\MRN>0$.
On the other hand, taking into account that in the model, $\mathsf{r}-$allele can mean an allele different from $\R$, $\beta^\mathsf{sim}$ could be null. Moreover,  $\mathsf{m_r^{sim}}$ could also be null even being $\MrN>0$ (in this case $\beta^\mathsf{sim}>0$, see (\ref{eqmales})). Therefore, we generate the parameters $\beta^\mathsf{sim}$ and $\mathsf{m_r^{sim}}$ from prior distributions which are mixture of distributions: one degenerated at 0 (in order to consider the possibility that $\beta^\mathsf{sim}$ and $\mathsf{m_r^{sim}}$ takes exactly the value 0)
and the other one $\pi(\beta)$ on $(0,1)$ and $\pi(\mathsf{m_r})$ on $(0,\infty)$, with weights given by ($\gamma^\mathsf{sim}$, $1-\gamma^\mathsf{sim}$) and ($\phi^\mathsf{sim}$,$1-\phi^\mathsf{sim}$), respectively.
Since we do not have information about the possible value of these parameters, we consider $\gamma^\mathsf{sim}$ and $\phi^\mathsf{sim}$ following a uniform distribution on $(0,1)$.

\section{A simulated example based on the observed sample $\mathcal{FM_\mathsf{N}}$}

Now, the previous algorithm is implemented using as observed data a sample which has been obtained by simulation. We analyze first the case where the relation between the parameters is $\mathsf{m_r}\geq(1-\beta)\mathsf{m_R}$ although it is worth to remind here that we are searching for a general method which works independently of the relation between the parameters.

\subsection{Case $\mathsf{m_r}\geq(1-\beta)\mathsf{m_R}$}\label{ejemplo1}

Our objective is to approximate the posterior distribution
$\theta|\mathcal{FM}_\mathsf{N}$, where $\mathcal{FM}_\mathsf{N}$ is
an observed sample which has been
simulated from a Y-BBP with mutation with parameter vector
$\theta\!=\!(\alpha,\beta,\mathsf{m_R},\mathsf{m_r})\!=\!(0.46,0.005,3.2,4)$
(notice that with those values the relation
$\mathsf{m_r}\geq(1-\beta)\mathsf{m_R}$ is satisfied)
and initial vector ($\mathsf{F}_0$, $\mathsf{M_0^R}$,
$\mathsf{M_0^r}$)= ($10,5,5$). For such a model with
this set of parameters and initial values, we proved in
\cite{ggm2012a} that there exists a positive probability of
survival of both genotypes.

\begin{table*}[!hbt]
\begin{center}
\caption{Reproduction laws for both genotypes, with $p_k$ the probability that a couple generates $k$ individuals, with $k\in\{0,\ldots,7\}$.}
\label{E0}
\begin{tabular}{lllllllll}
\hline\noalign{\smallskip}
 & $p_0$ & $p_1$ & $p_2$ & $p_3$ & $p_4$ & $p_5$ & $p_6$ & $p_7$ \\
\noalign{\smallskip}\hline\noalign{\smallskip}
$\mathsf{R}$-genotype &        0.0139& 0.0819& 0.2069& 0.2904& 0.2445& 0.1236& 0.0347& 0.0041\\
  $\mathsf{r}$-genotype   &    0.0027& 0.0248& 0.0991& 0.2203& 0.2938& 0.2350& 0.1044& 0.0199\\
\noalign{\smallskip}\hline
\end{tabular}
\end{center}
\end{table*}

We simulate 15 generations of this Y-BBP with mutations assuming
that reproduction laws of both genotypes follow the non-parametric offspring distributions with finite support given in Table \ref{E0}, with means $\mathsf{m_R}=3.2$ and
$\mathsf{m_r}=4$, respectively. The observed data can be seen in Table
\ref{E1++} and are denoted by $\mathcal{FM}_{15}$.

\begin{table*}[!hbt]
\begin{center}
\caption{The observed sample $\mathcal{FM}_{15}$ for the case $\mathsf{m_r}\geq(1-\beta)\mathsf{m_R}$, with $(\mathsf{M^R_{15}},\mathsf{M^r_{15}})=(1043,45850)$. This sample has been generated from the parameter vector $\theta=(\alpha,\beta,\mathsf{m_R},\mathsf{m_r})=(0.46,0.005,3.2,4)$.}
\label{E1++}
\begin{tabular}{llllllllllllllll}
\hline\noalign{\smallskip}
$\mathsf{n}$ & 1 & 2 & 3 & 4 & 5 & 6 & 7 & 8 & 9 & 10 & 11 & 12 & 13 & 14 & 15\\
\noalign{\smallskip}\hline\noalign{\smallskip}
$\mathsf{F_n}$ & 16&   21&   33&   53&  112&  188&  342&  609& 1112&  1985&  3563&  6547& 11980& 21904& 40101 \\
  $\mathsf{M_n}$   &    23 &  36&   46&   75&  103&  215&  397&  731& 1275&  2340&  4233&  7716& 13983& 25441& 46893  \\
\noalign{\smallskip}\hline
\end{tabular}
\end{center}
\end{table*}

We apply the Tolerance Rejection-ABC Algorithm generating the parameter vector
assuming independent non-informative prior distributions. In
particular, for $\alpha^{\mathsf{sim}}$ and $\beta^{\mathsf{sim}}$ (when it is positive) uniform distributions in the interval $(0,1)$ and for $\mathsf{m_R^{sim}}$ and $\mathsf{m_r^{sim}}$ (when it is positive), uniform distributions in the interval $(0,10)$. We have chosen, obviously, 0 as the minimum value for the support of the latter uniform distributions and 10 as the maximum value because we consider that number high enough for the number of offspring of many animal species although this
number could be adapted to any specific situation. After that, we simulate Y-BBPs with mutations until generation 15 using, as $\R$ and $\mathsf{r}$ reproduction laws, Poisson distributions with parameters, respectively, $\mathsf{m_R^{sim}}$ and $\mathsf{m_r^{sim}}$ (recall we use this generic type of distribution for the offspring laws because we know nothing about the true reproduction laws). We generate a pool of 50 millions of simulated paths. To compare the observed sample and the simulated ones we consider a tolerance level equal to the 0.00002 quantile of the sample
of the distances, so that the size of ABC samples
to approximate the posterior distribution $\theta|\mathcal{FM}_\mathsf{15}$ is 1000.

\begin{figure*}[!hbt]
\centering \scalebox{0.24}{\includegraphics{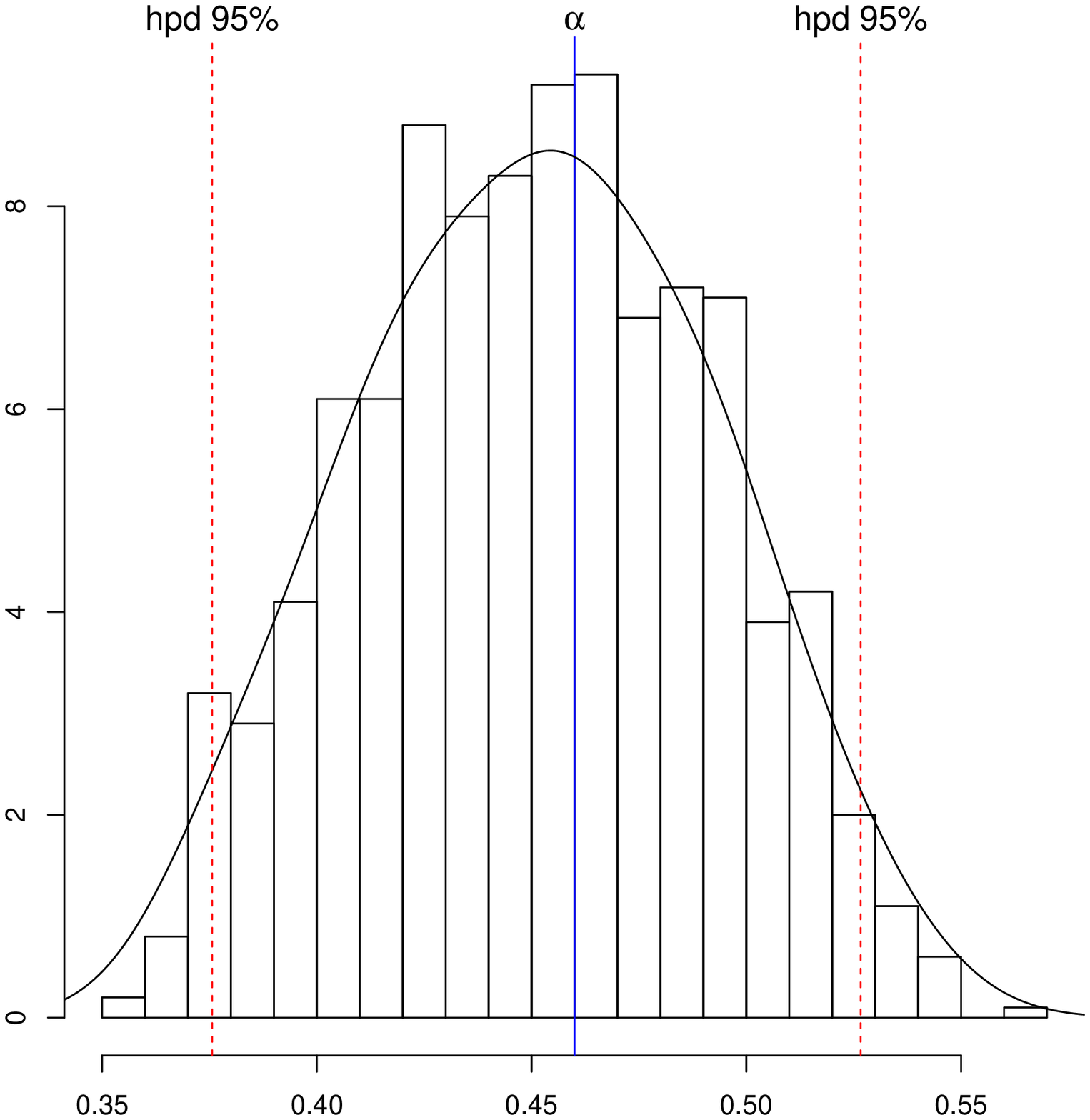}
\includegraphics{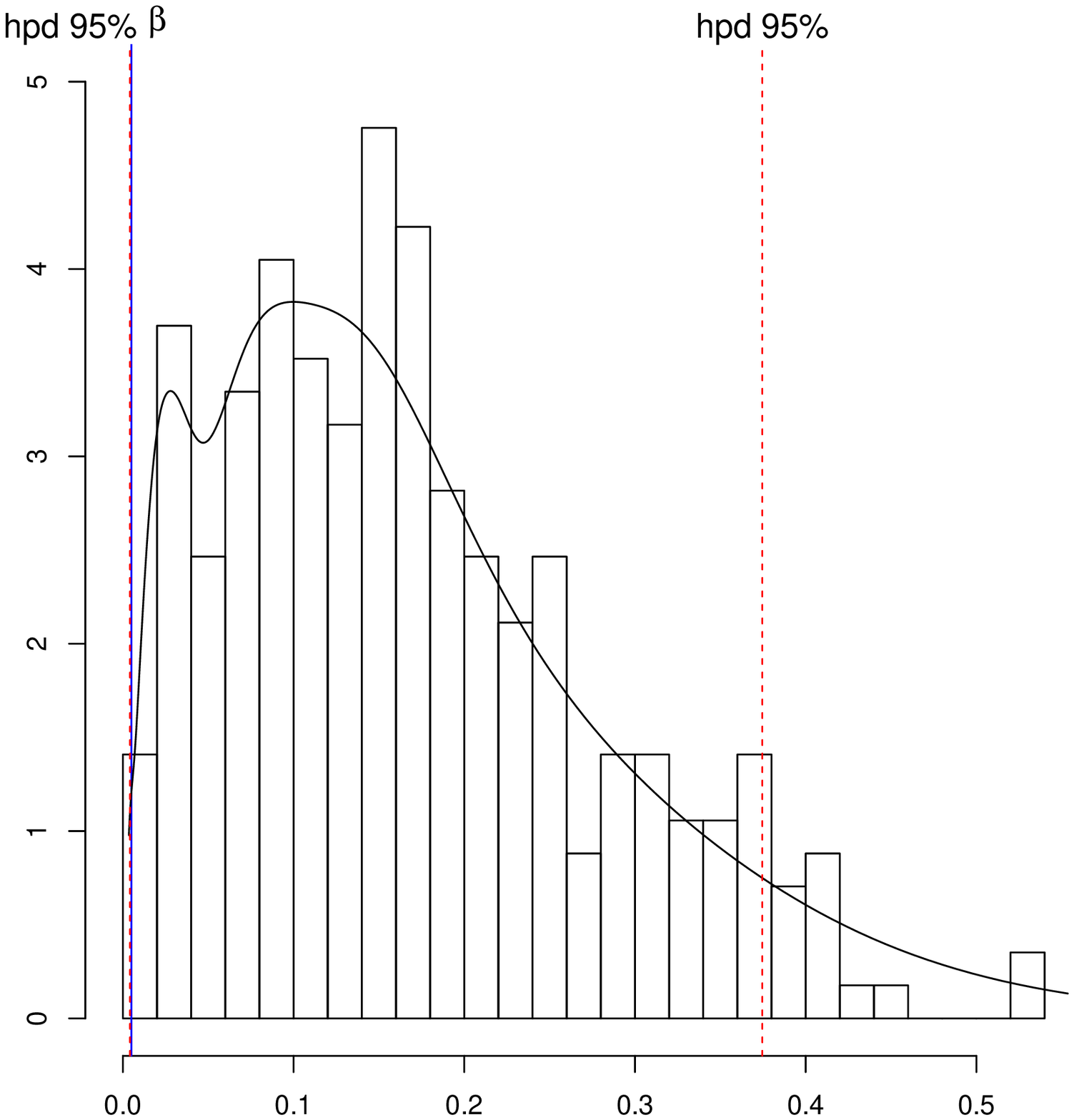}\includegraphics{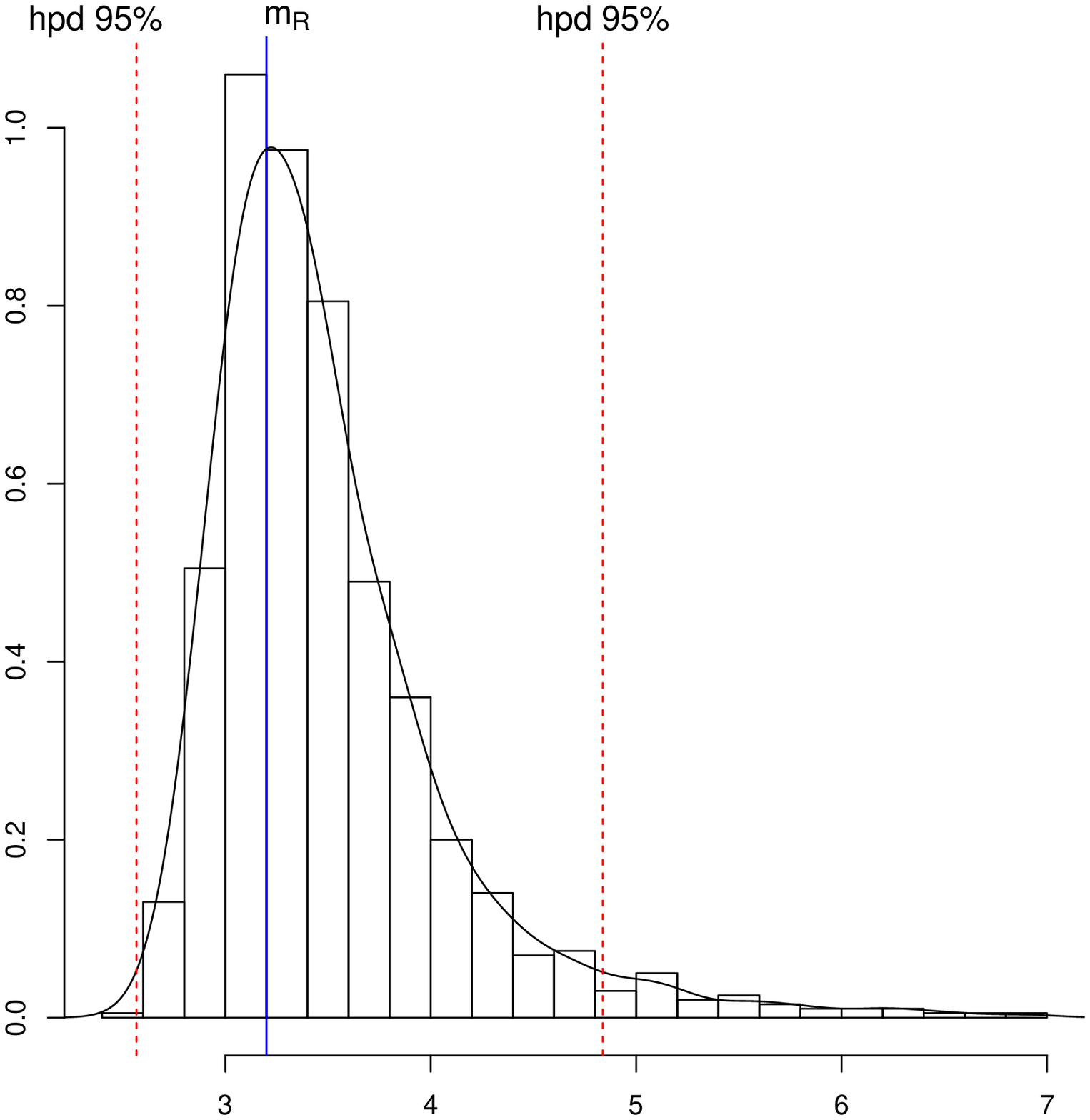}
\includegraphics{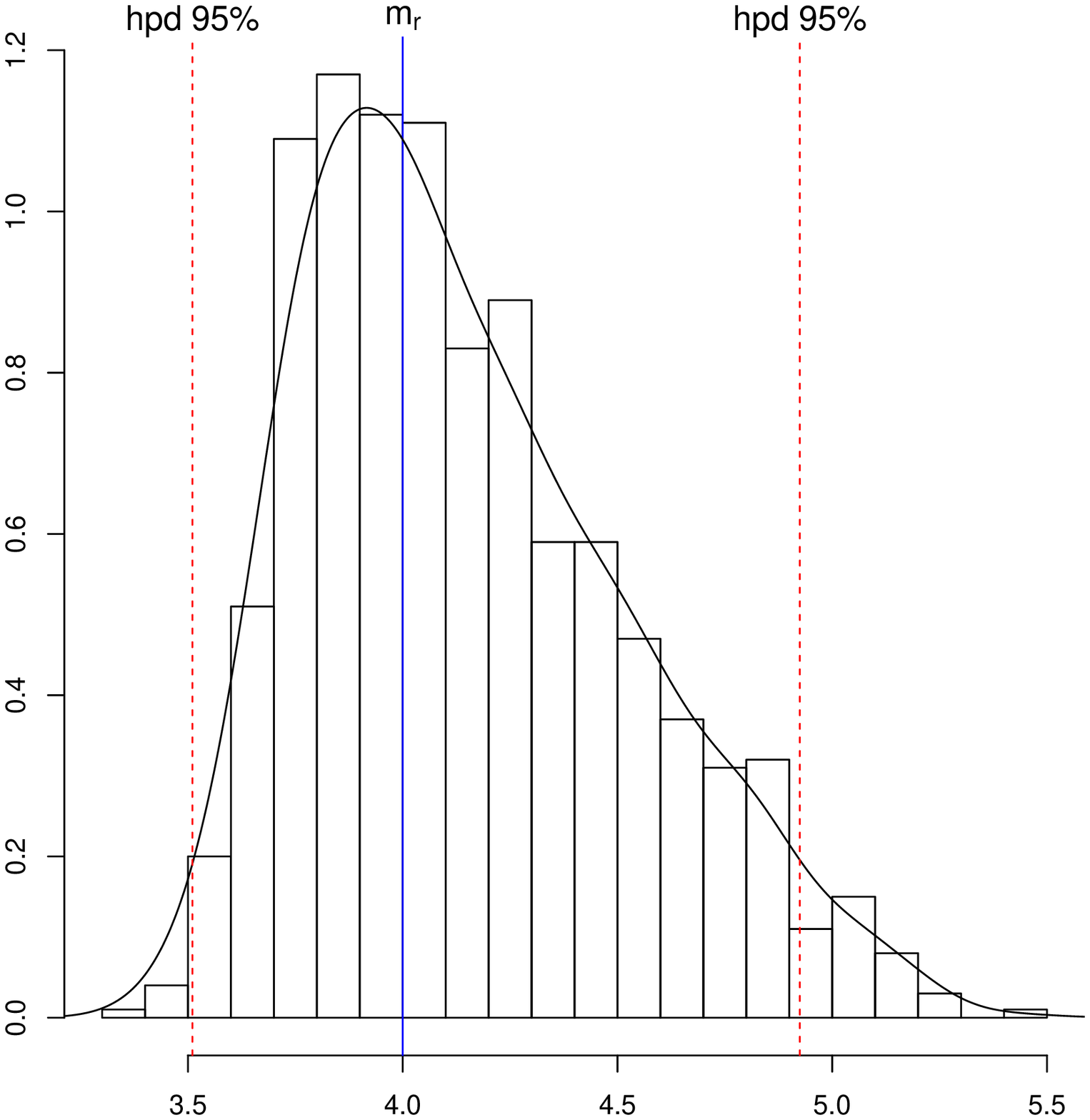}}
\caption{Approximate posterior densities, with 95\%
HPD sets, of the parameters $\alpha$, $\beta$ (in this case only considering paths where $\beta^{\mathsf{sim}}>0$), $\mathsf{m_R}$ and $\mathsf{m_r}$, respectively, given $\mathcal{FM}_{\mathsf{15}}$ in the case $\mathsf{m_r\geq(1-\beta)m_R}$. Vertical solid lines represent the true value of the parameters.} \label{comden}
\end{figure*}

In Figure \ref{comden}, we present the approximate posterior
distribution of every parameter, that
is,
$$\mathsf{\delta}|\rho(
\mathcal{FM}_{15}^{\mbox{\tiny{sim}}},\mathcal{FM}_{15})\leq \varepsilon,$$ with
$\delta$ equal to $\alpha$, $\beta$ (in this case, only paths where $\beta^{\mathsf{sim}}>0$ have been considered), $\mathsf{m_R}$ and
$\mathsf{m_r}$, the corresponding in every case, together with the
true value of the parameters (vertical solid line) and 95\%
HPD sets (vertical dotted lines). We can appreciate first that the approximate posterior distribution for $\alpha$ is very accurate. Actually, this happens in
every example we present in the paper and it is due to the
fact that the quotient between the total number of females and
the total number of individuals, which are observed, converges precisely to $\alpha$ when the number of generations tends to infinity (see \cite{cgutierrez}), therefore, the similitude between the chosen simulated paths and the observed sample
makes the estimation for $\alpha$ good enough.

\begin{figure*}[!hbt]
\centering \scalebox{0.25}{\includegraphics{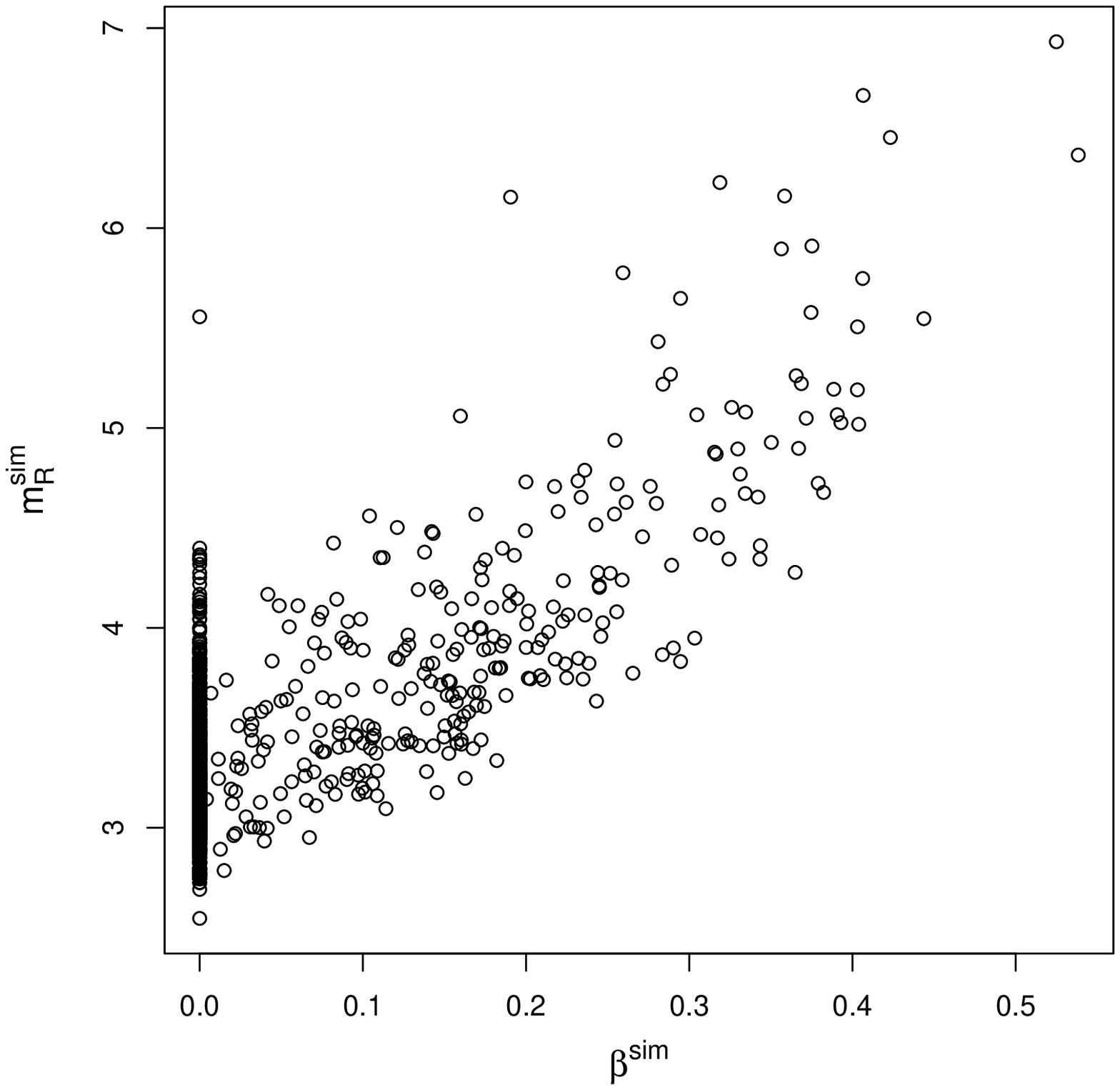}
\includegraphics{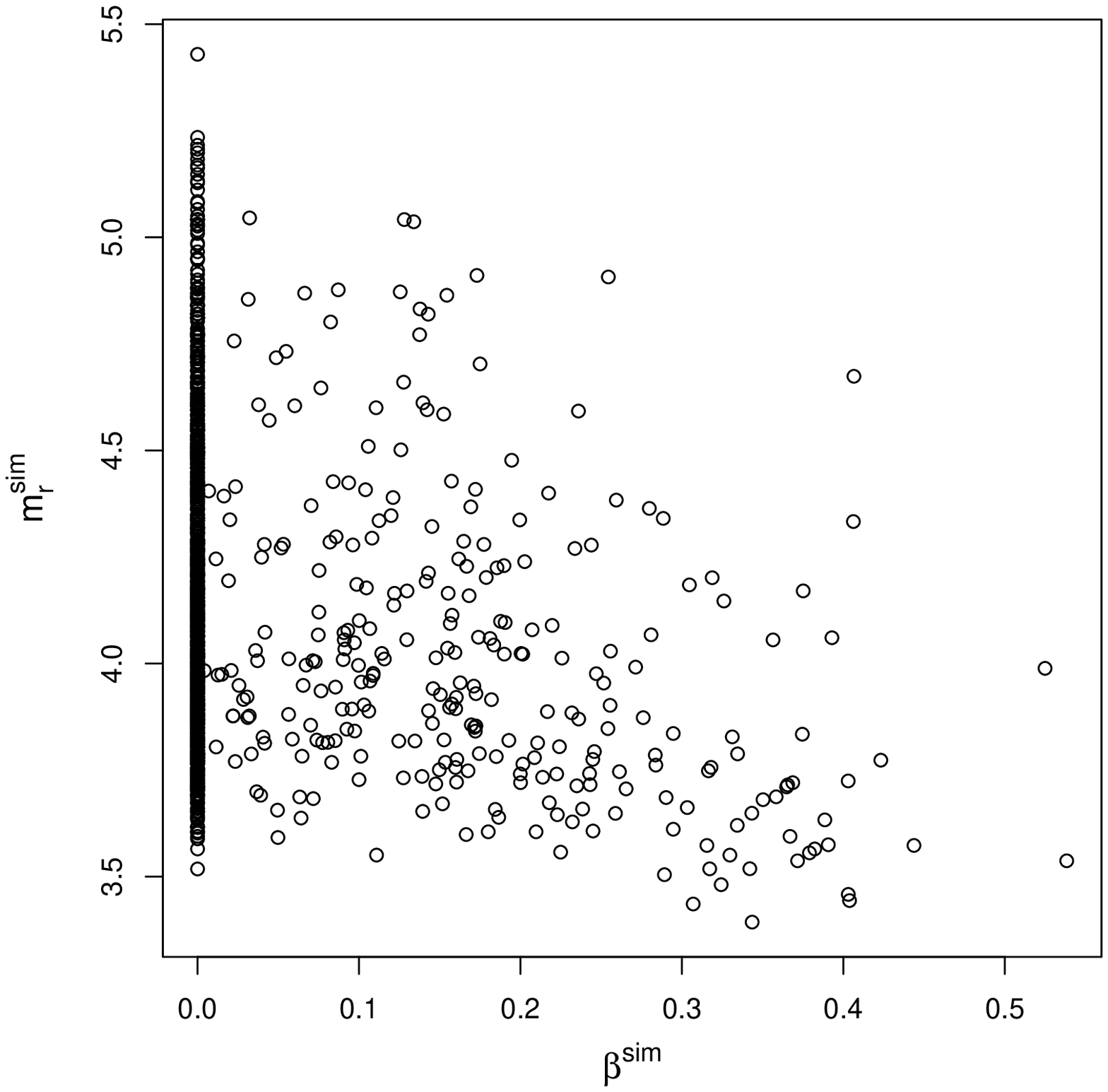}\includegraphics{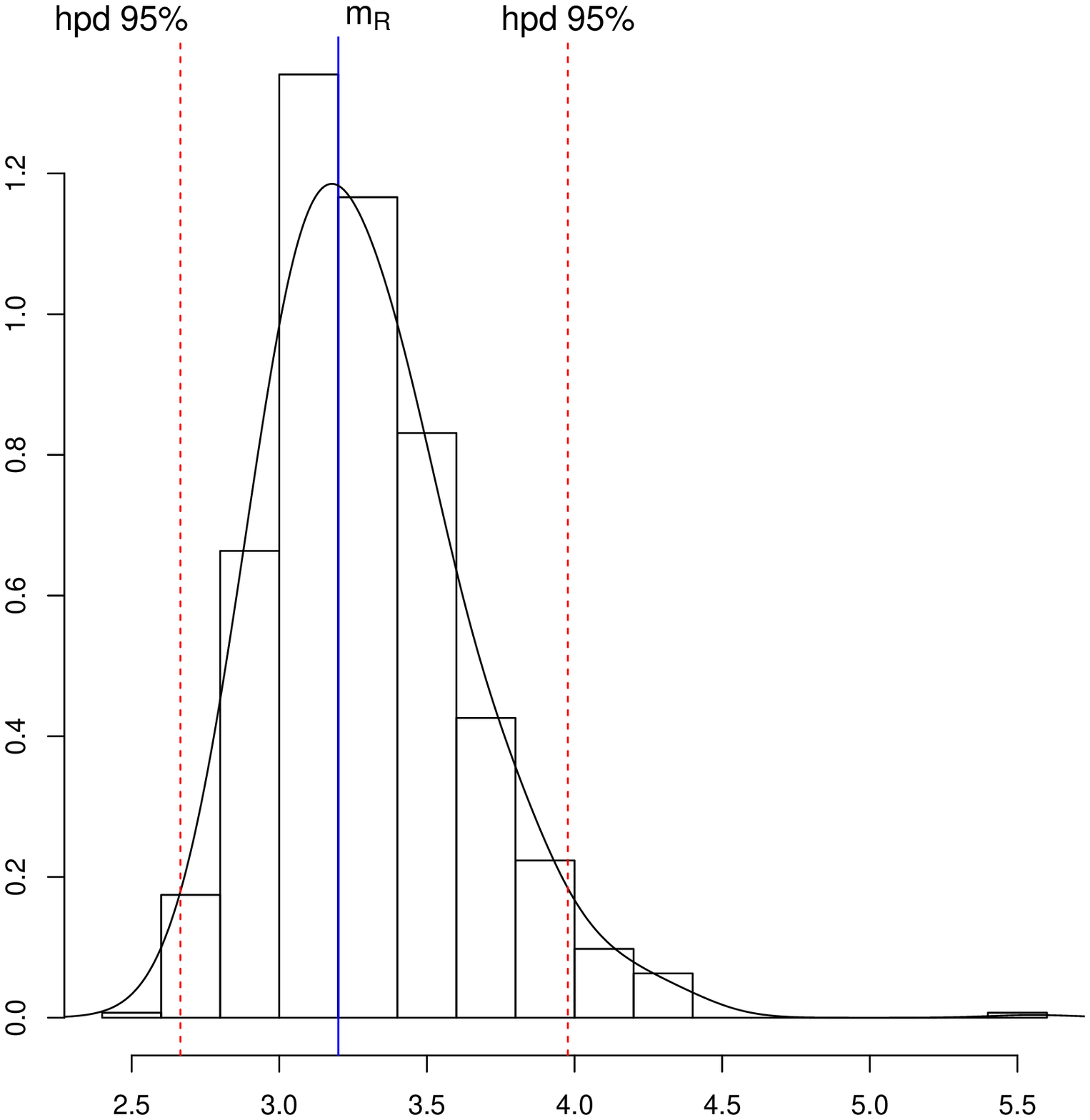}}
\caption{Scatter plots showing the relation between $\mathsf{\beta}$ and $\mathsf{m_R}$ and $\mathsf{m_r}$, respectively, and approximate posterior density, with 95\%
HPD sets, of $\mathsf{m_R}$ using paths with $\mathsf{\beta^{\mathsf{sim}}}=0$, given $\mathcal{FM}_{15}$.} \label{betarelation}
\end{figure*}

However, we can observe in Figure \ref{comden} that the estimation of
the posterior distribution of the parameter $\beta$ is not very
accurate because $P(\beta=0|\mathcal{FM}_{15})$ is very high, estimated by 0.716, despite the real value of $\beta$ is strictly greater than 0. Note at this point that, to estimate $\beta$ is a difficult task. First, because in general, its value, although positive, is very small in real situations (0.005 in
our example), close to zero (which represents the non-mutation). Secondly, due to the fact that from the total number of males with $\mathsf{r}-$allele, $\mathsf{M_n^r}$, is not possible to know, without some additional information, how many of those come from
mutations, $\mathsf{M_n^{R\to r}}$. On the other hand, the corresponding estimates of $\mathsf{m_R}|\mathcal{FM}_{15}$ and $\mathsf{m_r}|\mathcal{FM}_{15}$  are enough accurate, with the last one better than the first. This is due to the fact that we are in the case in which the $\mathsf{r}-$allele is the dominant one
($\mathsf{m_r}\geq(1-\beta)\mathsf{m_R}$) and therefore one has more
information about males with the mutant allele, in spite of the noise produced by the non-observed variable $\mathsf{M_n^{R\to r}}$.
Anyway, if one goes more in deep analyzing these data, one has to consider that both approximations of the posterior distributions (of $\mathsf{m_R}$ and $\mathsf{m_r}$) are related with $\mathsf{\beta}$, as it is shown in Figure \ref{betarelation}, and therefore the estimates of these parameters inherit in some sense the inaccuracy of $\beta$.
Actually, if we consider only the simulated paths where $\beta^{\mathsf{sim}}=0$, the kernel density estimate of $\mathsf{m_R}$ is really accurate (of course, since the true value of $\beta$ is close to zero).

To give a measure of the accuracy of the method for the different
parameters, we consider the relative mean square error (RMSE), which
was also proposed in \cite{beaumont} and \cite{ggmp2013b},
calculated by
$$\frac{1}n\sum_{k=1}^{n}\frac{(\delta^{\mathsf{sim}}_k-\delta)^2}{\delta^2},$$ with
$n=1000$, $\delta$ the true value of $\alpha$, $\beta$,
$\mathsf{m_R}$ or $\mathsf{m_r}$, the corresponding in each case,
and $\delta^{\mathsf{sim}}_k$ the corresponding value of $\delta$
on the $k$th simulated path chosen by the method.

\begin{table*}
\caption{RMSE for the estimates of $\mathsf{\alpha}$,
$\mathsf{\beta}$, $\mathsf{m_R}$ and $\mathsf{m_r}$ given by the Tolerance Rejection-ABC Algorithm when the sample $\mathcal{FM}_{15}$ is observed.} \label{MSE1}
\begin{center}
\begin{tabular}{l|c|c|c|c|}
 & $\alpha$ & $\beta$& $\mathsf{m_R}$ & $\mathsf{m_r}$ \\
\hline
Considering all simulated paths &  0.0080 & 0.0443 &   0.0408  &  0.0100 \\
Considering only simulated paths where $\beta^{\mathsf{sim}}=0$ &  0.0081 & &   0.0117  &  0.0110 \\
Considering only simulated paths where $\beta^{\mathsf{sim}}>0$ & 0.0077 &1557.8&  0.1141&  0.0076\\
\hline
\end{tabular}
\end{center}
\end{table*}

In particular, Table \ref{MSE1} shows the RMSE of the estimates of $\alpha$,
$\beta$, $\mathsf{m_R}$ and $\mathsf{m_r}$ given by the Tolerance
Rejection-ABC Algorithm when the sample $\mathcal{FM}_{15}$ is observed and considering
all chosen simulated paths (i.e. those simulated paths such that $\rho(\mathcal{FM}_{15}^{\mathsf{sim}},\mathcal{F\!M}_{15})\!\leq \!\varepsilon$), all chosen simulated paths where $\beta^{\mathsf{sim}}=0$ and all chosen simulated paths where $\beta^{\mathsf{sim}}>0$. One can appreciate that, in general, the RMSE for $\alpha$, $\mathsf{m_R}$ and $\mathsf{m_r}$ are very similar in all cases and very close to 0.

However, the RMSE for $\beta$ when only simulated paths where $\beta^{\mathsf{sim}}>0$ are considered, takes a high value, considerably greater than the value when all simulated paths are considered, even being the first one the real situation. This is due to the fact that $P(\beta=0|\mathcal{FM}_{15})$ is very high and that the true value of $\beta$ is very close to 0.

\begin{figure*}[!hbt]
\centering \scalebox{0.25}{\includegraphics{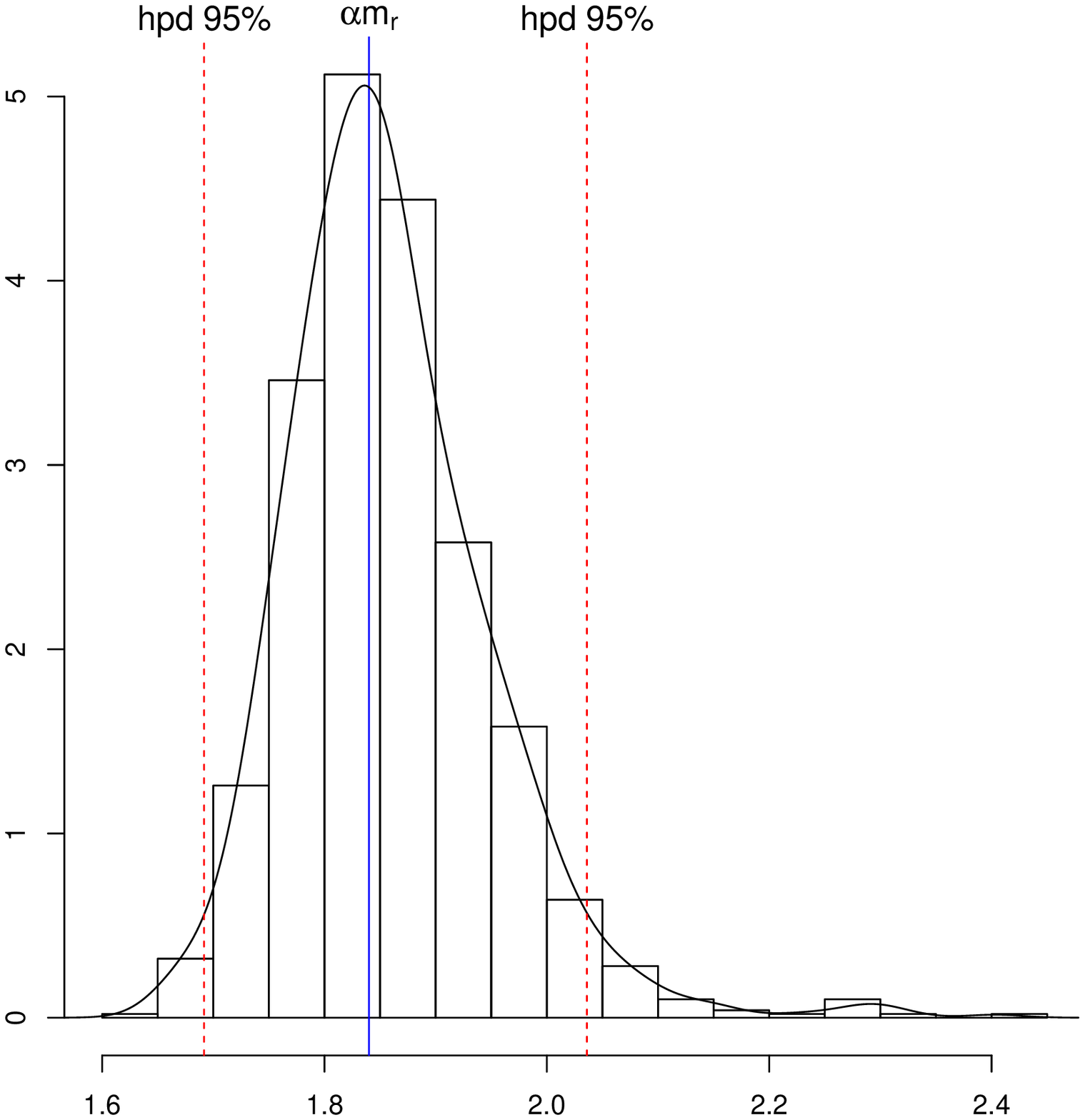}
\includegraphics{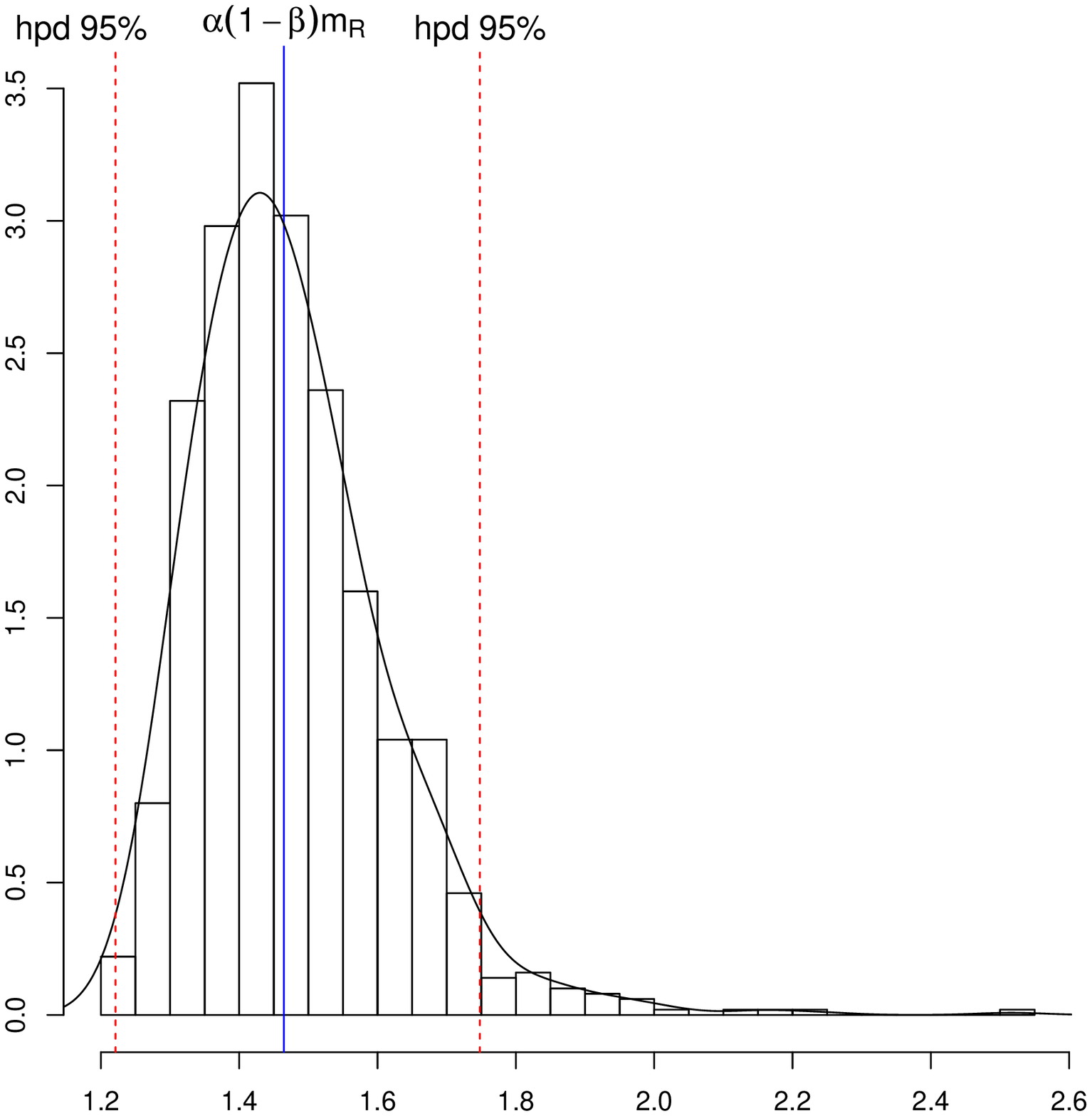}}
\caption{Approximate posterior densities, with 95\%
HPD sets, of the rates of growth of
the mutant allele and $\mathsf{R}-$allele, given $\mathcal{FM}_{15}$
in the case $\mathsf{m_r}\geq(1-\beta)\mathsf{m_R}$. Vertical solid lines represent the
true value of the rates of growth.} \label{ratios}
\end{figure*}

Anyway, note that although the methodology cannot
provide an adequate approximate posterior distribution of the
parameter $\beta$ and consequently of the parameters $\mathsf{m_R}$ and $\mathsf{m_r}$ either, it can
provide very accurate approximate posterior distributions of the
rates of growth of both alleles, see Figure \ref{ratios}, since the total number of males of each genotype is observed. In particular, as it was indicated in Section 2, when $\alpha<0.5$ and $\mathsf{m_r>(1-\beta)m_R}$ (as it is the
case of our example), on the set of coexistence of both alleles, the
rate of growth of the mutant allele is equal to
$\alpha\mathsf{m_r}$, while the rate of growth of the
$\mathsf{R}-$allele is equal to $\alpha(1-\beta)\mathsf{m_R}$.

Notice here that, although we do not know, a priori, whether $\mathcal{FM}_{\mathsf{N}}$ will provide or not accurate estimates of the parameters of the model in other cases different from $\mathsf{m_r}\geq (1-\beta)\mathsf{m_R}$, we are looking for a unified estimation procedure whose behavior does not depend on the parameters relation. This is why in the next section we will modify the previous sample scheme including additional information. This new sample scheme will be use in the rest of the paper.

\section{Introducing additional information: a new sample scheme}

Up to now, we have used the ABC algorithm to estimate the main
parameters of the model given $\mathcal{FM}_{\mathsf{N}}$. However, we have seen in the example given in Section \ref{ejemplo1} that
the estimate of the parameter $\beta$ is not very accurate,
having the approximate posterior distribution huge variability. In the example we have observed an atom at zero of size 0.716 when actually the true parameter is really small but not null.

Therefore, it seems reasonable to think that it is necessary to get some information about the number of mutant alleles stemming from $\mathsf{R}-$fathers, at least in
some generation, in order to obtain a more accurate approximation of the posterior distribution of $\beta$. In particular, we introduce this kind of information for the last generation. Hence, we assume from now on that the available
sample consists of the sample given by (\ref{muestrainicial}) as well as the total number of $\mathsf{r}-$males stemming from $\mathsf{R}-$fathers in the
last generation, that is, $\mathsf{M^{R\to r}_N}$. Moreover, as $\mathsf{M^r_N}$ is known, the total number of $\mathsf{r}-$males stemming from $\mathsf{r}-$fathers in the last generation, $\mathsf{M^{r\to r}_N}$, is also derived (see (\ref{eqmales})). Notice that, to obtain $\mathsf{M^{R\to r}_N}$ and $\mathsf{M^{r\to r}_N}$, it would be necessary to know who is every $\mathsf{r}-$male's father of the generation $\mathsf{N}$.
Therefore, it is plausible to assume that $\mathsf{M^R_{N-1}}$ and $\mathsf{M^r_{N-1}}$ are also observed, including males whom do not produce descendants. From now on, we denote this sample as $\overline{\mathcal{FM}}_\mathsf{N}$. Therefore,  $$\overline{\mathcal{FM}}_\mathsf{N}=\{\mathcal{FM}_\mathsf{N},\mathsf{M^R_{N-1}},\mathsf{M^r_{N-1}},\mathsf{M^{R\to r}_N},\mathsf{M^{r\to r}_N}\}.$$

\section{A series of simulated examples based on the observed sample $\overline{\mathcal{FM}}_\mathsf{N}$}

In the following subsections, we will illustrate, by means of simulated examples, how the Tolerance Rejection-ABC Algorithm works to approximate the posterior distribution $\theta|\overline{\mathcal{FM}}_\mathsf{N}$. We will consider different situations depending on whether some variables of the sample $\overline{\mathcal{FM}}_\mathsf{N}$ are positive or null.

\subsection{Observing $\mathsf{M^R_N}>0$, $\mathsf{M^{R\to r}_N}>0$ and $\mathsf{M^{r\to r}_N}>0$}\label{6.1}

We first consider the situation in which $\mathsf{M^R_N}\!>\!0$, $\mathsf{M^{R\to r}_N}\!>\!0$ and $\mathsf{M^{r\to r}_N}>0$. This implies that $\mathsf{M^R_{N-1}}>0$ and $\mathsf{M^r_{N-1}}>0$.
Moreover, this assumption also implies that $\beta$ and $\mathsf{m_r}$ are strictly positive and then their posterior distributions are not concentrated at zero value, which simplifies the Tolerance Rejection-ABC Algorithm described in Subsection \ref{sub3.1} because only simulated paths where $\beta^{\mathsf{sim}}>0$ and $\mathsf{m_r^{sim}}>0$ will be considered. On the other hand, the metric is slight more complex, including the new observed variables $\mathsf{M^R_{N-1}}$, $\mathsf{M^r_{N-1}}$, $\mathsf{M^R_N}$, $\mathsf{M^{R\to r}_N}$ and $\mathsf{M^{r\to r}_N}$ in the same way as previously. In particular, the distance between the simulated path, $\overline{\mathcal{FM}}_\mathsf{N}^{\mbox{\tiny{sim}}}$, and the observed data, $\overline{\mathcal{FM}}_\mathsf{N}$, is defined as
\begin{eqnarray*}
\rho^*(\overline{\mathcal{FM}}_\mathsf{N}^{\mbox{\tiny{sim}}},\overline{\mathcal{FM}}_\mathsf{N})=&&\\
&&\hspace*{-2.65cm}\left(\displaystyle{\sum_{\mathsf{n}=1}^\mathsf{N}}\left(\frac{\F^{\mbox{\tiny{sim}}}}{\F}-
\frac{\F}{\F^{\mbox{\tiny{sim}}}}\right)^2+\displaystyle{\sum_{\mathsf{n}=1}^\mathsf{N-2}}\left(\frac{\M^{\mbox{\tiny{sim}}}}{\M}- \frac{\M}{\M^{\mbox{\tiny{sim}}}}\right)^2\right.\\
&&\hspace*{-3cm}+\left(\frac{\MRNN^{\mbox{\tiny{sim}}}}{\MRNN}- \frac{\MRNN}{\MRNN^{\mbox{\tiny{sim}}}}\right)^2+\left(\frac{\MrNN^{\mbox{\tiny{sim}}}}{\MrNN}-
\frac{\MrNN}{\MrNN^{\mbox{\tiny{sim}}}}\right)^2\\
&&\hspace*{-3cm}+\left(\frac{\MRN^{\mbox{\tiny{sim}}}}{\MRN}- \frac{\MRN}{\MRN^{\mbox{\tiny{sim}}}}\right)^2+\left(\frac{\MRrN^{\mbox{\tiny{sim}}}}{\MRrN}-
\frac{\MRrN}{\MRrN^{\mbox{\tiny{sim}}}}\right)^2\\
&&\hspace*{-3cm}+\left.\left(\frac{\MrrN^{\mbox{\tiny{sim}}}}{\MrrN}- \frac{\MrrN}{\MrrN^{\mbox{\tiny{sim}}}}\right)^2\right)^{1/2}\\
\end{eqnarray*}

\subsubsection{Case $\mathsf{m_r}\geq (1-\beta)\mathsf{m_R}$}

To illustrate how to approximate the posterior distribution
$\theta|\overline{\mathcal{FM}}_\mathsf{N}$, first we study again the case $\mathsf{m_r}\geq(1-\beta)\mathsf{m_R}$,
considering the same observed sample given in Table \ref{E1++} and also
assuming that now it is observed that $\mathsf{M^{R\to r}_{15}}=6$ (i.e. from 45850 males with $\mathsf{r}-$allele in generation 15, 6 of them come from mutations), that $\mathsf{M^{r\to r}_{15}}=45844$, that $\mathsf{M^R_{14}}=754$ and that $\mathsf{M^r_{14}}=24687$.
With this new
information, we apply the Tolerance Rejection-ABC Algorithm using the
metric $\rho^*(\cdot ,\cdot)$.

\begin{figure*}[!hbt]
\centering \scalebox{0.24}{\includegraphics{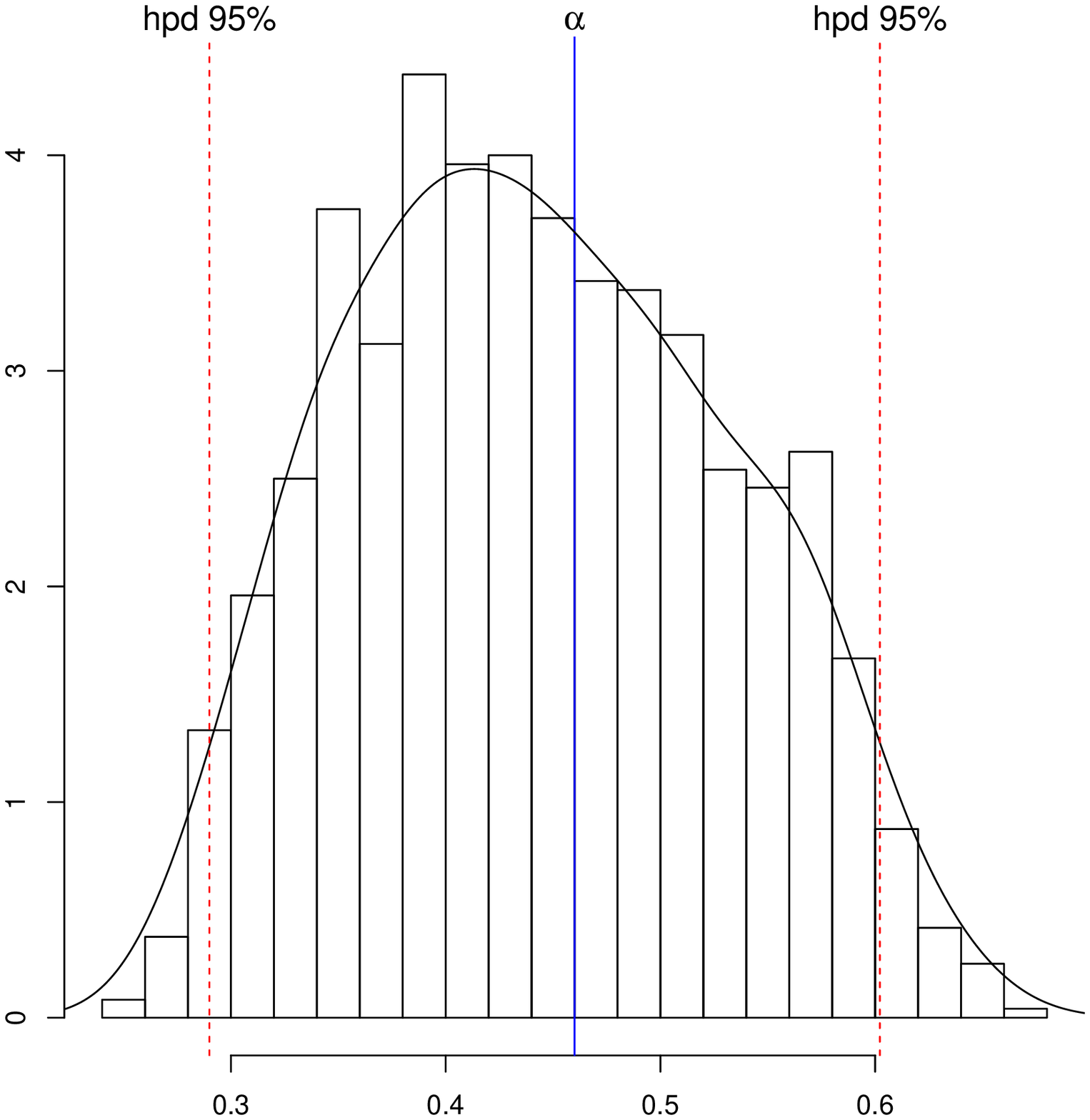}
\includegraphics{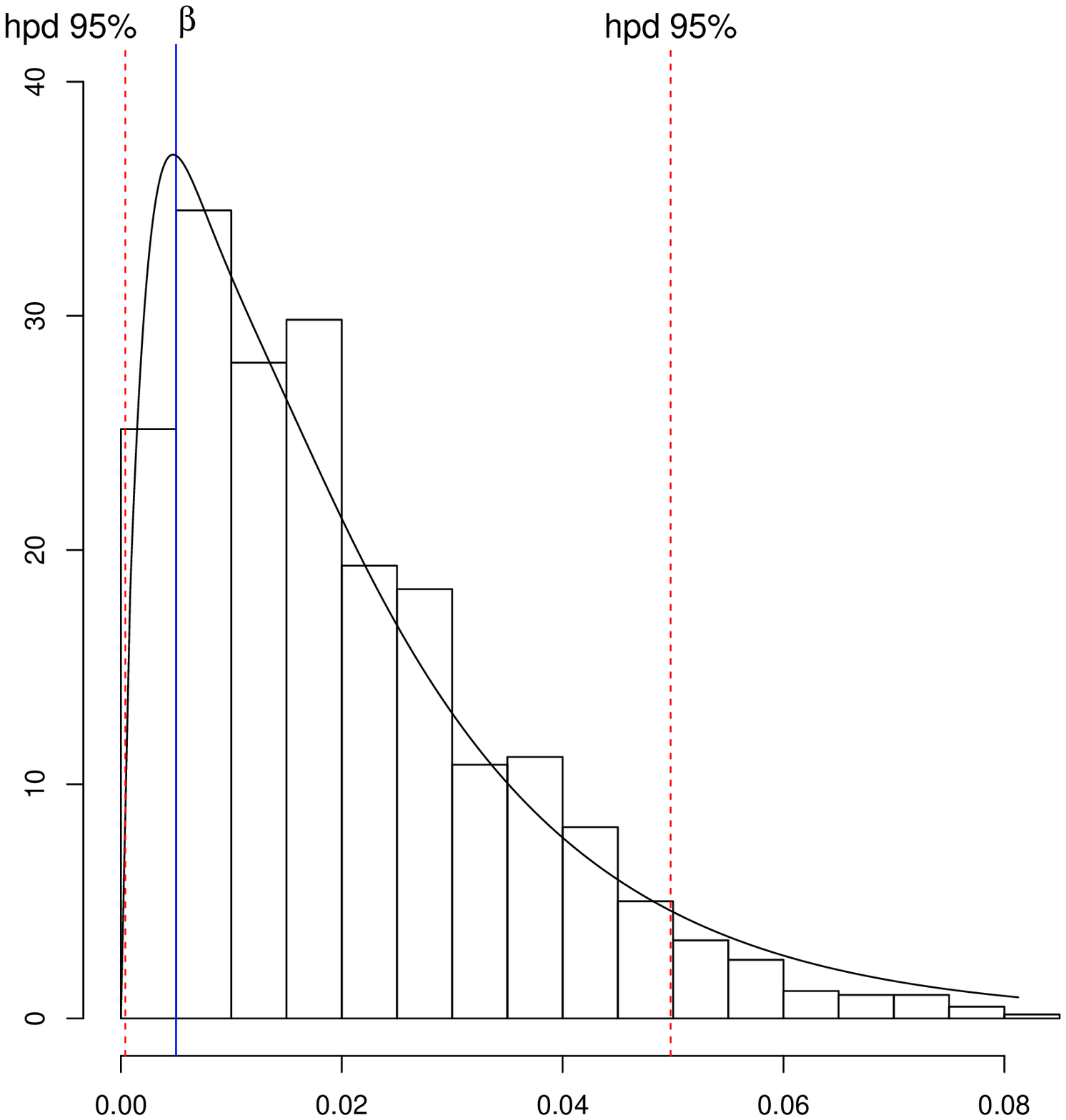}\includegraphics{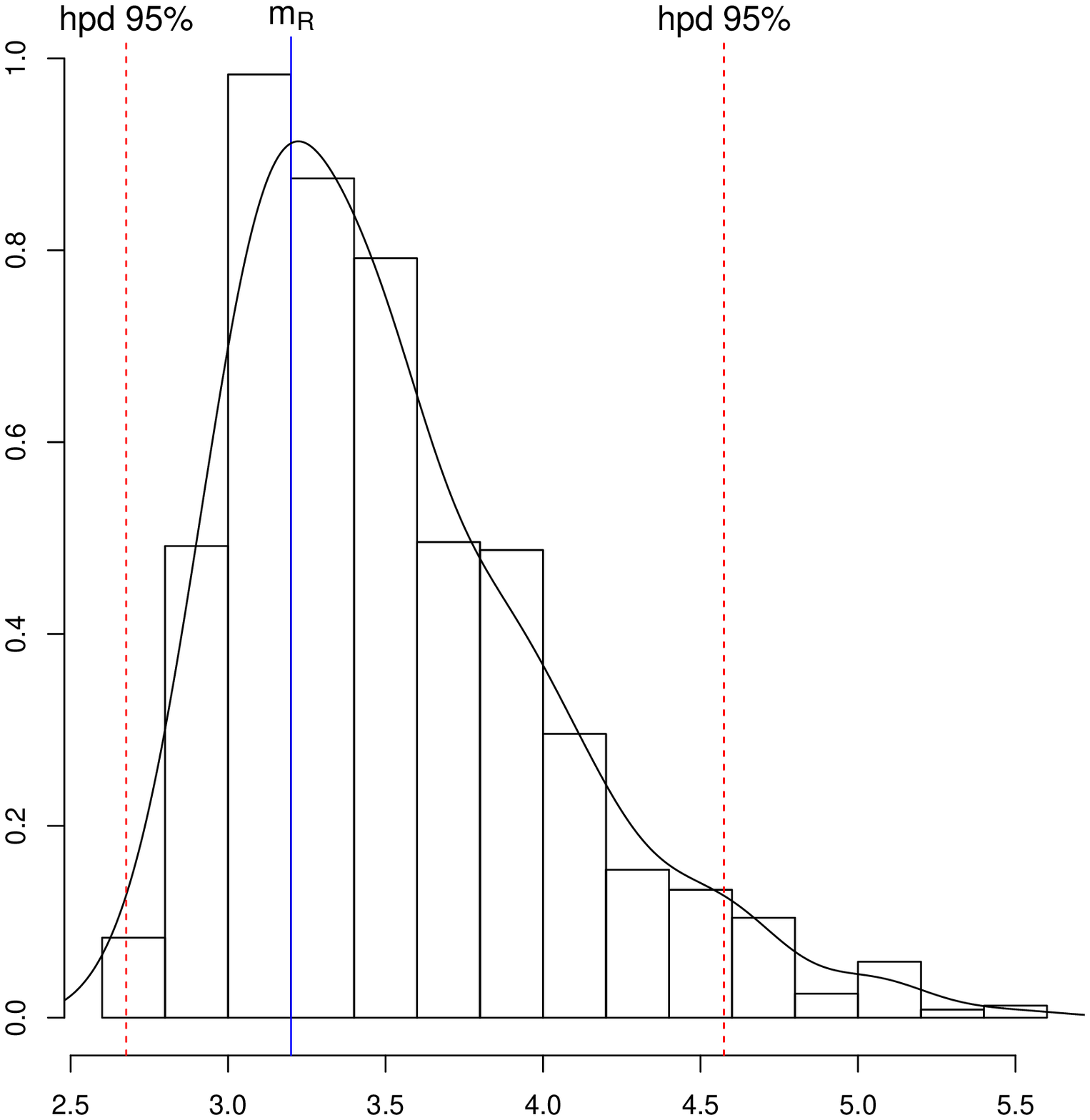}
\includegraphics{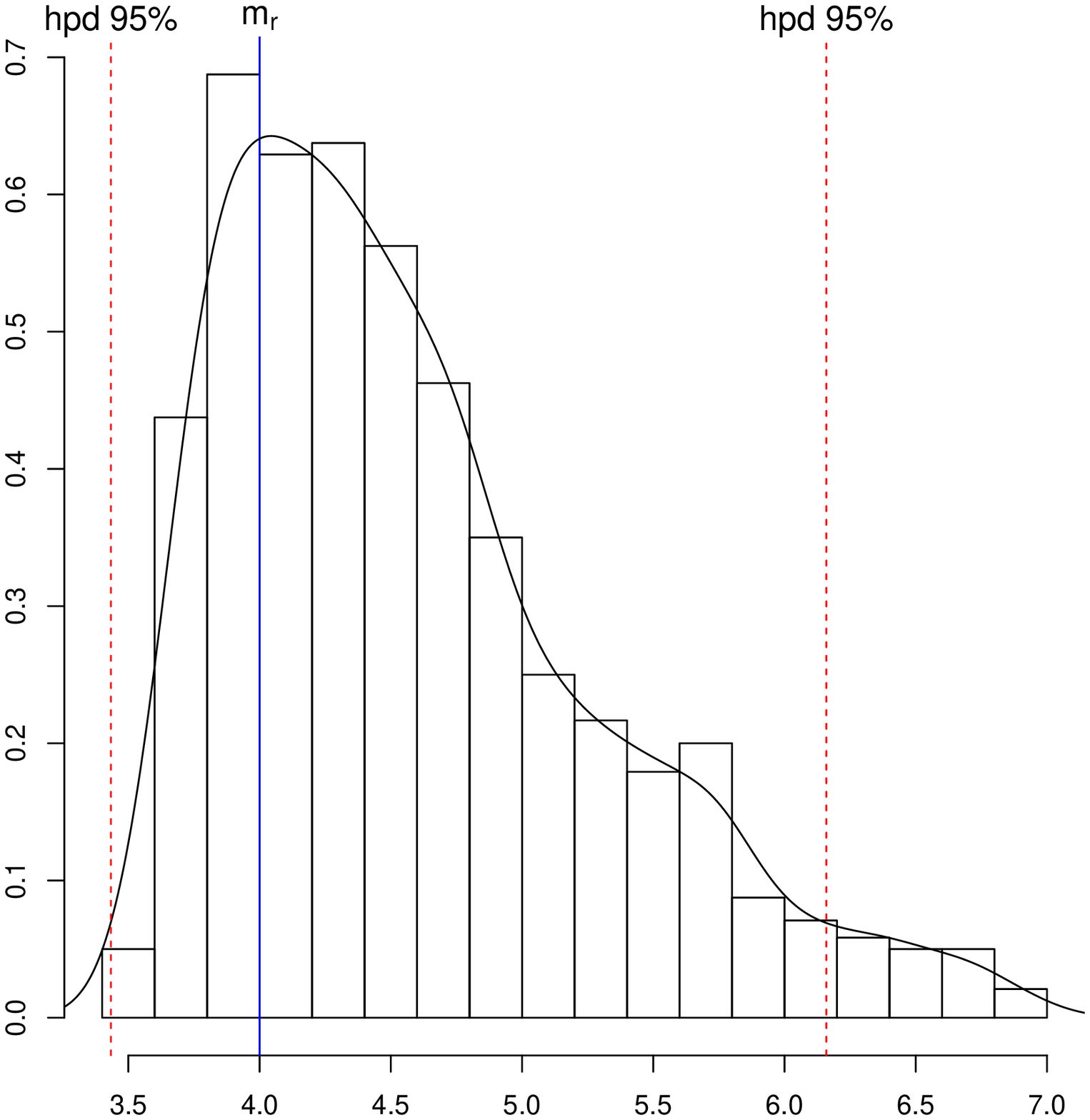}}
\caption{Approximate posterior densities, with 95\%
HPD sets, of the parameters $\alpha$, $\beta$, $\mathsf{m_R}$ and $\mathsf{m_r}$, respectively, given $\overline{\mathcal{FM}}_{15}$ in Table \ref{E1++} and $(\mathsf{M^R_{14}},\mathsf{M^r_{14}})=(754,24687)$ and $(\mathsf{M^{R}_{15}},\mathsf{M^{R\to r}_{15}},\mathsf{M^{r\to r}_{15}})=(1043,6,45844)$, in the case
$\mathsf{m_r}\geq(1-\beta)\mathsf{m_R}$. Vertical solid lines represent the
true value of the parameters.} \label{comden1}
\end{figure*}

\begin{figure*}[!hbt]
\centering \scalebox{0.25}{\includegraphics{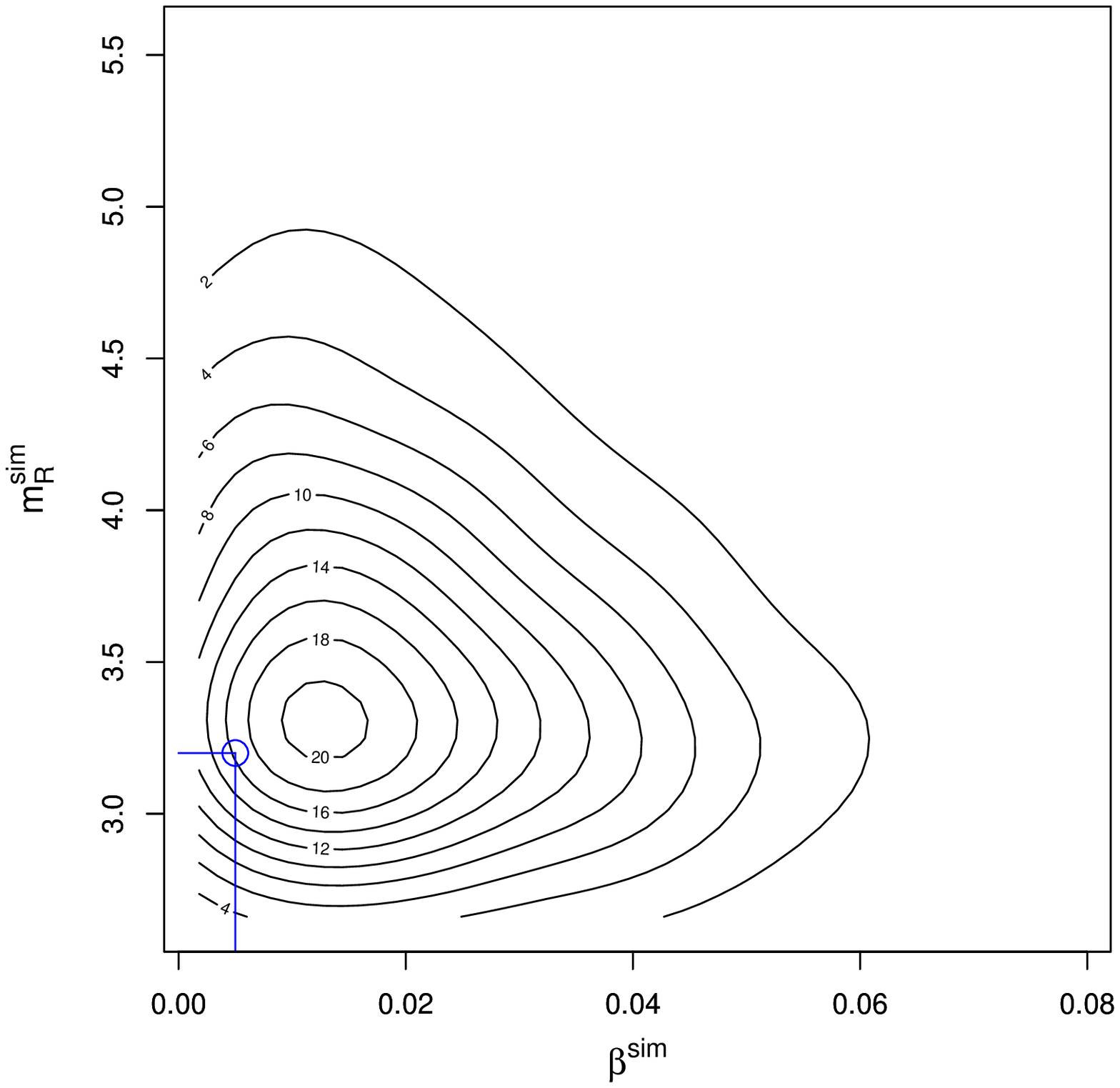}
\includegraphics{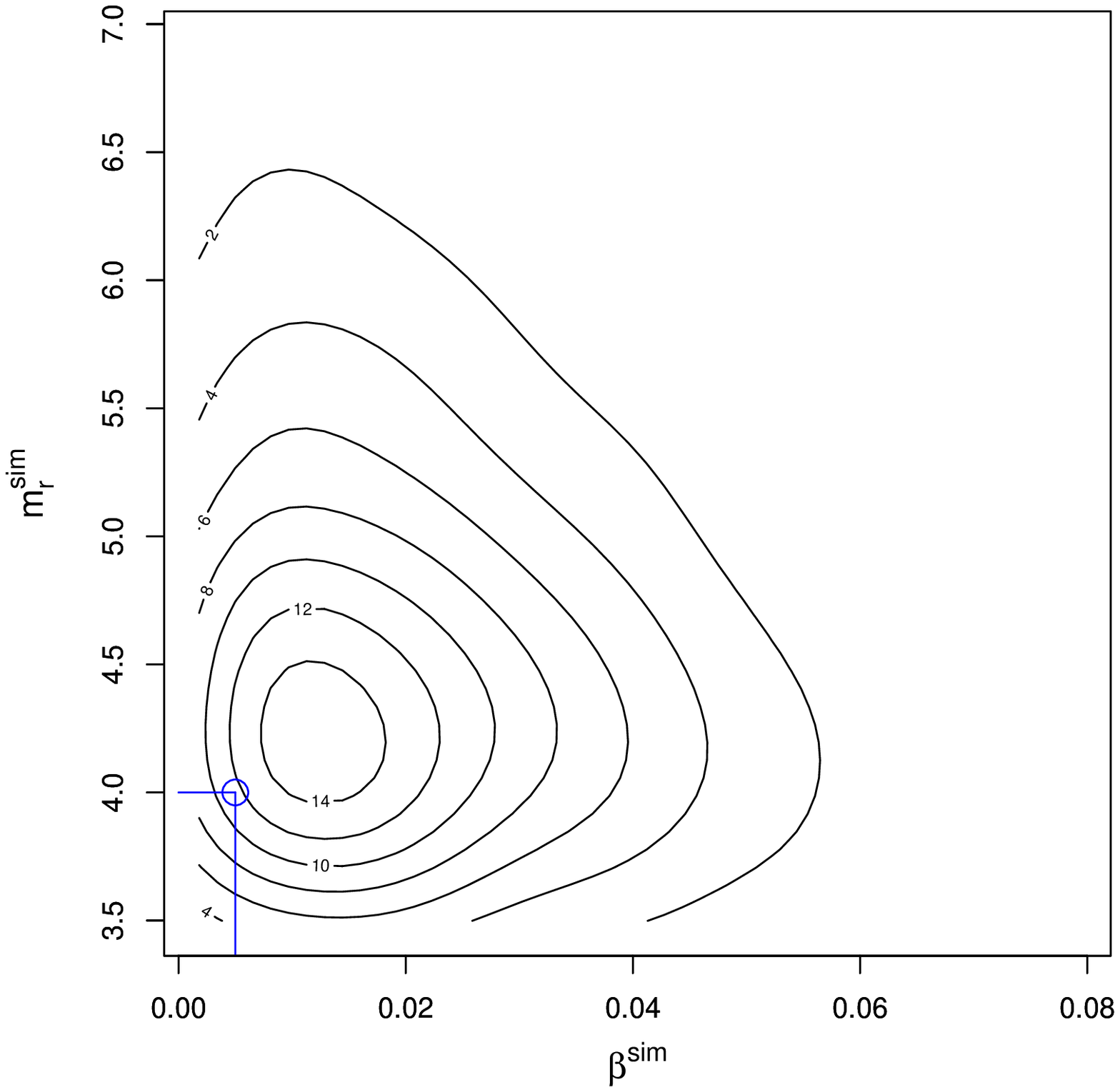}\includegraphics{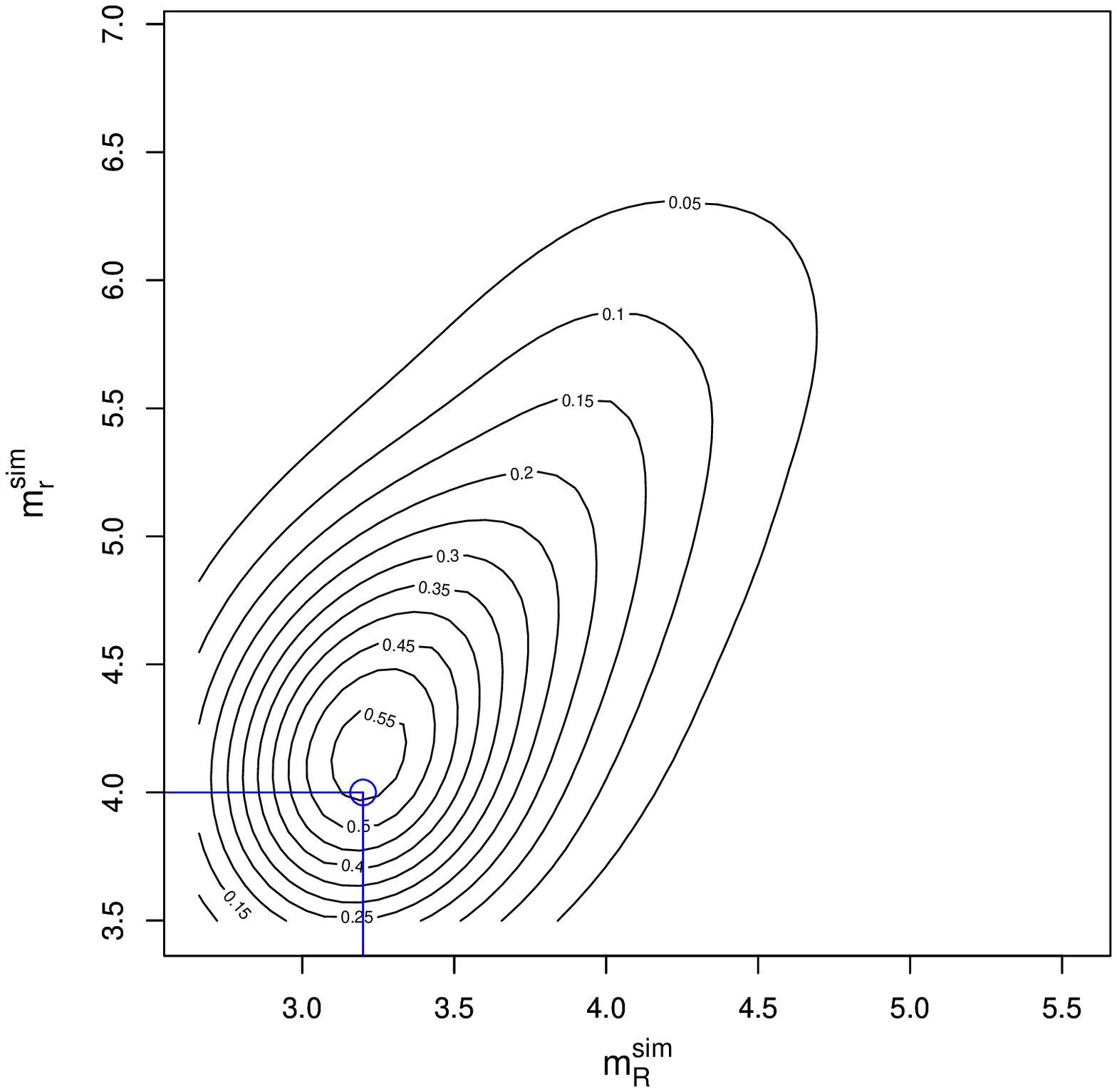}}
\caption{Contour plots showing the relation between $\mathsf{\beta}$ and $\mathsf{m_R}$ and $\mathsf{m_r}$ and the relation between $\mathsf{m_R}$ and $\mathsf{m_r}$ given $\overline{\mathcal{FM}}_{15}$ in Table \ref{E1++}, with $(\mathsf{M^R_{14}},\mathsf{M^r_{14}})=(754,24687)$ and $(\mathsf{M^{R}_{15}},\mathsf{M^{R\to r}_{15}},\mathsf{M^{r\to r}_{15}})=(1043,6,45844)$, in the case
$\mathsf{m_r}\geq(1-\beta)\mathsf{m_R}$. Solid lines represent the true values of the parameter vectors.} \label{comden1b}
\end{figure*}

In Figure \ref{comden1}, we present the approximate posterior
distributions of all parameters together with the true value of
the parameters (solid line). One can appreciate
how the approximate posterior distribution of $\beta$ has improved compared with the corresponding approximation given in Figure \ref{comden}. Now, the true value of all parameters are into the 95\% HPD sets and the corresponding RMSE for $\beta$ and $\mathsf{m_R}$ are, respectively, $18.392$ and $0.035$, considerably smaller than that given in Table \ref{MSE1} ($1557.8$ and $0.1141$, respectively) where only simulated paths with $\beta^{\mathsf{sim}}>0$ were considered, as it is now our case. For the rest of the parameters, the approximate posterior distributions in Figure \ref{comden1} are very similar to that given in Figure \ref{comden} being the corresponding RMSE for $\alpha$ and $\mathsf{m_r}$, $0.037$ and $0.053$, respectively, similar values to that given in Table \ref{MSE1} ($0.0077$ and $0.0076$, respectively) where only simulated paths with $\beta^{\mathsf{sim}}>0$ were considered.
Moreover, since the range of the posterior
distribution of $\beta$ is very small, its estimation does not affect to the estimation of neither $\mathsf{m_R}$ nor $\mathsf{m_r}$, which are positively correlated (see contour plots showed in Figure \ref{comden1b}).

\subsubsection{Case $0<\mathsf{m_r}<(1-\beta)\mathsf{m_R}$}

Next we illustrate how the algorithm works to approximate the posterior distribution
$\theta|\overline{\mathcal{FM}}_\mathsf{N}$, in the case
$0<\mathsf{m_r}<(1-\beta)\mathsf{m_R}$. To this end, we consider a second
simulated example with initial
vector ($\mathsf{F}_0$, $\mathsf{M_0^R}$, $\mathsf{M_0^r}$)=
($10,5,5$) as in the previous case, and parameter vector $\theta=(\alpha, \beta,
\mathsf{m_R}, \mathsf{m_r})=(0.45, 0.01, 3.5, 2.6)$. For a Y-BBP with mutations with this set of parameters
and initial values, we also proved in \cite{ggm2012a} that there
exists a positive probability of survival of both genotypes.

\begin{table*}[!hbt]
\begin{center}
\caption{Reproduction laws for both genotypes, with $p_k$ the probability that a couple generates $k$ individuals, with $k\in\{0,\ldots,7\}$.}
\label{E0b}
\begin{tabular}{lllllllll}
\hline\noalign{\smallskip}
 & $p_0$ & $p_1$ & $p_2$ & $p_3$ & $p_4$ & $p_5$ & $p_6$ & $p_7$ \\
\noalign{\smallskip}\hline\noalign{\smallskip}
$\mathsf{R}$-genotype &        0.0078 &0.0547& 0.1641& 0.2734& 0.2734& 0.1641& 0.0547& 0.0078\\
  $\mathsf{r}$-genotype   &    0.0388 &0.1604& 0.2843& 0.2800& 0.1654& 0.0586& 0.0115& 0.0010\\
\noalign{\smallskip}\hline
\end{tabular}
\end{center}
\end{table*}

We simulate 15 generations of this Y-BBP with mutations assuming
that reproduction laws of both genotypes follow the non-parametric offspring distributions with finite support given in Table \ref{E0b}, with means $\mathsf{m_R}=3.5$ and
$\mathsf{m_r}=2.6$. The simulated data can be seen in Table
\ref{E2} and they are denoted by
$\overline{\mathcal{FM}}_{15}$.

\begin{table*}[!hbt]
\begin{center}
\caption{The observed
sample $\overline{\mathcal{FM}}_{15}$ for the case $0<\mathsf{m_r}<(1-\beta)\mathsf{m_R}$, with $(\mathsf{M^R_{14}},\mathsf{M^r_{14}})=(4113,172)$ and $(\mathsf{M^{R}_{15}},\mathsf{M^{R\to r}_{15}},\mathsf{M^{r\to r}_{15}})=(6351,62,196)$.
This sample has been generated from the parameter vector $\theta=(\alpha,\beta,\mathsf{m_R},\mathsf{m_r})=(0.45, 0.01, 3.5,2.6)$.}

\label{E2}
\begin{tabular}{llllllllllllllll}
\hline\noalign{\smallskip}
$\mathsf{n}$ & 1 & 2 & 3 & 4 & 5 & 6 & 7 & 8 & 9 & 10 & 11 & 12 & 13 & 14 & 15\\
\noalign{\smallskip}\hline\noalign{\smallskip}
$\mathsf{F_n}$ &  22&   13&   23&   42&   69&  107&  156&  246&  390&  630&  940& 1469& 2266& 3461& 5437 \\
  $\mathsf{M_n}$   &    12&   16&   25&   42 &  73&  125&  192&  302&  477&  739& 1219& 1763& 2876& 4285& 6609\\
\noalign{\smallskip}\hline
\end{tabular}
\end{center}
\end{table*}

\begin{figure*}[!hbt]
\centering \scalebox{0.24}{\includegraphics{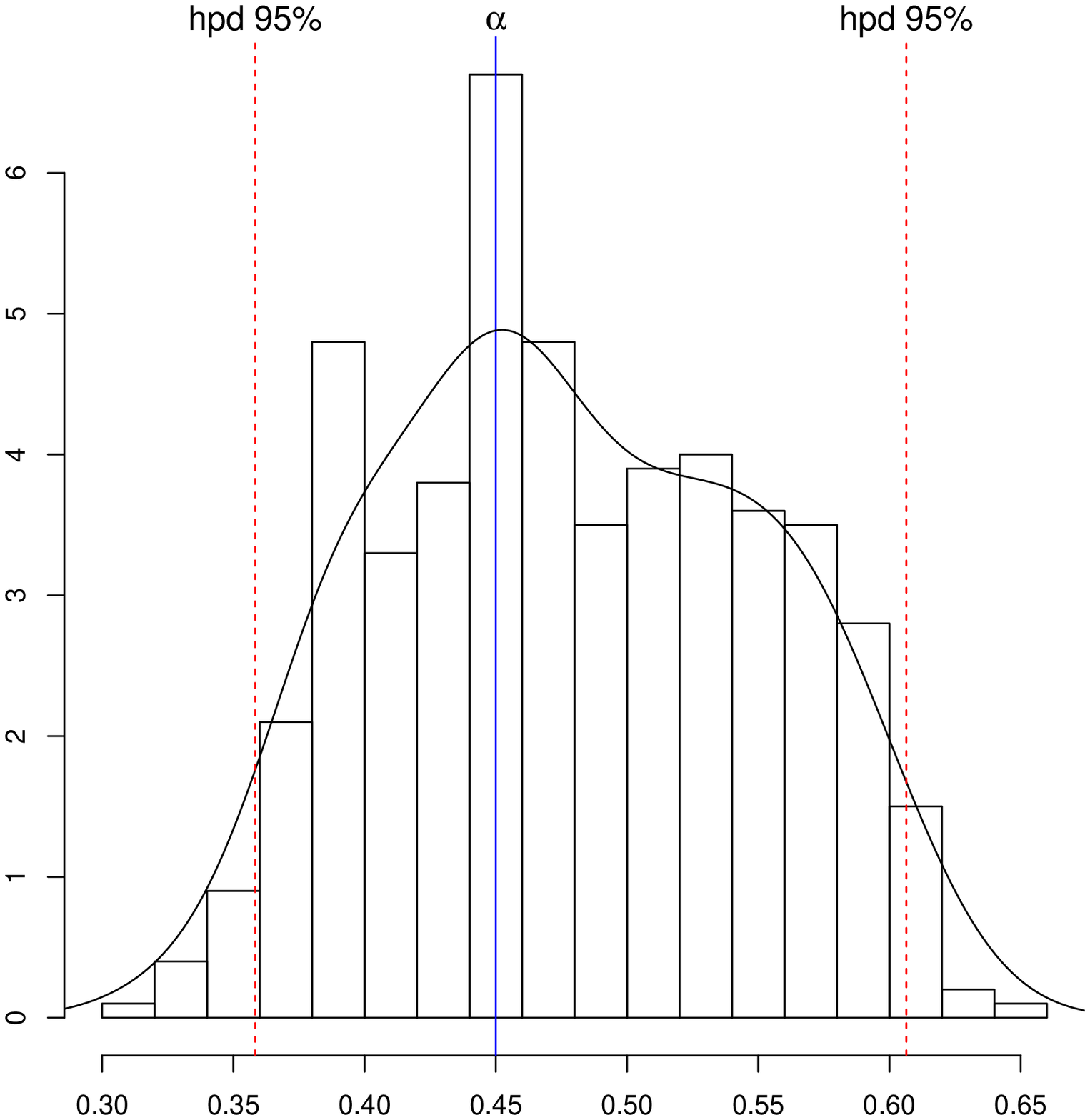}
\includegraphics{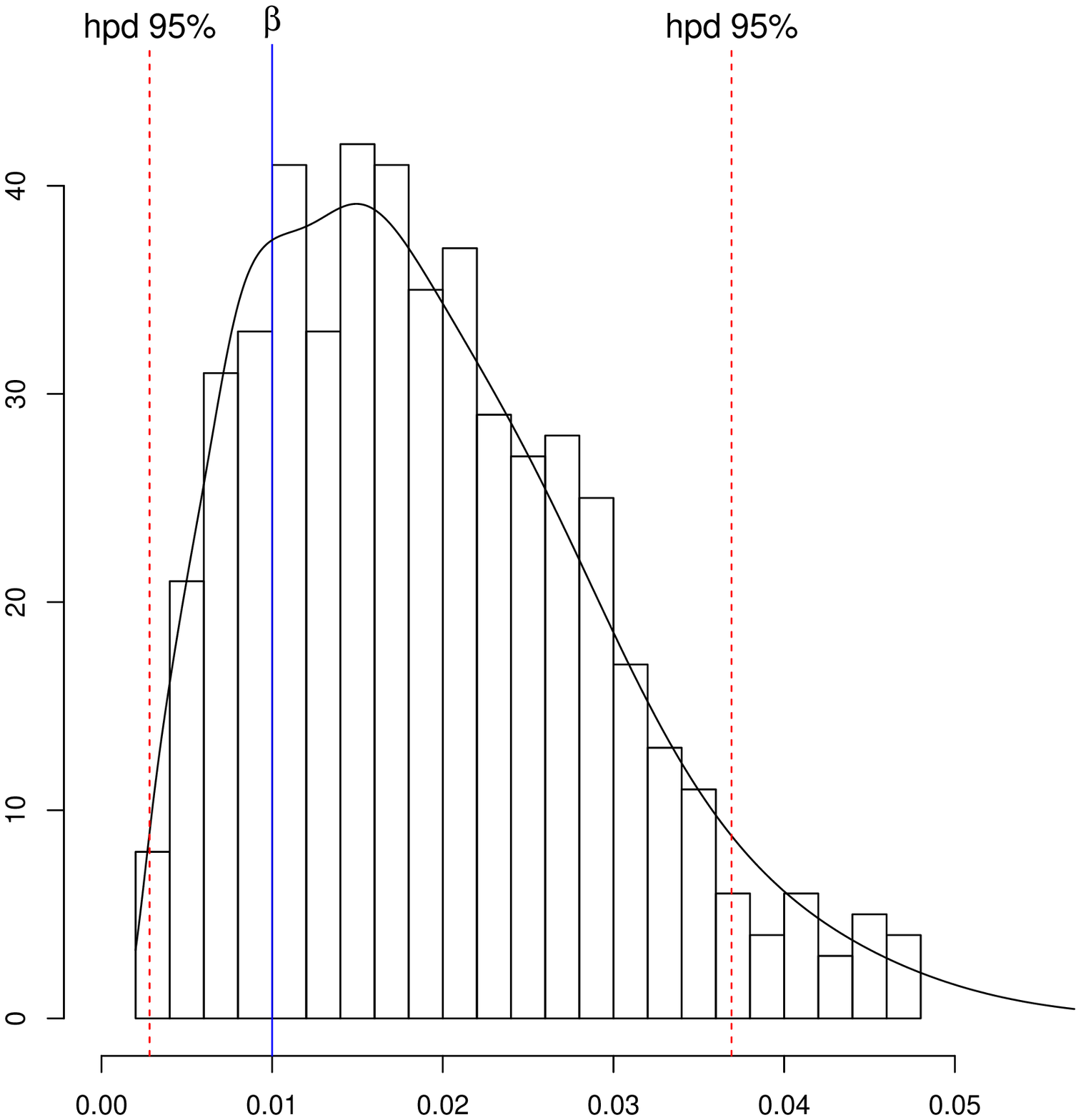}\includegraphics{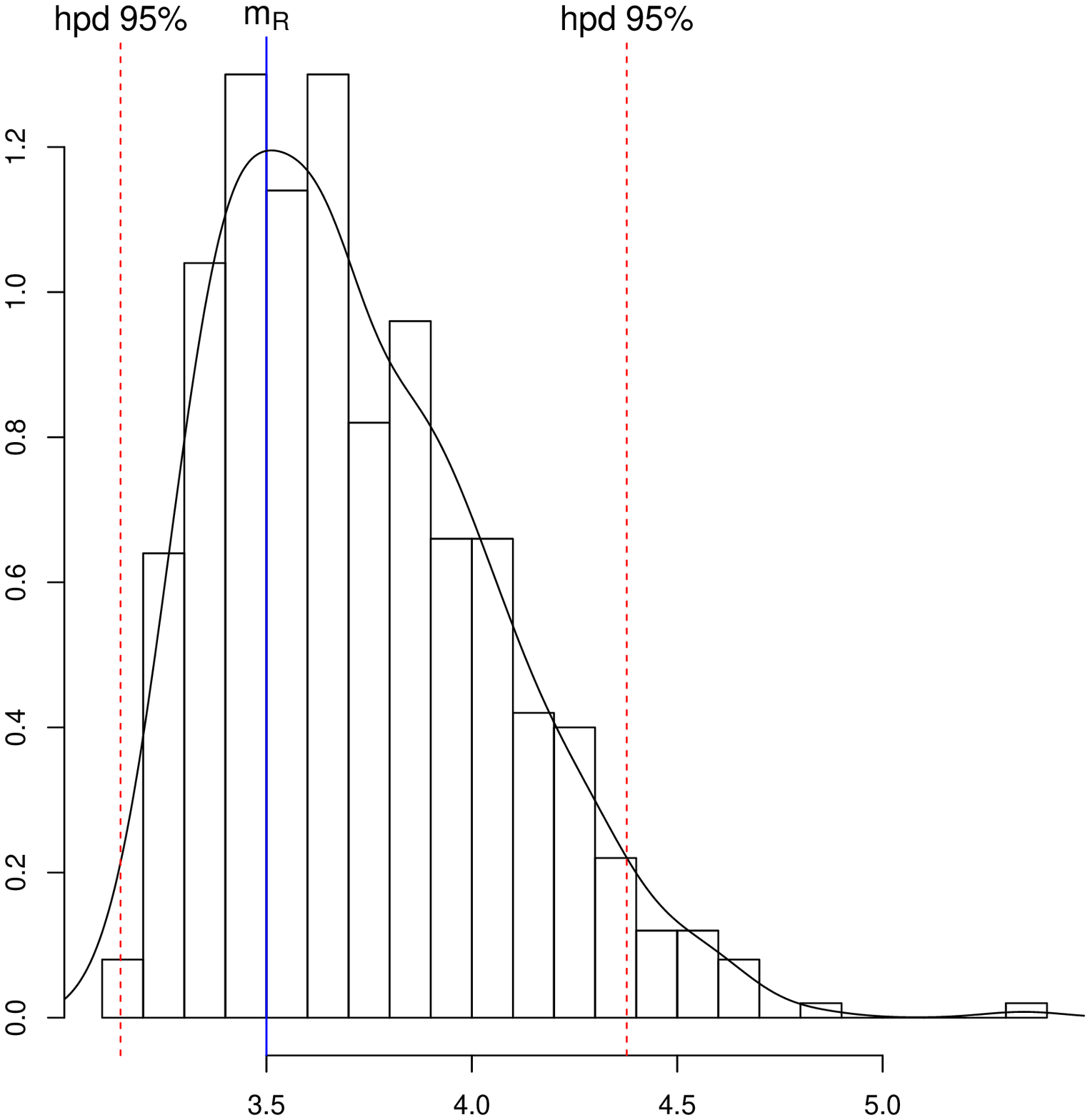}
\includegraphics{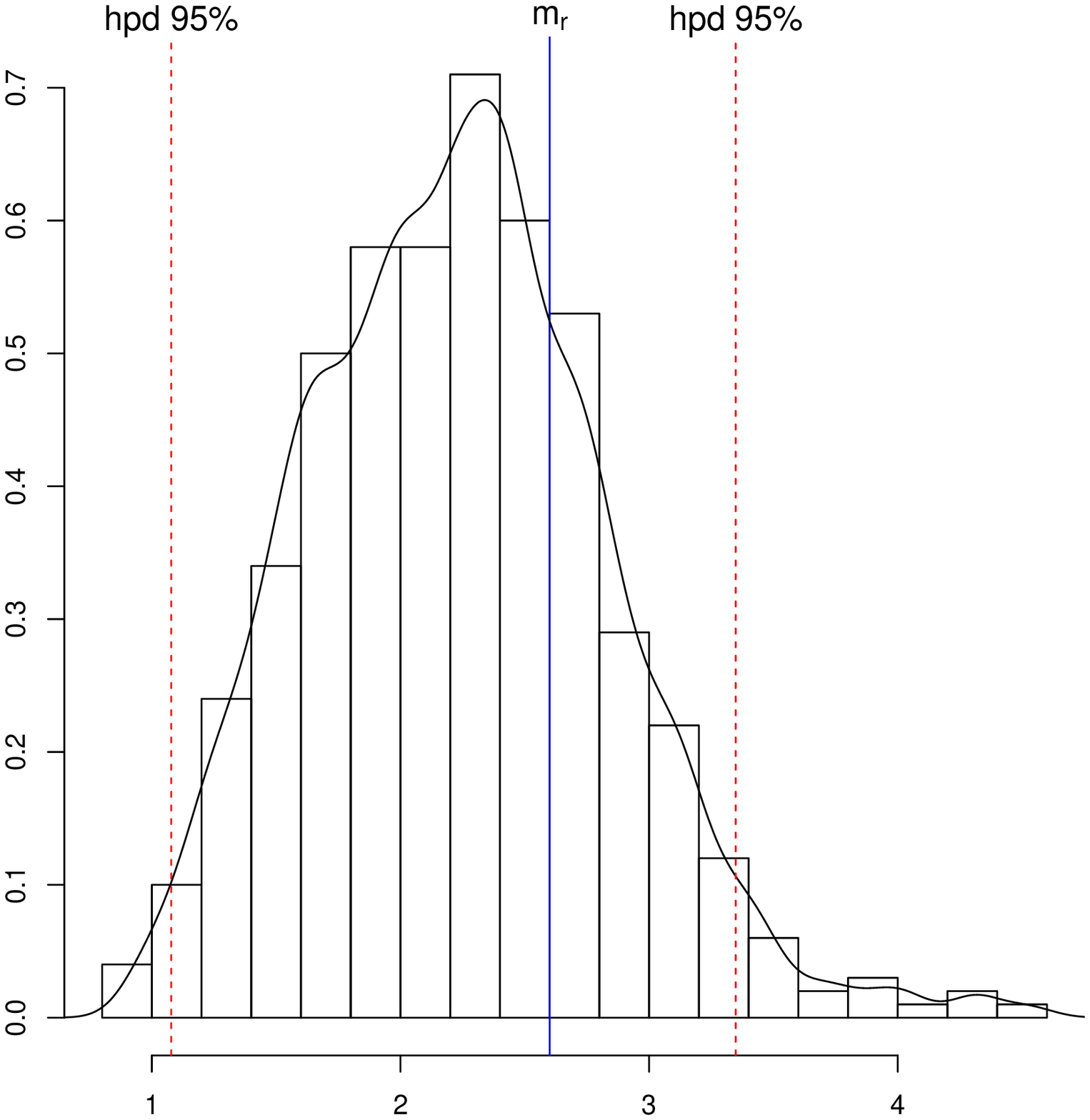}}
\caption{Approximate posterior densities, with 95\%
HPD sets, of the parameters $\alpha$, $\beta$, $\mathsf{m_R}$ and $\mathsf{m_r}$, given $\overline{\mathcal{FM}}_{15}$ in Table \ref{E2},
in the case $0<\mathsf{m_r}< (1-\beta)\mathsf{m_R}$. Vertical solid lines represent the
true value of the parameters.} \label{comden2}
\end{figure*}

We now plot (see Figure \ref{comden2}) the approximate posterior
distributions of the parameters, once the algorithm have been applied. Again, we can appreciate that the methodology provides accurate approximations to the posterior distributions of all parameters in this new context, being the RMSE for $\alpha$, $\beta$, $\mathsf{m_R}$ and $\mathsf{m_r}$, respectively, $0.0285$, $1.7918$, $0.0131$, $0.0719$. Notice that, in this case, the RMSE for $\beta$ and $\mathsf{m_R}$ are smaller than that given in the case $\mathsf{m_r}\geq (1-\beta)\mathsf{m_R}$ in the previous subsection, since in the case $\mathsf{m_r}<(1-\beta)\mathsf{m_R}$ we have more information on these parameters because the rate that define the growth is essentially $(1-\beta)\mathsf{m_R}$ that allows to obtain more accurate approximations of $\beta|\mathcal{FM}_{\mathsf{15}}$ and $\mathsf{m_R}|\mathcal{FM}_{\mathsf{15}}$.

Therefore, as final conclusion of this Subsection \ref{6.1}, we can establish that the proposed Tolerance Rejection-ABC Algorithm works adequately to estimate the parameters of a Y-BBP with mutations, given the information provided by the sample $\overline{\mathcal{FM}}_{\mathsf{N}}$ with $\mathsf{M_N^R}>0$, $\mathsf{M_N^{R\to r}}>0$ and $\mathsf{M_N^{r\to r}}>0$ whichever the relation between the parameters.

\subsection{Observing $\mathsf{M^R_N}>0$, $\MRrN>0$ and $\MrrN=0$}\label{6.2}

In the previous subsection it was considered that $\mathsf{m_r}>0$ since $\MrrN$ was assumed to be non-null. Now, we research the situation where $\MrrN=0$. Obviously, this event occurs in models with $\mathsf{m_r}=0$ but it can be also observed in models with $\mathsf{m_r}>0$. Due to this fact, the estimation of $\mathsf{m_r}$ in this case can be a difficult task. Most probably the approximate posterior distribution of $\mathsf{m_r}|\overline{\mathcal{FM}}_{\mathsf{N}}$ will present an atom at zero with non-null probability. This kind of problems are usual in branching process theory. For example, based on the observation of a Galton-Watson process it is difficult to make inference on whether the extinction or explosion of such process will occur (see \cite{Guttorp2013} or \cite{Guttorp2015}).

The algorithm works in the same way as it was described in Subsection \ref{sub3.1}, now
using the metric $\rho^*(\cdot,\cdot)$. In this case we only consider simulated paths $\overline{\mathcal{FM}}_{\mathsf{N}}^{\mbox{\tiny{sim}}}$ such that  $\mathsf{M^{r\to r}_{N}}^{\mbox{\tiny sim}}=0$. Therefore, the last sum term of $\rho^*(\cdot,\cdot)$ is deleted.

To illustrate this particular case, we fix the para\-meter vector
$\theta=(\alpha, \beta, \mathsf{m_R}, \mathsf{m_r})=(0.45, 0.10,
3, 0)$ and initial vector ($\mathsf{F}_0$, $\mathsf{M_0^R}$,
$\mathsf{M_0^r}$)= ($10,5,5$). For a Y-BBP with mutations with
this set of parameters and initial values, we proved in
\cite{ggm2012a} that there exists a positive probability of
survival of the $\mathsf{R}$-genotype and therefore also of the
$\mathsf{r}$-genotype.

\begin{table*}[!hbt]
\begin{center}
\caption{Reproduction laws for  $\mathsf{R}$-genotype, with $p_k$ the probability that a couple generates $k$ individuals, with $k\in\{0,\ldots,7\}$.}
\label{E0c}
\begin{tabular}{lllllllll}
\hline\noalign{\smallskip}
 & $p_0$ & $p_1$ & $p_2$ & $p_3$ & $p_4$ & $p_5$ & $p_6$ & $p_7$ \\
\noalign{\smallskip}\hline\noalign{\smallskip}
$\mathsf{R}$-genotype &0.0199 &0.1044& 0.2350 &0.2938 &0.2203& 0.0991& 0.0248& 0.0027\\
\noalign{\smallskip}\hline
\end{tabular}
\end{center}
\end{table*}

We simulate 15 generations of this Y-BBP with mutations assuming
that the reproduction law of $\mathsf{R}$-geno\-type follows non-parametric offspring distribution with finite support given in Table \ref{E0c}, with mean $\mathsf{m_R}=3$. The simulated data can be seen in Table \ref{E3} and they are denoted, as in the
previous cases, by $\overline{\mathcal{FM}}_{15}$.

\begin{table*}[!hbt]
\begin{center}
\caption{The observed
sample $\overline{\mathcal{FM}}_{15}$ for the case $\MrrN=0$, with $(\mathsf{M^R_{14}},\mathsf{M^r_{14}})=(96,12)$ and $(\mathsf{M^{R}_{15}},\mathsf{M^{R\to r}_{15}},\mathsf{M^{r\to r}_{15}})=(99,16,0)$. This sample has been generated from the parameter vector $\theta=(\alpha,\beta,\mathsf{m_R},\mathsf{m_r})=(0.45, 0.10, 3,0)$.}
\label{E3}
\begin{tabular}{llllllllllllllll}
\hline\noalign{\smallskip}
$\mathsf{n}$ & 1 & 2 & 3 & 4 & 5 & 6 & 7 & 8 & 9 & 10 & 11 & 12 & 13 & 14 & 15\\
\noalign{\smallskip}\hline\noalign{\smallskip}
$\mathsf{F_n}$ &  6   & 7 &  13&    8 &   9&   11&   15&   23&   27&   34&   52&   56&   70&   81&   97 \\
  $\mathsf{M_n}$   &    7  &  7 &   9 &  13 &   7 &   8&   20&   22&   34&   48&   48&   73&   79&  108&  115 \\
\noalign{\smallskip}\hline
\end{tabular}
\end{center}
\end{table*}

\begin{figure*}[!hbt]
\centering \scalebox{0.24}{\includegraphics{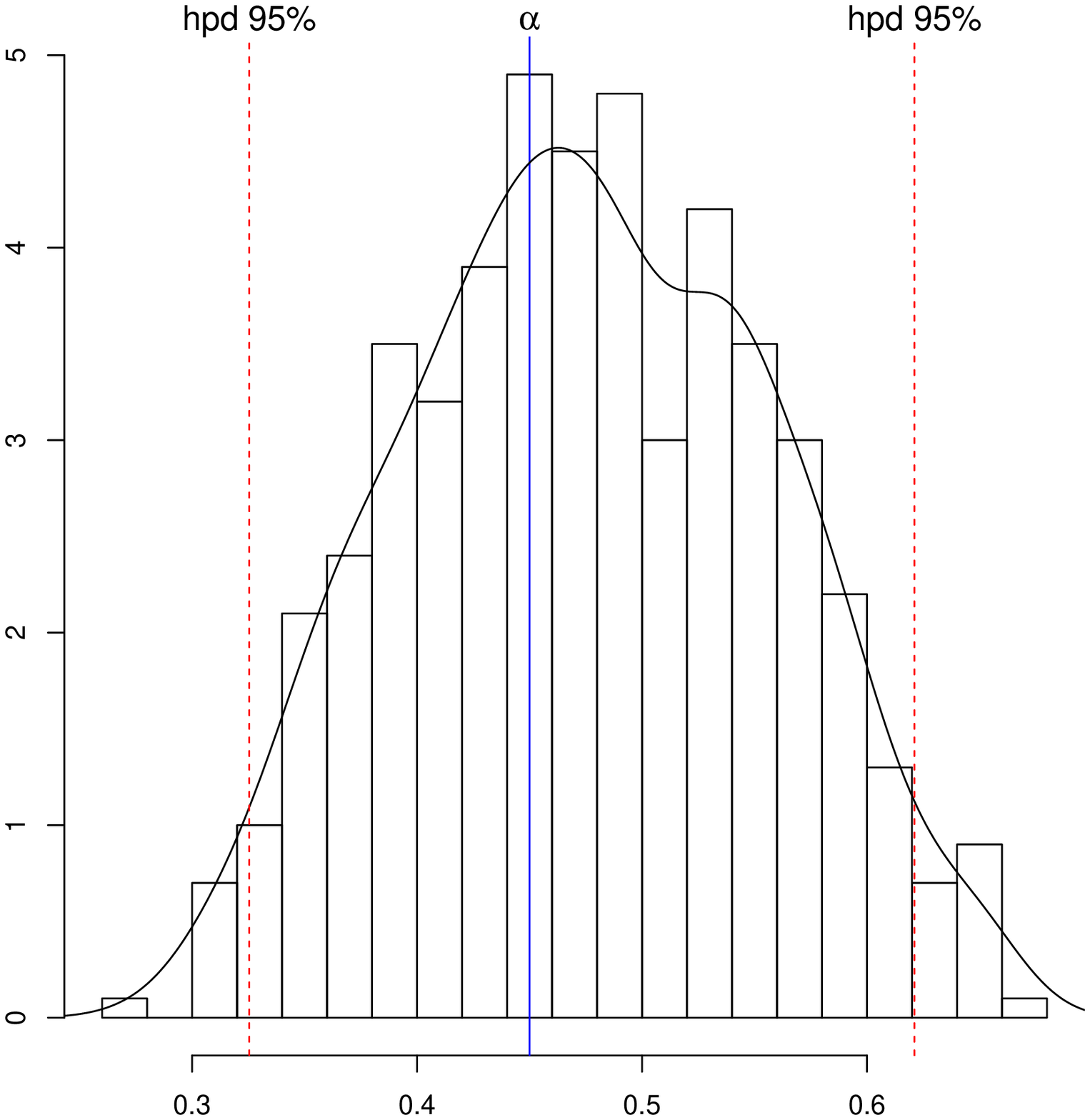}
\includegraphics{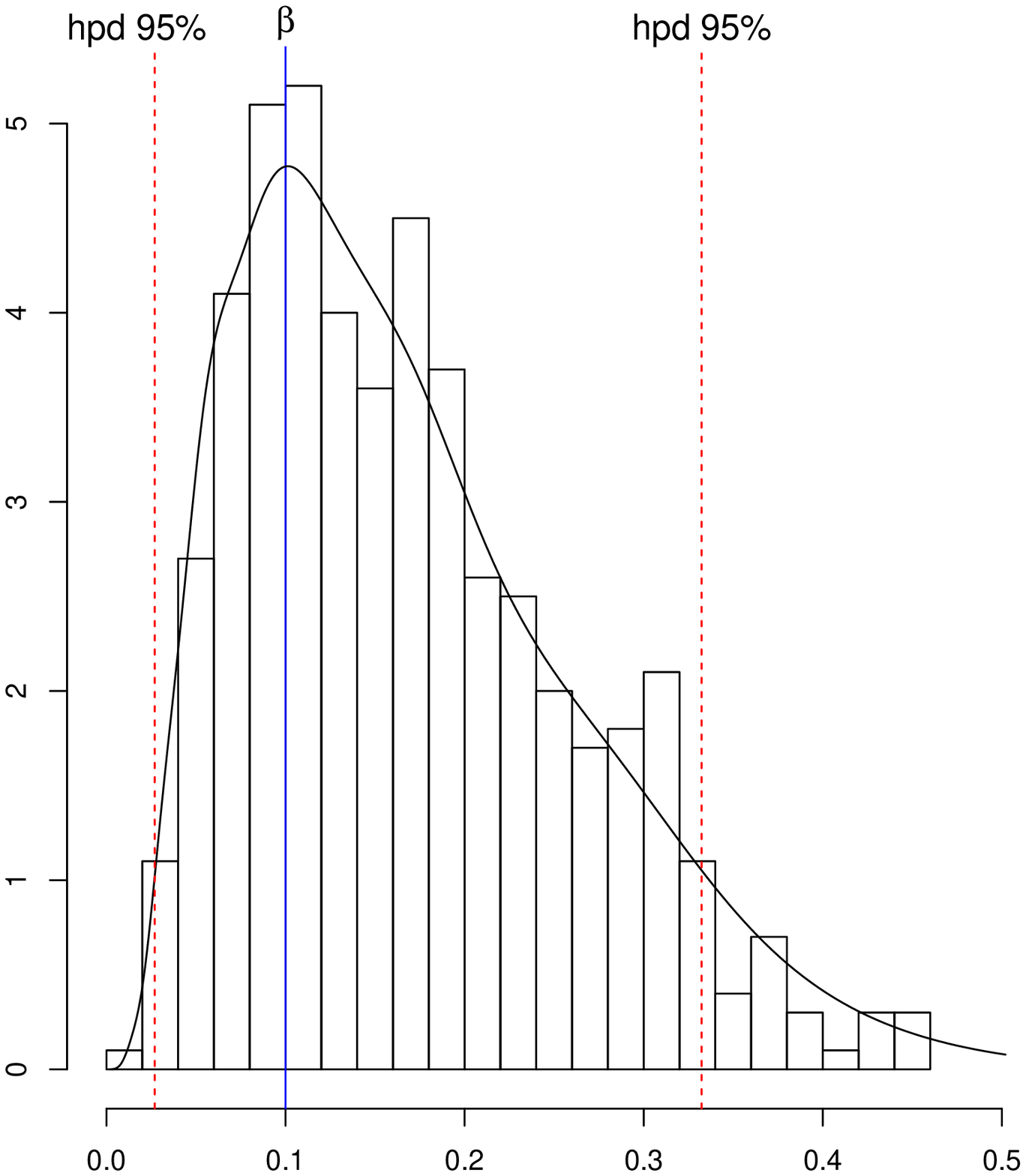}\includegraphics{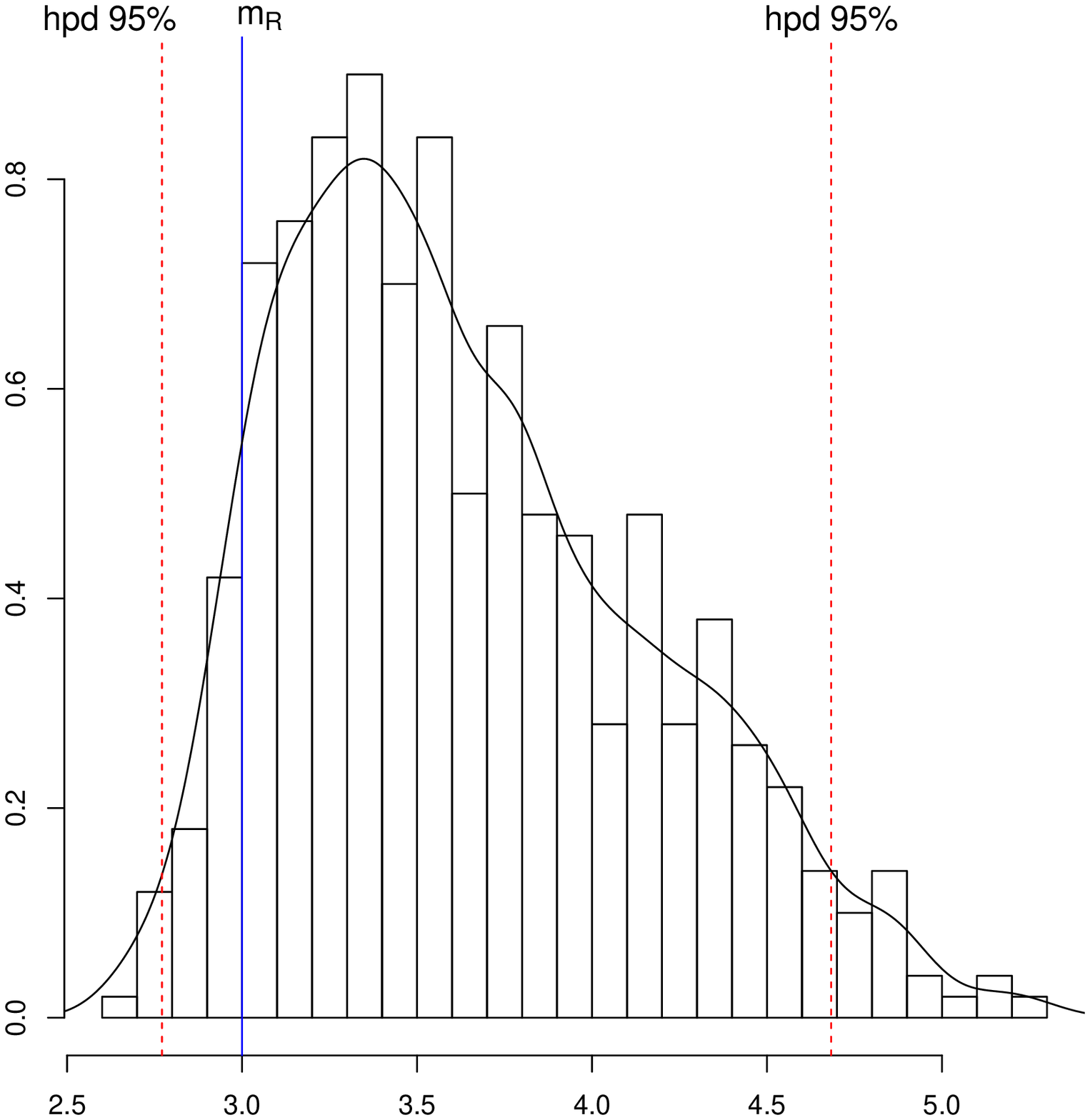}
\includegraphics{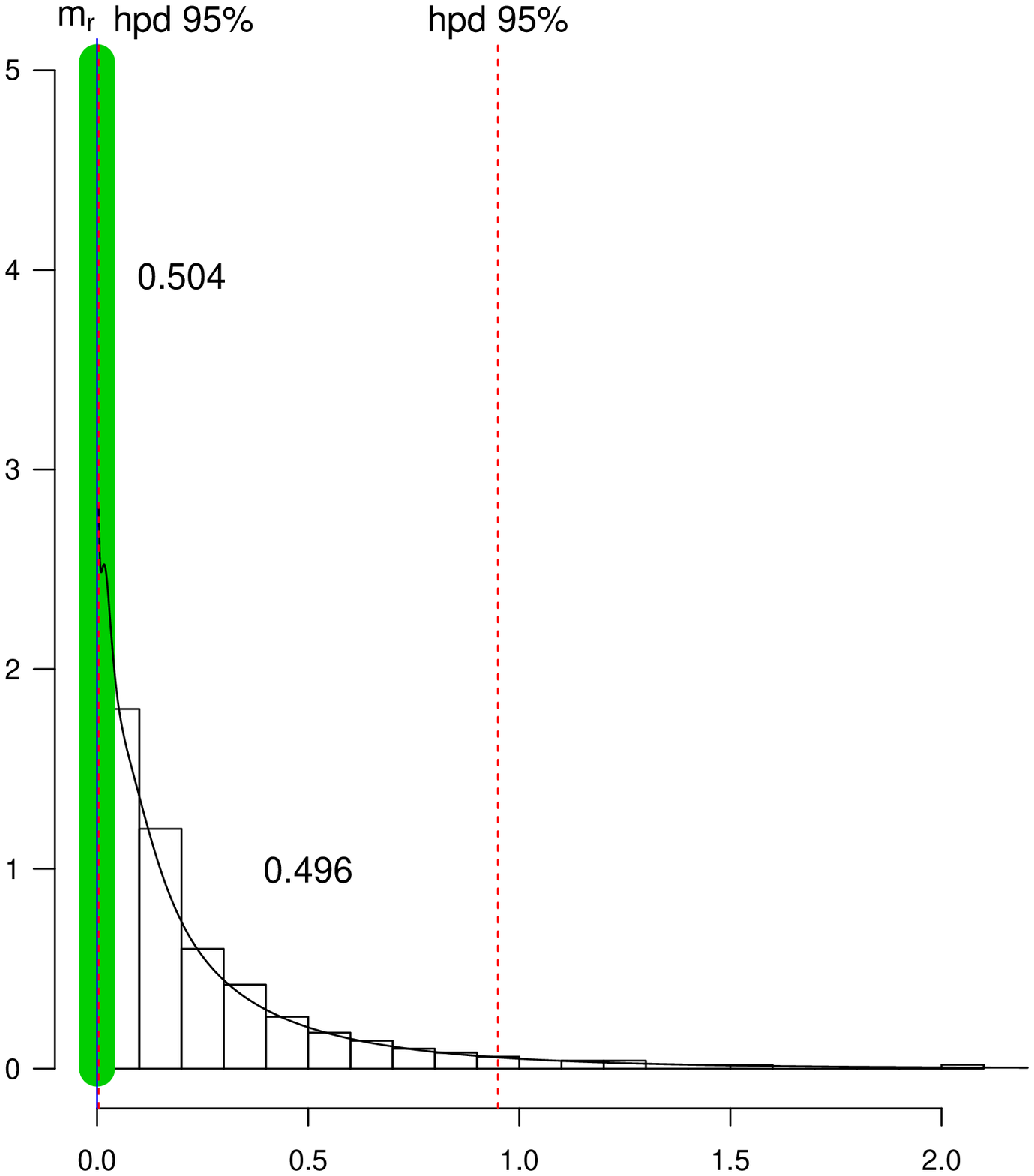}}
\caption{Approximate posterior densities, with 95\%
HPD sets, of the parameters $\alpha$, $\beta$, $\mathsf{m_R}$ and $\mathsf{m_r}$, respectively, given $\overline{\mathcal{FM}}_{15}$ in Table \ref{E3}
where $\MrrN=0$ and considering all simulated paths except for the parameter $\mathsf{m_r}$ for which only simulated path with $\mathsf{m_r}^{\mathsf{sim}}>0$ have been considered. Vertical solid lines represent the true value of the parameters and solid bar represents the estimate of $P(\mathsf{m_r}=0|\overline{\mathcal{FM}}_{15})$.} \label{comden3b}
\end{figure*}

In Figure \ref{comden3b}, we present the approximate posterior densities of every parameter. The approximations of $\alpha|\overline{\mathcal{FM}}_{15}$, $\beta|\overline{\mathcal{FM}}_{15}$ and $\mathsf{m_R}|\overline{\mathcal{FM}}_{15}$ have been calculated considering all chosen simulated paths. In those cases, the algorithm provides accurate approximations to the posterior densities of all parameters with the 95\% HPD sets containing their true values and with small values of their RMSE (see Table \ref{MSE2}). On the other hand, the approximation of $\mathsf{m_r}|\overline{\mathcal{FM}}_{15}$ has been obtained considering only chosen simulated paths where $\mathsf{m_r}^{\mathsf{sim}}>0$. In this case, we also represent in such figure the $P(\mathsf{m_r}=0|\overline{\mathcal{FM}}_{15})$ (area of the vertical solid bar) which is estimated by 0.504.

\begin{figure*}[!hbt]
\centering \scalebox{0.25}{\includegraphics{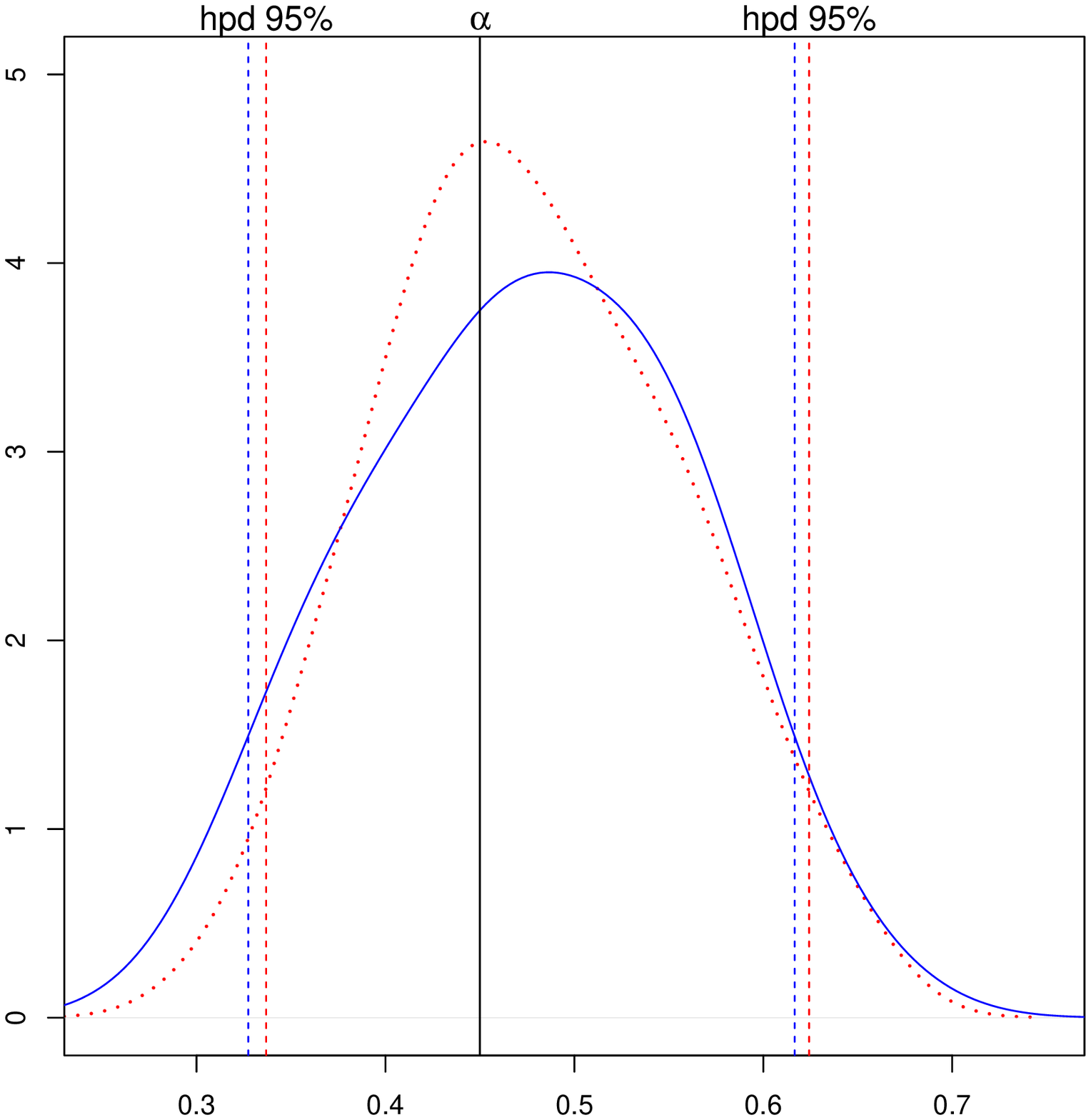}
\includegraphics{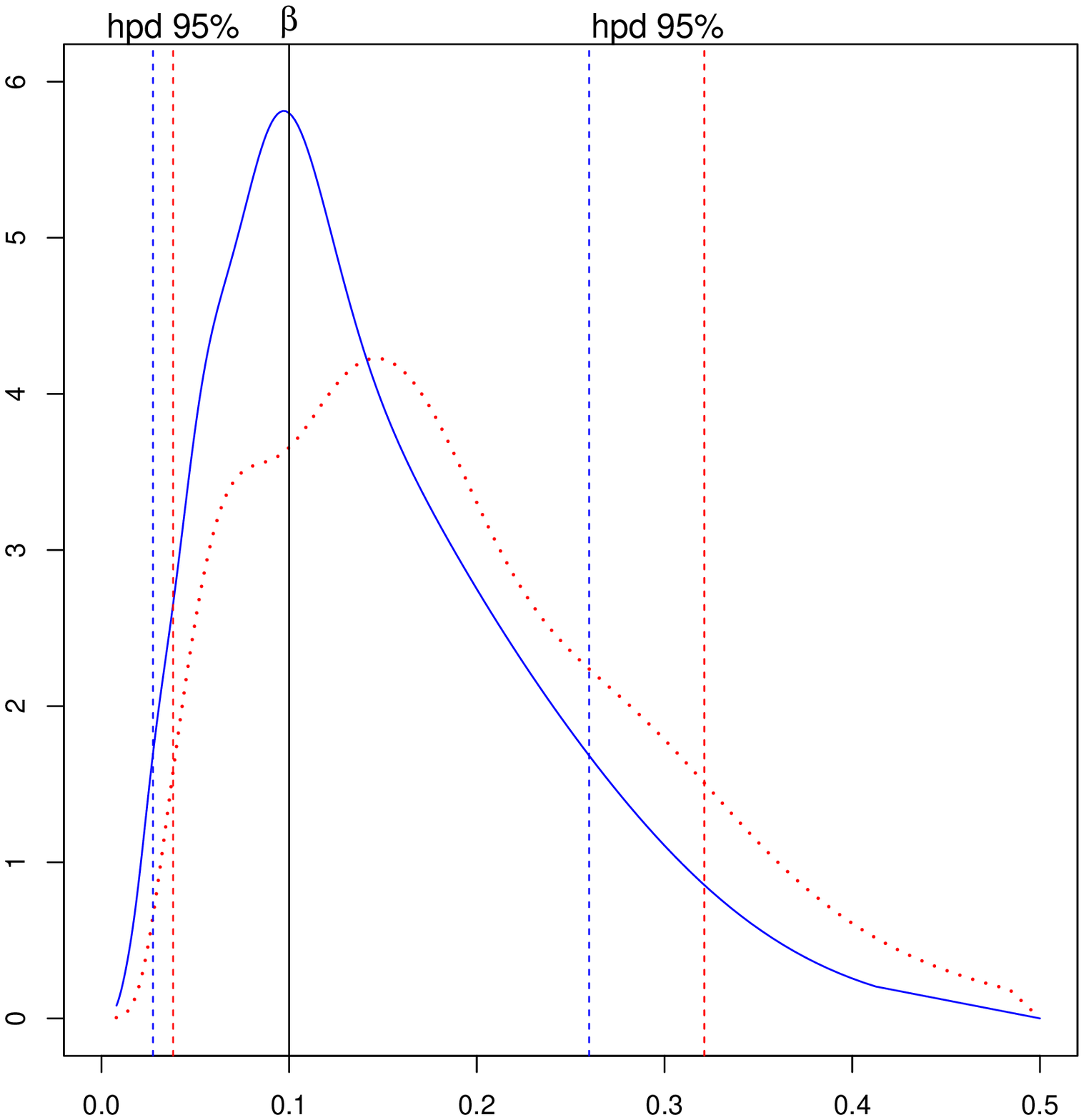}\includegraphics{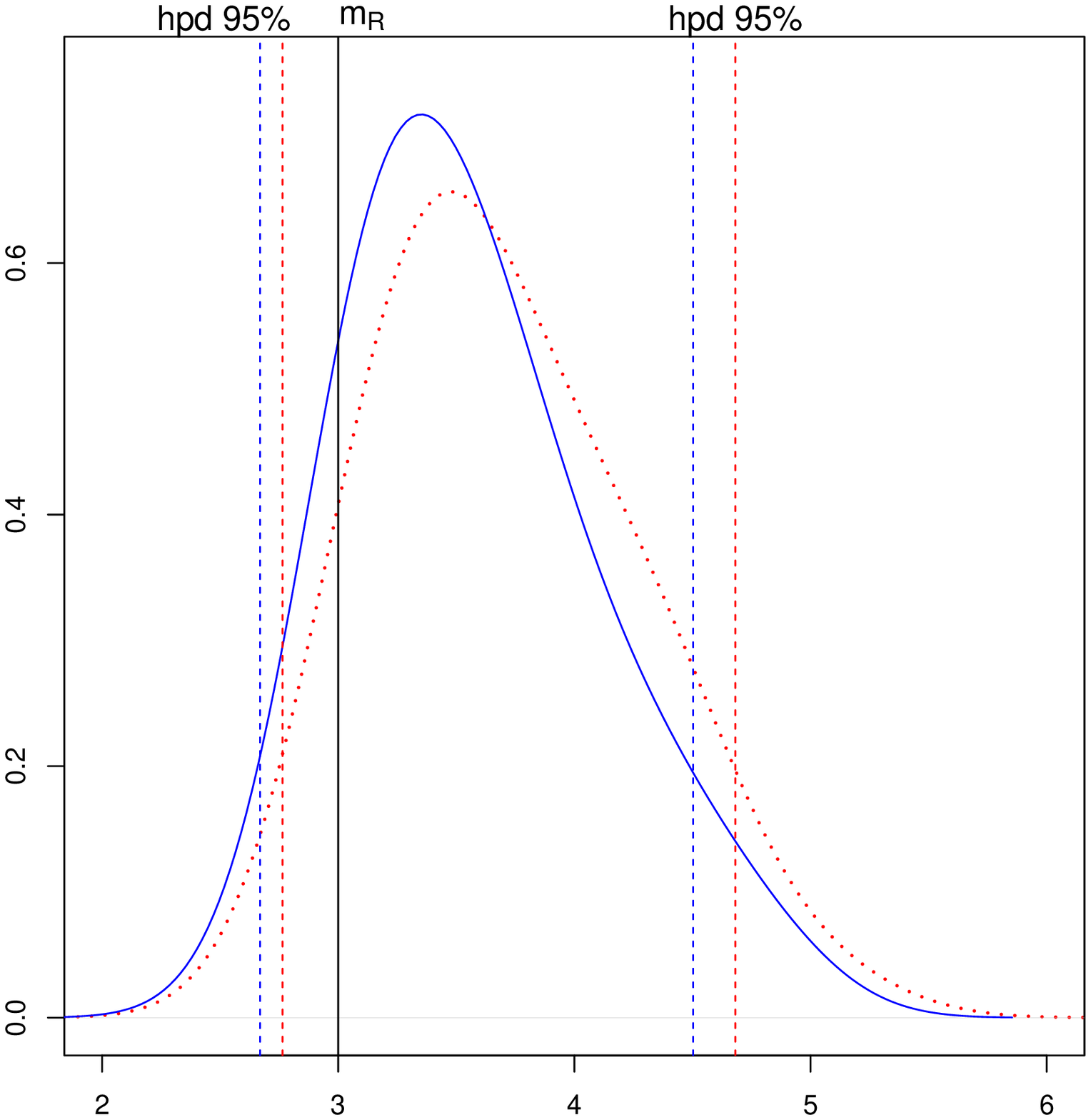}}
\caption{Comparison of the approximate posterior densities, with 95\% HPD sets, of the
parameters $\alpha$, $\beta$ and $\mathsf{m_R}$, given $\overline{\mathcal{FM}}_{15}$ in Table \ref{E3} when only simulated paths with $\mathsf{m_r^{sim}}=0$ have been considered (dotted line) and when only simulated paths with $\mathsf{m_r^{sim}}>0$ have been considered (solid line), in the case $\MrrN=0$. Vertical solid lines represent the true value of the parameters.} \label{comden3bmrsn0}
\end{figure*}

At this point, from the estimates and the observed sample, one can wonder about the following hypothesis test:
\begin{equation}\label{hypmr}H_0:\mathsf{m_r}=0 \ \mbox{vs.} \ H_1: \mathsf{m_r}>0.\end{equation} Considering that we have assumed in the implementation of the algorithm that $\mathsf{m_r^{sim}}$ could take the value 0 with probability $\phi^{\mathsf{sim}}$, being $\phi^{\mathsf{sim}}\sim U(0,1)$, we consider its expected value at calculating the Bayes factor, $K$, and therefore, it is verified that $P(\mathsf{m_r}>0)=P(\mathsf{m_r}=0)$ and then $$K=\frac{P(\mathsf{m_r}=0|\overline{\mathcal{FM}}_{15})P(\mathsf{m_r}>0)}
{P(\mathsf{m_r}>0|\overline{\mathcal{FM}}_{15})P(\mathsf{m_r}=0)}=\frac{0.504}{0.496}=1.06.$$
Although the Bayes factor is greater than 1 and this leads us to conclude that $\mathsf{m_r}=0$ is supported by the observed sample, it is also true that the value of $K$ is very close to 1 and then, the acceptance of $H_0$ is not strongly supported.
For that reason, in Figure \ref{comden3bmrsn0} we present a comparison of the approximate posterior densities of the parameters $\alpha$, $\beta$ and $\mathsf{m_R}$ considering simulated paths where $\mathsf{m_r^{sim}}=0$ (dotted line) and simulated paths where $\mathsf{m_r^{sim}}>0$ (solid line). Notice that the true values of the three parameters are into 95\% HPD sets in both cases. Moreover, in Table \ref{MSE2} are presented the RMSE for the estimates of all parameters for these cases. One can appreciate that in all cases the RMSE for $\alpha$ is very similar and  close to 0. Moreover, the RMSE for $\beta$ and $\mathsf{m_R}$ take their smaller values when only simulated paths where $\mathsf{m_r^{sim}}>0$ are considered. This is due to the close relation of these parameters so, when $\mathsf{m_r^{sim}}>0$ the values of $\beta^{\mathsf{sim}}$ and $\mathsf{m_R^{sim}}$ are smaller than in the case $\mathsf{m_r^{sim}}=0$ since the $\mathsf{r}$-males do not stem only from mutations.

\begin{table*}
\caption{RMSE for the estimates of $\mathsf{\alpha}$,
$\mathsf{\beta}$, $\mathsf{m_R}$ and $\mathsf{m_r}$ given by the Tolerance Rejection-ABC Algorithm when the sample $\overline{\mathcal{FM}}_{15}$ \mbox{in Table \ref{E3}} is observed.} \label{MSE2}
\begin{center}
\begin{tabular}{l|c|c|c|c|}
 & $\alpha$ & $\beta$& $\mathsf{m_R}$ & $\mathsf{m_r}$ \\
\hline
Considering all simulated paths &  0.0349 & 1.2934 & 0.0751 & 4.8186$^*$ \\
Considering only simulated paths where $\mathsf{m_r}^{\mathsf{sim}}=0 $ &  0.0327& 1.6791 &0.0864  &   \\
Considering only simulated paths where $\mathsf{m_r}^{\mathsf{sim}}>0$ & 0.0371 &0.9014& 0.0636   & 2.3900$^*$\\
\hline
\multicolumn{5}{l}{$^*$\mbox{RMSE proposed in \cite{knuth}, when the true value is zero}}
\end{tabular}
\end{center}
\end{table*}

We finally estimate the difference in means of the approximate posterior densities of each parameter between these two groups (we name Group A1 to the set of all chosen simulated paths where $\mathsf{m_r^{sim}}=0$ and name Group A2 to the set of all chosen simulated paths where $\mathsf{m_r^{sim}}>0$) using the Bayesian alternative to the \emph{t test} (see \cite{BEST}). We obtain that the $95\%$ HPD for $\beta$ and $\mathsf{m_R}$ are, respectively, $(0.02236,$ $0.0536)$ and $(0.0472,0.2286)$ which do not include zero. The Bayes factors are, respectively, $4704.56$ and $7.15$, and the probabilities that the true values of the differences are greater than zero are, respectively, $100\%$ and $99.8\%$ which leads us to conclude that there exist significant differences in the means of the approximate posterior densities of $\beta|\overline{\mathcal{FM}}_{15}$ and $\mathsf{m_R}|\overline{\mathcal{FM}}_{15}$ between Groups A1 and A2. However, the $95\%$ HPD for $\alpha$ is $(-0.0122, 0.0162)$ which includes zero. The Bayes factor in this case is $0.104$ and the probability that the true value of the differences is greater than zero is $60\%$. This leads us to conclude that there are no significant differences in the means of the approximate posterior density of $\alpha|\overline{\mathcal{FM}}_{15}$ between Groups A1 and A2. Visually, one can appreciate such differences in Figure \ref{comden3bmrsn0}.

\subsection{Observing $\mathsf{M^R_N}>0$, $\MRrN=0$ and $\MrrN>0$}\label{6.3}

In a similar way than in previous subsection, next we describe the algorithm when it has been observed that $\mathsf{M^R_N}>0$, $\MRrN=0$ and $\mathsf{M^{r\to r}_N}>0$. Obviously, this event occurs in models with $\beta=0$, but it can be also observed in models with $\beta>0$. Due to this fact, as we pointed out previously, the estimation of $\beta$ in this case can be a difficult task. Now, the approximate posterior distribution of $\beta|\overline{\mathcal{FM}}_{\mathsf{N}}$ will present an atom at zero with non-null probability. The algorithm works in the same way as it was described in Subsection \ref{sub3.1}, now using the metric $\rho^*(\cdot,\cdot)$. In this case we only consider simulated paths $\overline{\mathcal{FM}}_{\mathsf{N}}^{\mbox{\tiny{sim}}}$ such that  $\mathsf{M^{R\to r}_{N}}^{\mbox{\tiny sim}}=0$. Therefore, the next-to-last sum term of $\rho^*(\cdot,\cdot)$ is deleted.

To illustrate this particular case, we fix the para\-meter vector
$\theta=(\alpha, \beta, \mathsf{m_R}, \mathsf{m_r})=(0.65, 0.01,
3, 3.5)$ and initial vector ($\mathsf{F}_0$, $\mathsf{M_0^R}$,
$\mathsf{M_0^r}$)= ($10,5,5$). For a Y-BBP with mutations with
this set of parameters and initial values, we proved in
\cite{ggm2012a} that there exists a positive probability of
survival of the $\mathsf{R}$-genotype and therefore also of the
$\mathsf{r}$-genotype.

\begin{table*}[!hbt]
\begin{center}
\caption{Reproduction laws for  both genotypes, with $p_k$ the probability that a couple generates $k$ individuals, with $k\in\{0,\ldots,7\}$.}
\label{E4c}
\begin{tabular}{lllllllll}
\hline\noalign{\smallskip}
 & $p_0$ & $p_1$ & $p_2$ & $p_3$ & $p_4$ & $p_5$ & $p_6$ & $p_7$ \\
\noalign{\smallskip}\hline\noalign{\smallskip}
$\mathsf{R}$-genotype &        0.0199 &0.1044& 0.2350 &0.2938 &0.2203& 0.0991& 0.0248& 0.0027\\
$\mathsf{r}$-genotype &        0.0078 &0.0547& 0.1641& 0.2734& 0.2734& 0.1641& 0.0547& 0.0078\\
\noalign{\smallskip}\hline
\end{tabular}
\end{center}
\end{table*}

We simulate 15 generations of this Y-BBP with mutations assuming
that reproduction laws of  $\mathsf{R}$ and $\mathsf{r}$-genotypes follow non-parametric offspring distributions with finite support given in Table \ref{E4c}, with means $\mathsf{m_R}=3$ and $\mathsf{m_r}=3.5$. The simulated data can be seen in Table
\ref{E4} and they are denoted, as in the previous cases, by $\overline{\mathcal{FM}}_{15}$.

\begin{table*}[!hbt]
\begin{center}
\caption{The observed
sample $\overline{\mathcal{FM}}_{15}$ for the case $\MRrN=0$, with $(\mathsf{M^R_{14}},\mathsf{M^r_{14}})=(11,77)$ and $(\mathsf{M^{R}_{15}},\mathsf{M^{R\to r}_{15}},\mathsf{M^{r\to r}_{15}})=(10,0,90)$.
This sample has been generated from the parameter vector $\theta=(\alpha,\beta,\mathsf{m_R},\mathsf{m_r})=(0.65, 0.01, 3, 3.5)$.}
\label{E4}
\begin{tabular}{llllllllllllllll}
\hline\noalign{\smallskip}
$\mathsf{n}$ & 1 & 2 & 3 & 4 & 5 & 6 & 7 & 8 & 9 & 10 & 11 & 12 & 13 & 14 & 15\\
\noalign{\smallskip}\hline\noalign{\smallskip}
$\mathsf{F_n}$ & 24& 18&  32 &23& 28& 25& 45& 76& 90& 112& 135& 157& 185& 202& 204 \\
  $\mathsf{M_n}$   &    10 &14& 11& 14& 16& 30& 35& 41& 50& 62& 73& 78& 92& 88& 100 \\
\noalign{\smallskip}\hline
\end{tabular}
\end{center}
\end{table*}

\begin{figure*}[!hbt]
\centering \scalebox{0.24}{\includegraphics{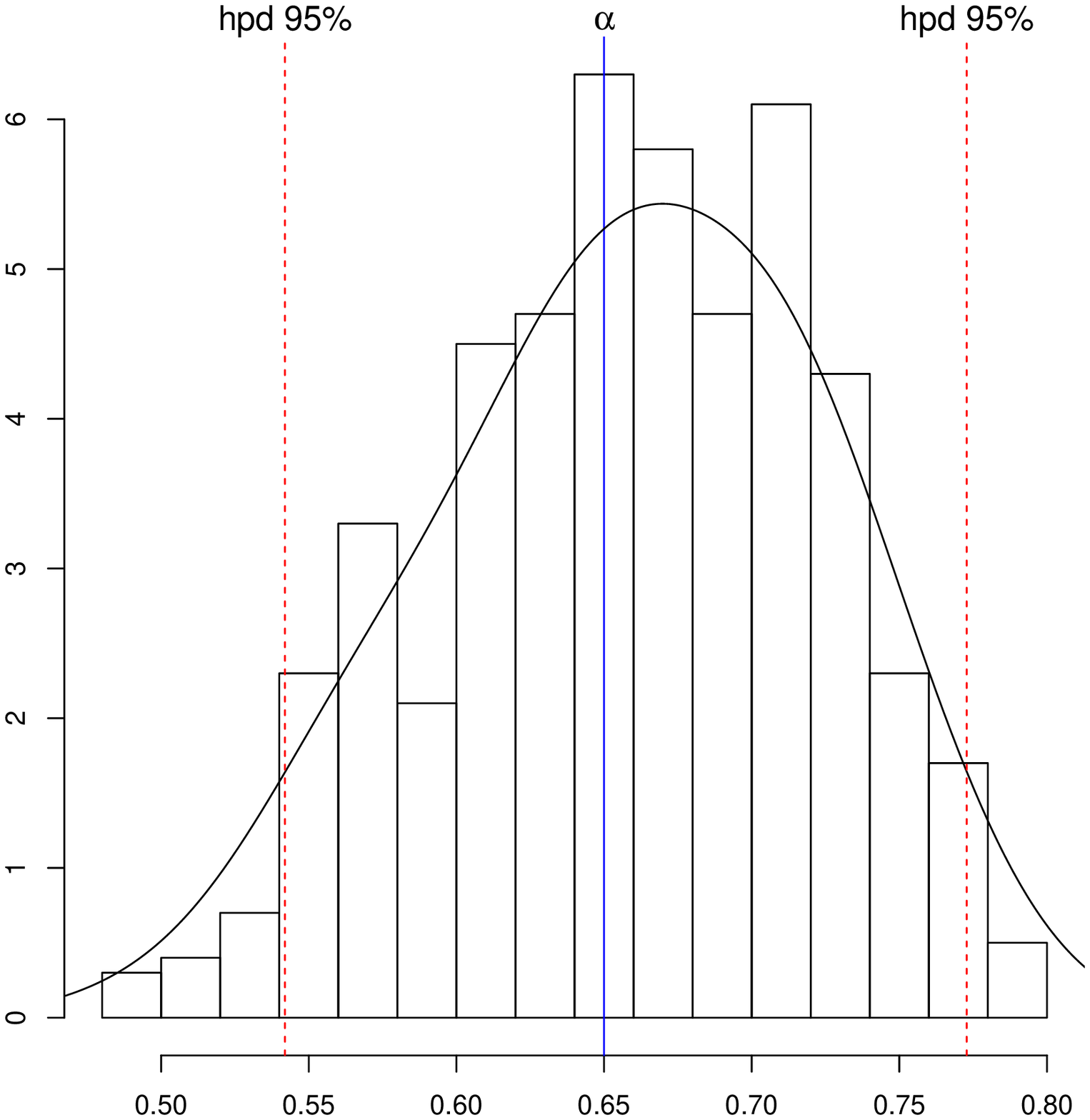}
\includegraphics{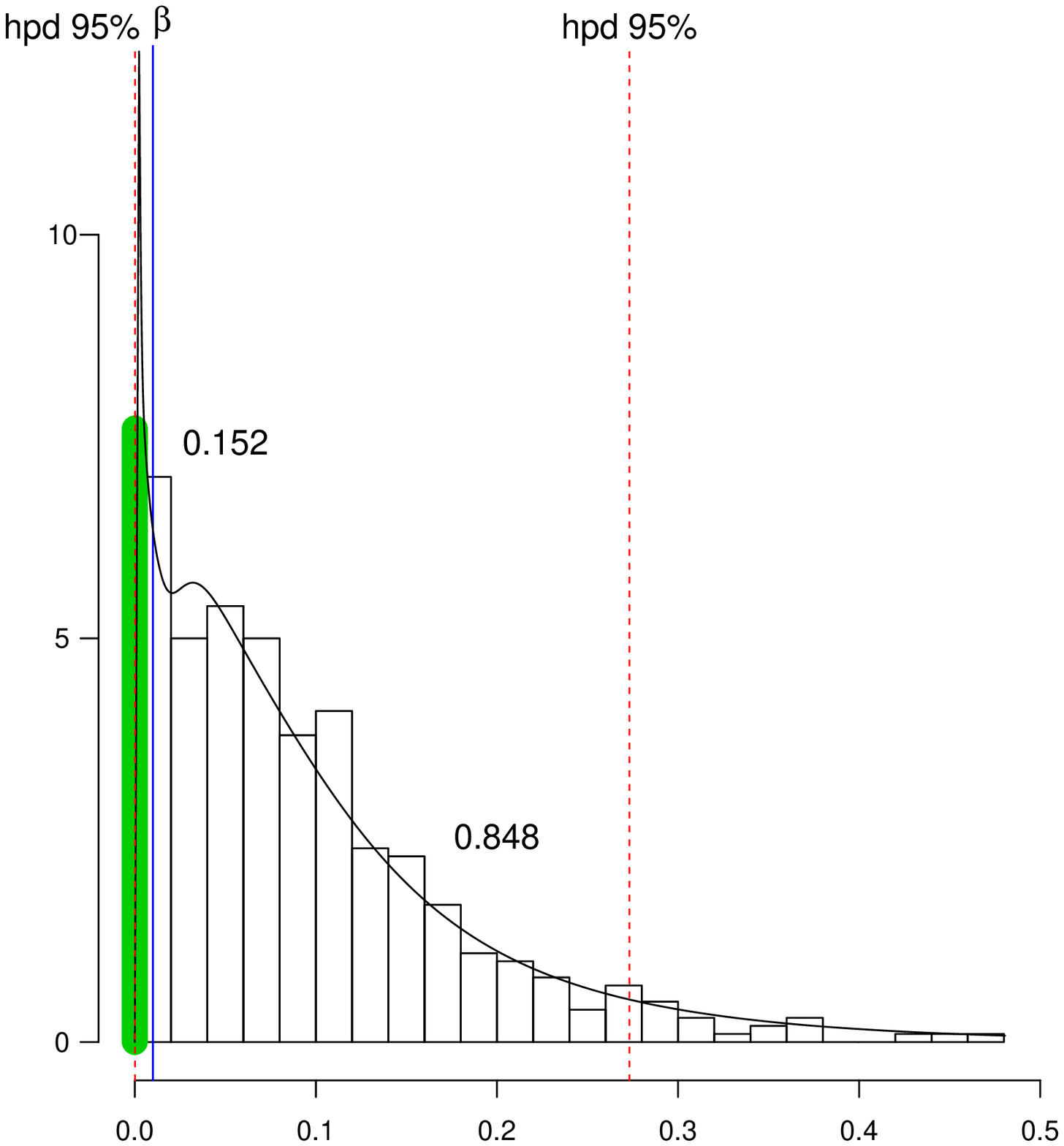}\includegraphics{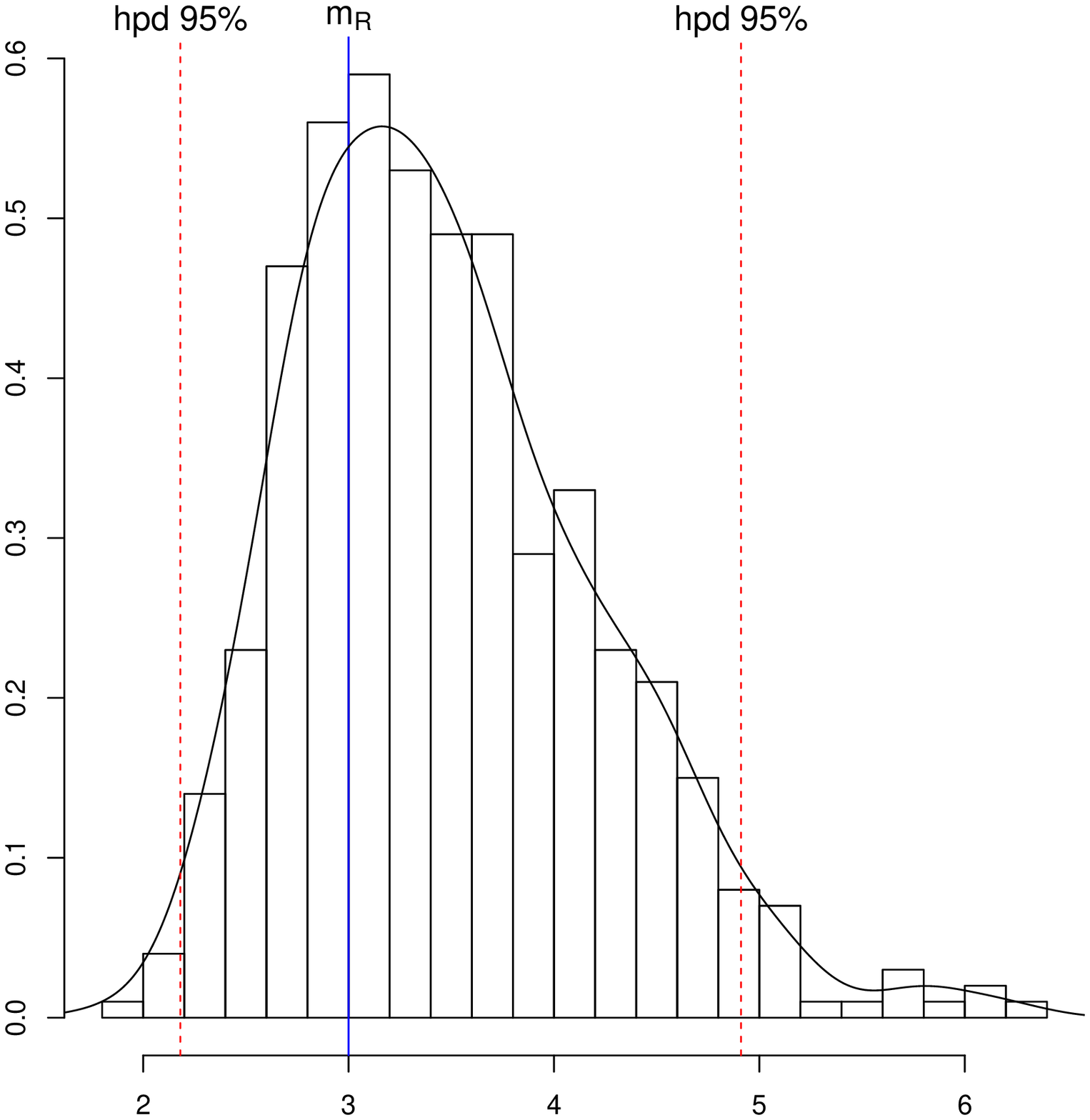}
\includegraphics{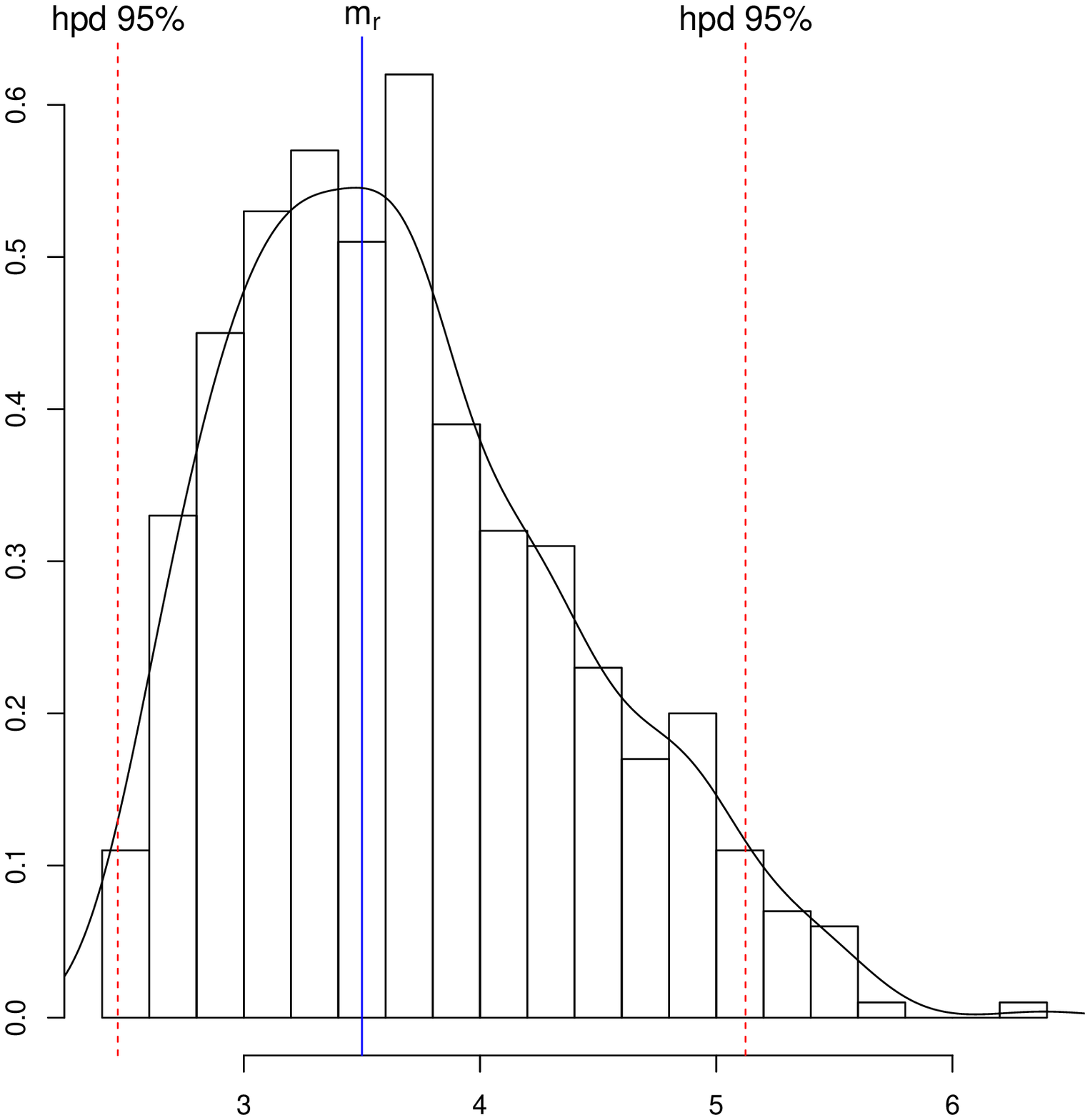}}
\caption{Approximate posterior densities, with 95\%
HPD sets, of the parameters $\alpha$, $\beta$, $\mathsf{m_R}$ and $\mathsf{m_r}$, given $\overline{\mathcal{FM}}_{15}$ in Table \ref{E4} in the case
$\MRrN=0$ considering all simulated paths except for $\beta$ where only simulated paths with $\beta^{\mathsf{sim}}>0$ have been considered. Vertical solid lines represent the
true value of the parameters and solid bar represents the estimate of $P(\beta=0|\overline{\mathcal{FM}}_{15})$.} \label{comden4}
\end{figure*}

\begin{figure*}[!hbt]
\centering \scalebox{0.25}{\includegraphics{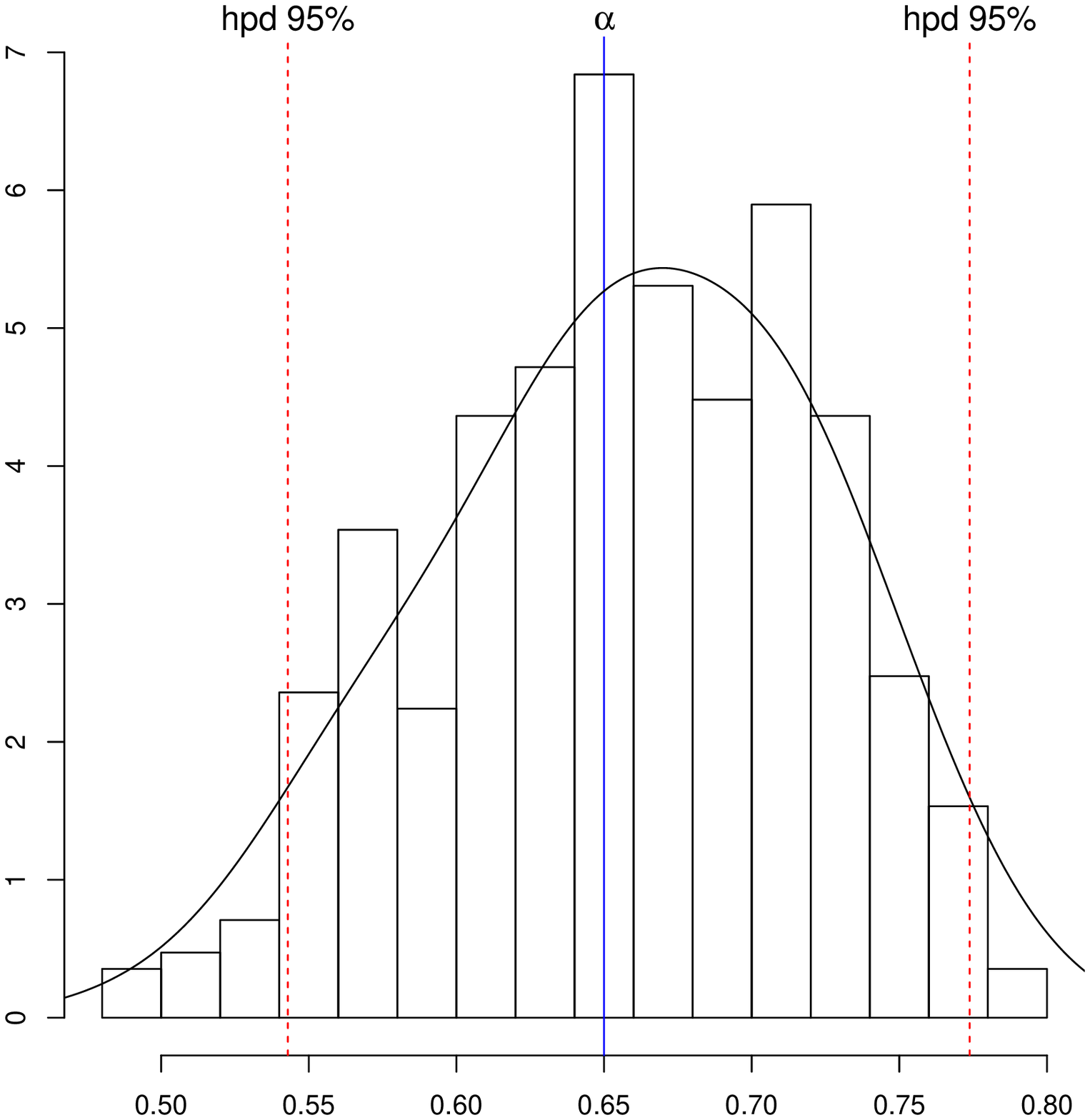}
\includegraphics{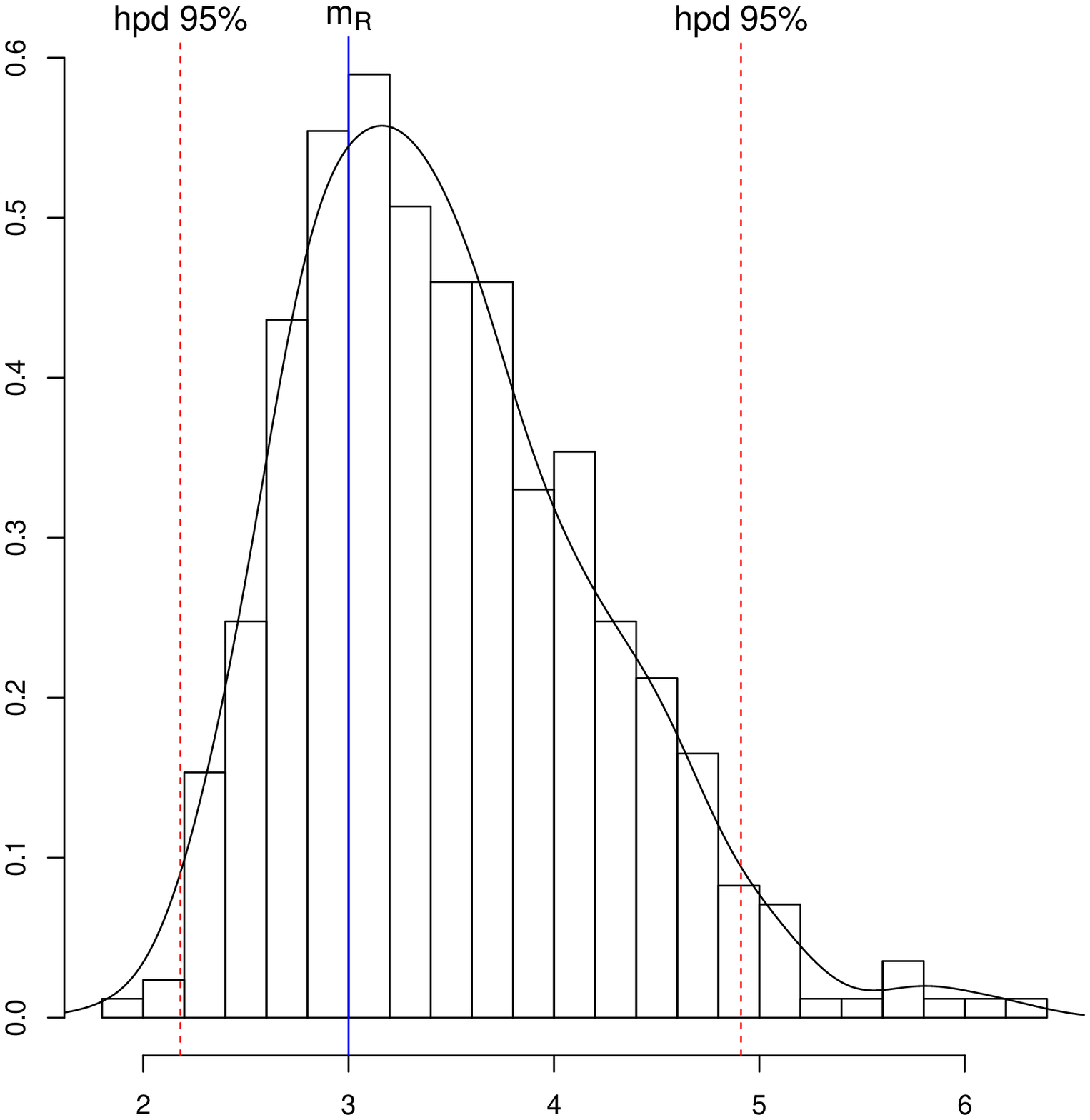}
\includegraphics{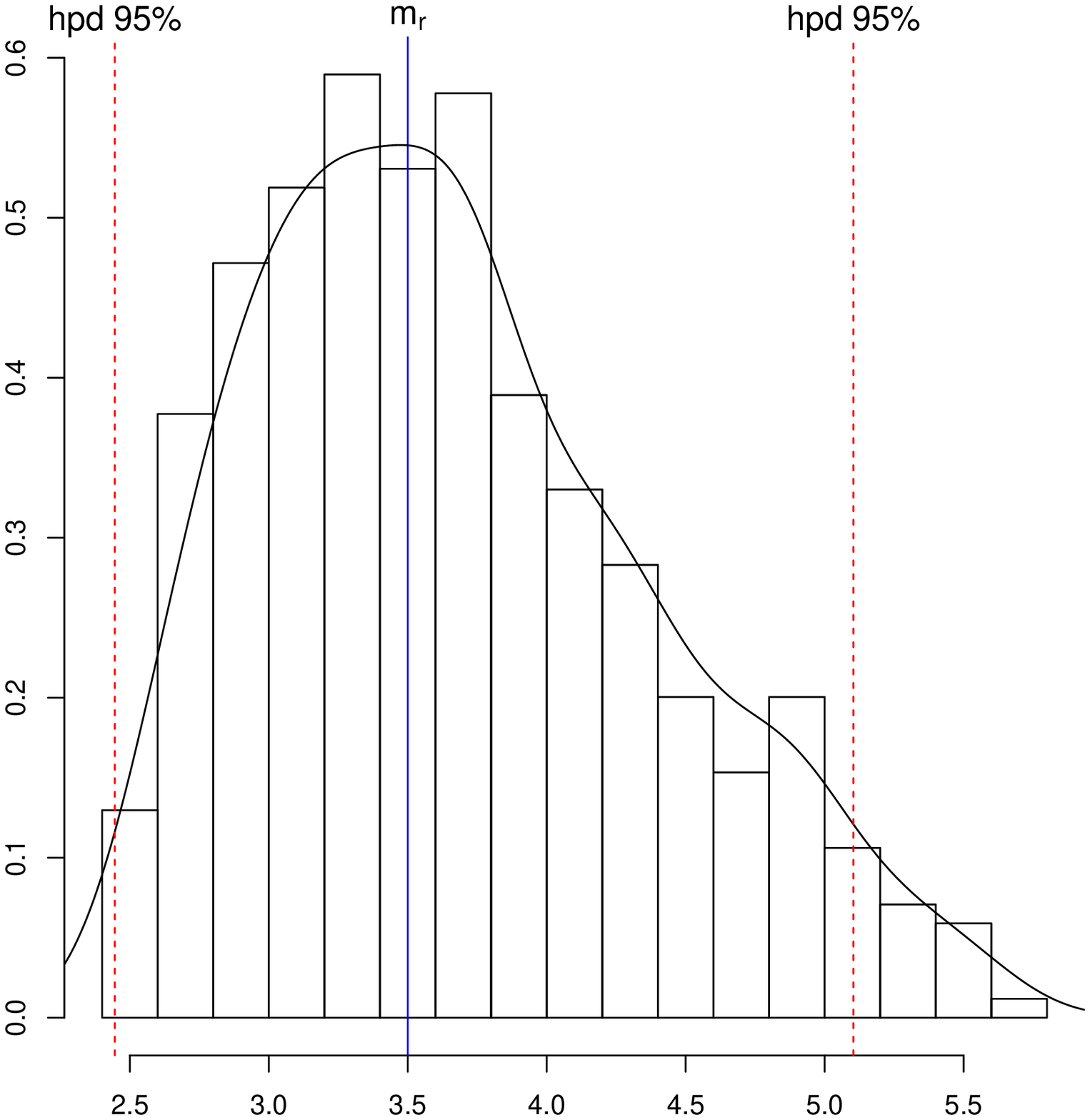}}
\caption{Approximate posterior densities, with 95\%
HPD sets, of the parameters $\alpha$, $\mathsf{m_R}$ and $\mathsf{m_r}$, given $\overline{\mathcal{FM}}_{15}$ in Table \ref{E4} in the case
$\MRrN=0$ considering only simulated paths where $\beta^{\mathsf{sim}}>0$. Vertical solid lines represent the
true value of the parameters.} \label{comden4betan0}
\end{figure*}

In Figure \ref{comden4}, we show the approximate posterior densities of every parameter. The approximations of $\alpha|$ $\overline{\mathcal{F\!M}}_{15}$, $\mathsf{m_R}|\overline{\mathcal{FM}}_{15}$ and $\mathsf{m_r}|\overline{\mathcal{FM}}_{15}$ have been obtained considering all chosen simulated paths. In all those cases, the algorithm provides accurate approximations to the posterior densities of all parameters with the $95\%$ HPD sets containing their true values and with small values of their RMSE (see Table \ref{MSE3}). On the other hand, the approximation of $\beta|\overline{\mathcal{FM}}_{15}$ has been obtained considering only chosen simulated paths where $\beta^{\mathsf{sim}}>0$. In this case the $95\%$ HPD set also contains the true value of the parameter so the approximation to the posterior density is also considered enough accurate. Moreover, we also represent the $P(\beta=0|\overline{\mathcal{FM}}_{15})$ (area of the vertical solid bar) which is estimated by 0.152.

As in the previous subsection, from the estimates and the observed sample, one can wonder about the following hypothesis test:
\begin{equation}\label{hypbeta}H_0:\beta=0 \ \mbox{vs.} \ H_1: \beta>0.\end{equation} Considering that we have assume in the implementation of the algorithm that $\beta^{\mathsf{sim}}$ could take the value 0 with probability $\gamma^{\mathsf{sim}}$, being $\gamma^{\mathsf{sim}}\sim U(0,1)$, we consider its expected value at calculating the Bayes factor, $K$, and then, it is verified that $P(\beta>0)=P(\beta=0)$, thus $$K=\frac{P(\beta=0|\overline{\mathcal{FM}}_{15})P(\beta>0)}
{P(\beta>0|\overline{\mathcal{FM}}_{15})P(\beta=0)}=\frac{0.152}{0.848}=0.18.$$
That value of the Bayes factor leads us to conclude (see \cite{jef}) that there are substantial evidences against the null hypothesis, and then $\beta>0$ is more supported by the observed sample, which is the real situation.

In Figure \ref{comden4betan0} we present the approximate posterior densities of $\alpha|\overline{\mathcal{FM}}_{15}$, $\mathsf{m_R}|\overline{\mathcal{FM}}_{15}$ and $\mathsf{m_r}|\overline{\mathcal{FM}}_{15}$ calculated considering only paths where $\beta^{\mathsf{sim}}>0$. Notice that, the true values of the all three parameters are into 95\% HPD sets. Moreover, in Table \ref{MSE3} are presented the RMSE for the estimates of those parameters. Notice that the RMSE are very similar to those calculated when all chosen simulated paths are considered and they are very close to 0.

\begin{table*}
\caption{RMSE for the estimates of $\mathsf{\alpha}$,
$\mathsf{\beta}$, $\mathsf{m_R}$ and $\mathsf{m_r}$ given by the Tolerance Rejection ABC Algorithm when the sample $\overline{\mathcal{FM}}_{15}$ \mbox{in Table \ref{E4}} is observed.} \label{MSE3}
\begin{center}
\begin{tabular}{l|c|c|c|c|}
 & $\alpha$ & $\beta$& $\mathsf{m_R}$ & $\mathsf{m_r}$ \\
\hline
Considering all simulated paths &  0.0096 & 120.72 & 0.0858 & 0.0448 \\
Considering only simulated paths where $\beta^{\mathsf{sim}}>0$ & 0.0097 & 142.18 & 0.0890 & 0.0432\\
\hline
\end{tabular}
\end{center}
\end{table*}

As final conclusion of Subsections \ref{6.2} and \ref{6.3} we establish that, if in the observed sample, one of the random variables $\mathsf{M^{R\to R}_N}$ or $\mathsf{M^{R\to r}_N}$ is equal to 0 then we apply the Tolerance Rejection-ABC Algorithm and solve the corresponding hypothesis test considering the approximate posterior distributions conditioned to the decision given by this test.

\begin{obs}In order not to extend the paper, we have not considered explicitly in subsection \ref{6.2} an example where $\MrN=0$ and $\mathsf{m_r}>0$, however an example of this kind of situation is considered in subsection \ref{6.3} where $\MRN=0$ and $\beta>0$. Analogously, it has not been considered explicitly in subsection \ref{6.3} an example where $\MRN=0$ and $\beta=0$, however an example of this kind of situation is consider in subsection \ref{6.2} where $\MrN=0$ and $\mathsf{m_r}=0$. In both cases, the results are analogous to those shown in the paper.\end{obs}

\begin{obs}The case $\MrN=0$, i.e. $\MRrN=0$ and $\mathsf{M^{r\to r}_N}=0$, is not illustrated in the paper. A sample where $\MrN=0$ is observed could belong to a coexistence path although it would not be guaranteed. Anyway, to make inference about the parameters in this situation, the Tolerance Rejection-ABC Algorithm would be applied and then both hypothesis test, in (\ref{hypmr}) and (\ref{hypbeta}), should be solved.\end{obs}

\begin{obs} The case $\MRN=0$ is not illustrated either in the paper, since this case represents the extinction of the $\mathsf{R}$-allele and then the behavior of $\mathsf{r}$-allele is described by a two-sex Galton-Watson process (see \cite{ggm2012a}).\end{obs}

\section{Sensitivity analysis}\label{sensitivity}

In this section we examine the sensitivity of inferences depending on the probability distribution used to generate the simulated paths. We apply  the Tolerance Rejec\-tion-ABC Algorithm to the examples in Subsections 6.1.1 and 6.1.2. but now generating the pool of simulated paths, instead from the Poisson distribution, from negative binomial distribution laws with different value of size ($\mathsf{k}$) since these kind of distributions have been also used in practical cases (see \cite{fg}, \cite{modesleemam} or \cite{pakes}).

In Tables \ref{HPDE1} and \ref{HPDE2} we present the point estimates of  $\alpha$, $\beta$, $\mathsf{m_R}$ and $\mathsf{m_r}$, under squared error loss as
well as their $95\%$ HPD sets for the two examples, respectively. It can be
seen that in all cases the HPD sets contain the true values of the parameters, being very similar among them for different distributions.
This allow us to conclude that this is a robust methodology against the probability distribution used to simulate the processes.

\begin{table*}
\caption{Point estimates, with 95\% HPD sets, for the parameters $\mathsf{\alpha}$,
$\mathsf{\beta}$, $\mathsf{m_R}$ and $\mathsf{m_r}$,  given the sample $\overline{\mathcal{FM}}_{15}$ in Table \ref{E1++}
with $(\mathsf{M^R_{14}},\mathsf{M^r_{14}})=(754,24687)$ and $(\mathsf{M^{R}_{15}},\mathsf{M^{R\to r}_{15}},\mathsf{M^{r\to r}_{15}})=(1043,6,45844)$, in the case $\mathsf{m_r}\geq(1-\beta)\mathsf{m_R}$,
where $(\alpha,\beta,\mathsf{m_R},\mathsf{m_r})=(0.46, 0.005, 3.2,4)$.}
\label{HPDE1}
\begin{center}
\begin{tabular}{c|lcc|ccc|ccc|ccc}
 &
\multicolumn{3}{c|}{${\alpha}|\overline{\mathcal{FM}}_{15}$}&
\multicolumn{3}{c|}{${\beta}|\overline{\mathcal{FM}}_{15}$}&
\multicolumn{3}{c|}{$\mathsf{{m}_R}|\overline{\mathcal{FM}}_{15}$}&
\multicolumn{3}{c}{$\mathsf{{m}_r}|\overline{\mathcal{FM}}_{15}$}\\ & \multicolumn{3}{c|}{}&\multicolumn{3}{c|}{}&\multicolumn{3}{c|}{}&\multicolumn{3}{c}{}\\
{Base distribution} & \multicolumn{1}{l}{Mean} & \multicolumn{2}{l|}{95\% HPD }
& \multicolumn{1}{l}{Mean} & \multicolumn{2}{l|}{95\% HPD }&  \multicolumn{1}{l}{Mean} & \multicolumn{2}{l|}{95\% HPD }
& \multicolumn{1}{l}{Mean} & \multicolumn{2}{l}{95\% HPD }\\
 \hline

\mbox{Poisson} &       0.443& 0.289&0.602&0.020&0.001&0.050&3.520 & 2.676 & 4.574 & 4.578 & 3.434 & 6.159            \\
\mbox{Negative binomial}     &          &       &       &       &       &       &       &       &       &    &       &     \\
  $\mathsf{k=1}$  &    0.449& 0.281&0.628&0.022&0.001&0.056&3.622&2.539&4.970&4.716&3.467&6.270      \\
 $\mathsf{k}=2$    &   0.436& 0.278&0.615&0.021&0.001&0.056&3.608&2.589&4.779&4.703&3.491&6.266         \\
$\mathsf{k}=5$    &    0.445& 0.287&0.615&0.020&0.001&0.051&3.563&2.689&4.686&4.620&3.492&6.110          \\
$\mathsf{k}=10$    &   0.445& 0.292&0.610&0.020&0.001&0.048&3.550&2.657&4.674&4.584&3.462&6.019  \\

\hline
\end{tabular}
\end{center}
\end{table*}

\begin{table*}
\caption{Point estimates, with 95\% HPD sets, for the parameters $\mathsf{\alpha}$,
$\mathsf{\beta}$, $\mathsf{m_R}$ and $\mathsf{m_r}$,  given the sample $\overline{\mathcal{FM}}_{15}$ in Table \ref{E2}, in the case $\mathsf{m_r}<(1-\beta)\mathsf{m_R}$, where $(\alpha,\beta,\mathsf{m_R},\mathsf{m_r})=(0.45, 0.01, 3.5,2.6)$.}
\label{HPDE2}
\begin{center}
\begin{tabular}{c|lcc|ccc|ccc|ccc}
 &
\multicolumn{3}{c|}{${\alpha}|\overline{\mathcal{FM}}_{15}$}&
\multicolumn{3}{c|}{${\beta}|\overline{\mathcal{FM}}_{15}$}&
\multicolumn{3}{c|}{$\mathsf{{m}_R}|\overline{\mathcal{FM}}_{15}$}&
\multicolumn{3}{c}{$\mathsf{{m}_r}|\overline{\mathcal{FM}}_{15}$}\\ & \multicolumn{3}{c|}{}&\multicolumn{3}{c|}{}&\multicolumn{3}{c|}{}&\multicolumn{3}{c}{}\\
{Base distribution} & \multicolumn{1}{l}{Mean} & \multicolumn{2}{l|}{95\% HPD }
& \multicolumn{1}{l}{Mean} & \multicolumn{2}{l|}{95\% HPD }&  \multicolumn{1}{l}{Mean} & \multicolumn{2}{l|}{95\% HPD }
& \multicolumn{1}{l}{Mean} & \multicolumn{2}{l}{95\% HPD }\\
 \hline

\mbox{Poisson} &       0.478&0.358&0.606&0.019&0.003&0.037&3.721 & 3.145 & 4.377 & 2.238 & 1.078 & 3.348        \\
\mbox{Negative binomial}     &          &       &       &       &       &       &       &       &       &    &       &     \\
  $\mathsf{k=1}$  &    0.477&0.341&0.615&0.020&0.002&0.041&3.766&3.075&4.596&2.231&0.800&3.590     \\
 $\mathsf{k}=2$    &   0.472&0.346&0.603&0.019&0.002&0.039&3.733&3.092&4.522&2.238&0.939&3.458          \\
$\mathsf{k}=5$    &    0.479&0.353&0.605&0.019&0.002&0.037&3.706&3.121&4.416&2.232&0.999&3.392        \\
$\mathsf{k}=10$  &     0.474&0.351&0.603&0.019&0.002&0.036&3.725&3.113&4.440&2.224&1.077&3.273           \\
\hline
\end{tabular}
\end{center}
\end{table*}

\section{Prediction of the future population size}

Finally, once that the algorithm has been proved to be a useful tool to obtain accurate approximations of the posterior distributions of the parameters, from them, we can also estimate others random variables related to the process. For instance, from a practical standpoint, it is  of interest to infer the size of future generations. Next, we apply a Monte Carlo procedure, proposed in \cite{ggmmp2016}, to approximate the predictive distributions.
In particular, for each $\theta^{(i)}$, $i=1,\ldots,m$ sampled from $\theta|\overline{\mathcal{FM}}_{\mathsf{N}}$, one can simulate $s$ process until the $\mathsf{l}$th generation, which started with $\FN$ females,
$\MRN$ $\mathsf{R}$-males and $\MrN$ $\mathsf{r}$-males, obtaining values to approximate the predictive posterior distributions $(\mathsf{F_{N+l}}$, $\mathsf{M^{R}_{N+l}}$, $\mathsf{M^{R\to r}_{N+l}}$, $\mathsf{M^{r\to r}_{N+l}})|\overline{\mathcal{FM}}_{\mathsf{N}}$
and $(\mathsf{Z^{R}_{N+l}}$, $\mathsf{Z^{r}_{N+l}})|\overline{\mathcal{FM}}_{\mathsf{N}}$ by Gaussian kernel estimators.

To illustrate this procedure, we consider the example given en subsection 6.1.2 considering the observed sample $\overline{\mathcal{FM}}_{15}$ given in Table \ref{E2},
$m=1000$, $s=2000$ and $\mathsf{l}=1$. Concretely, we simulate a generation of 2000 processes started with $(\mathsf{F_{15}},\mathsf{M^R_{15}},\mathsf{M^r_{15}})=(5437,6351,258)$, for each parameter $\theta^{(i)}$, $i=1,\ldots,1000$.

\begin{figure*}[!hbt]
\centering \scalebox{0.24}{\includegraphics{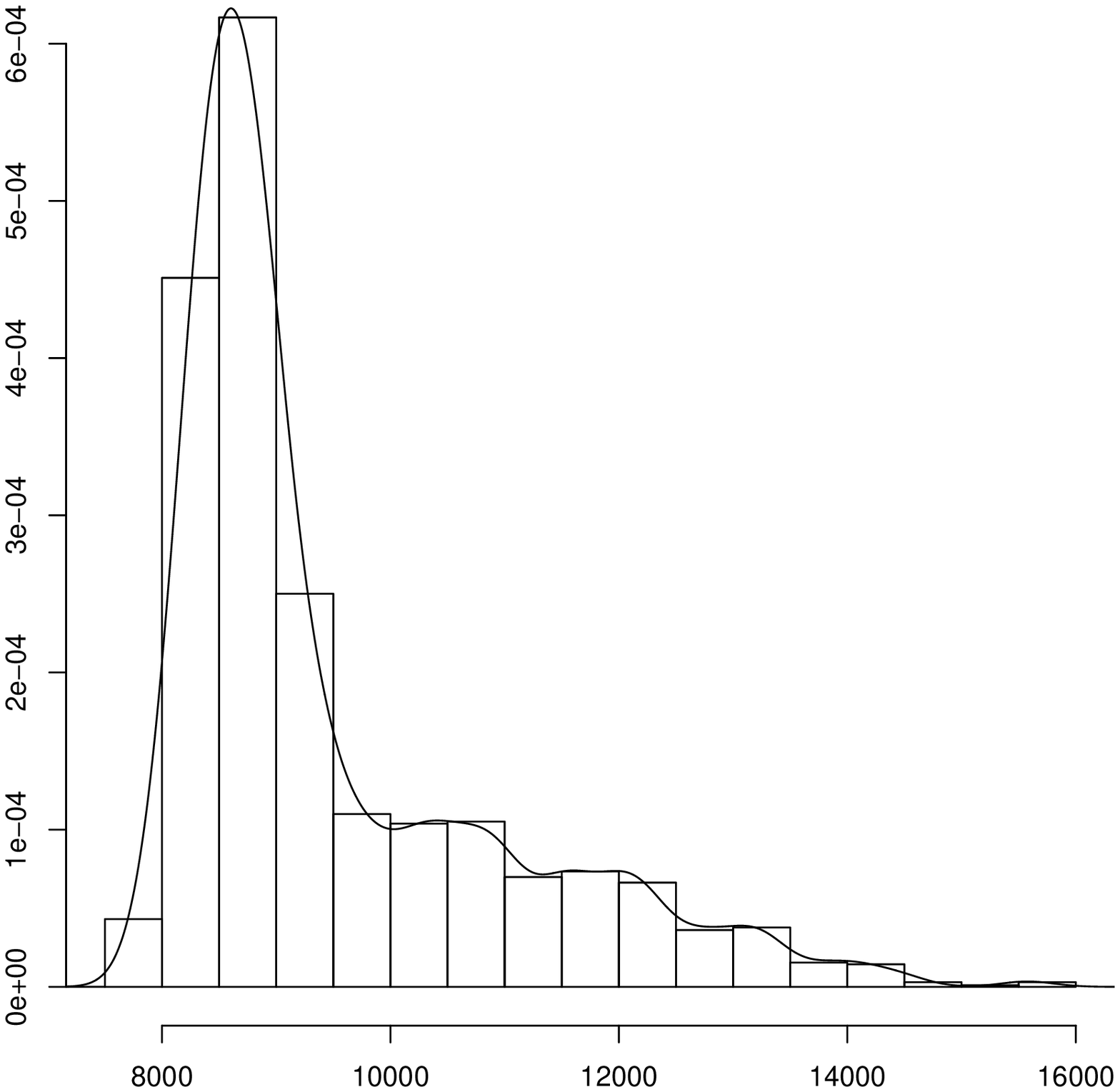}
\includegraphics{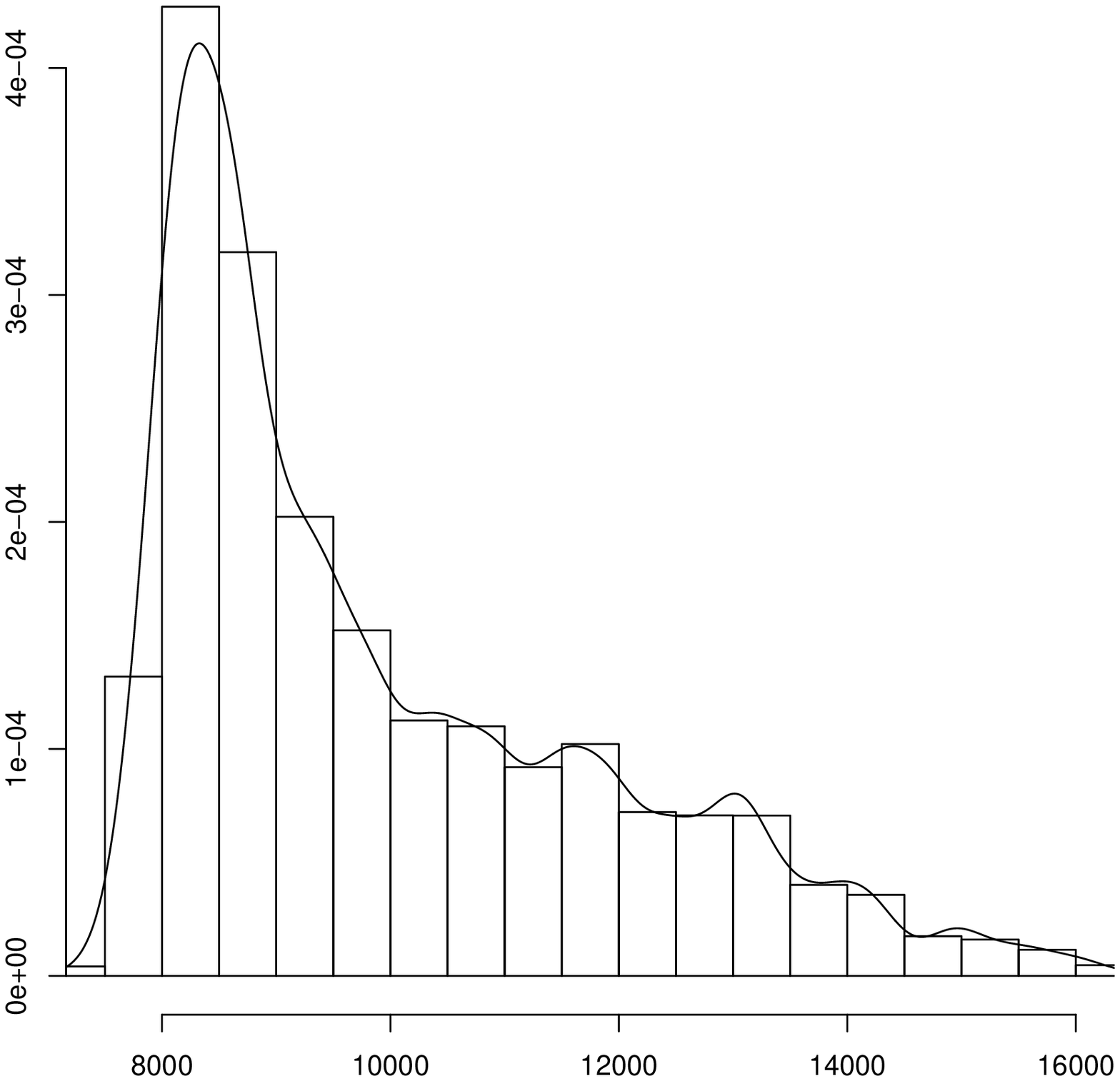}\includegraphics{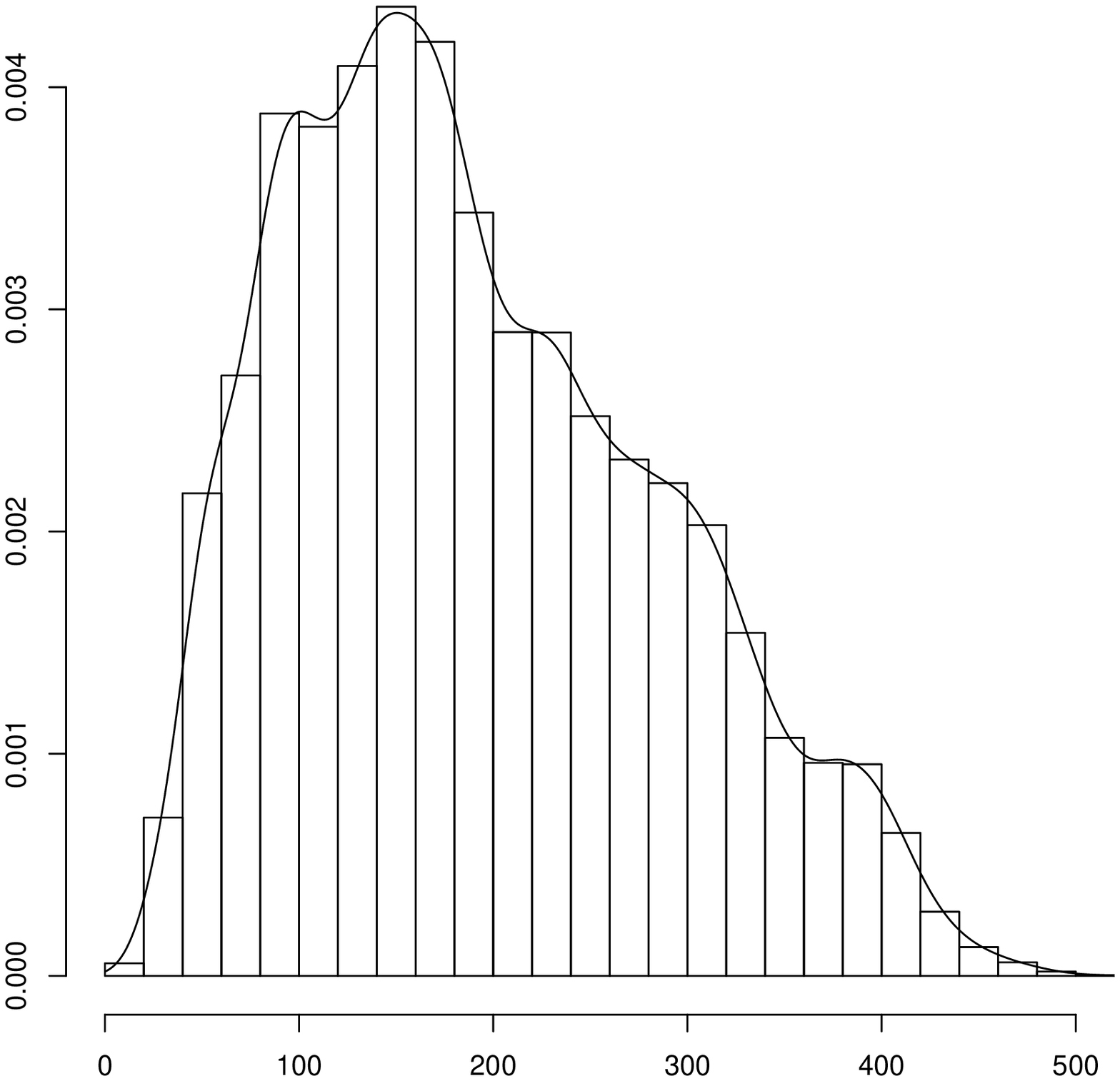}
\includegraphics{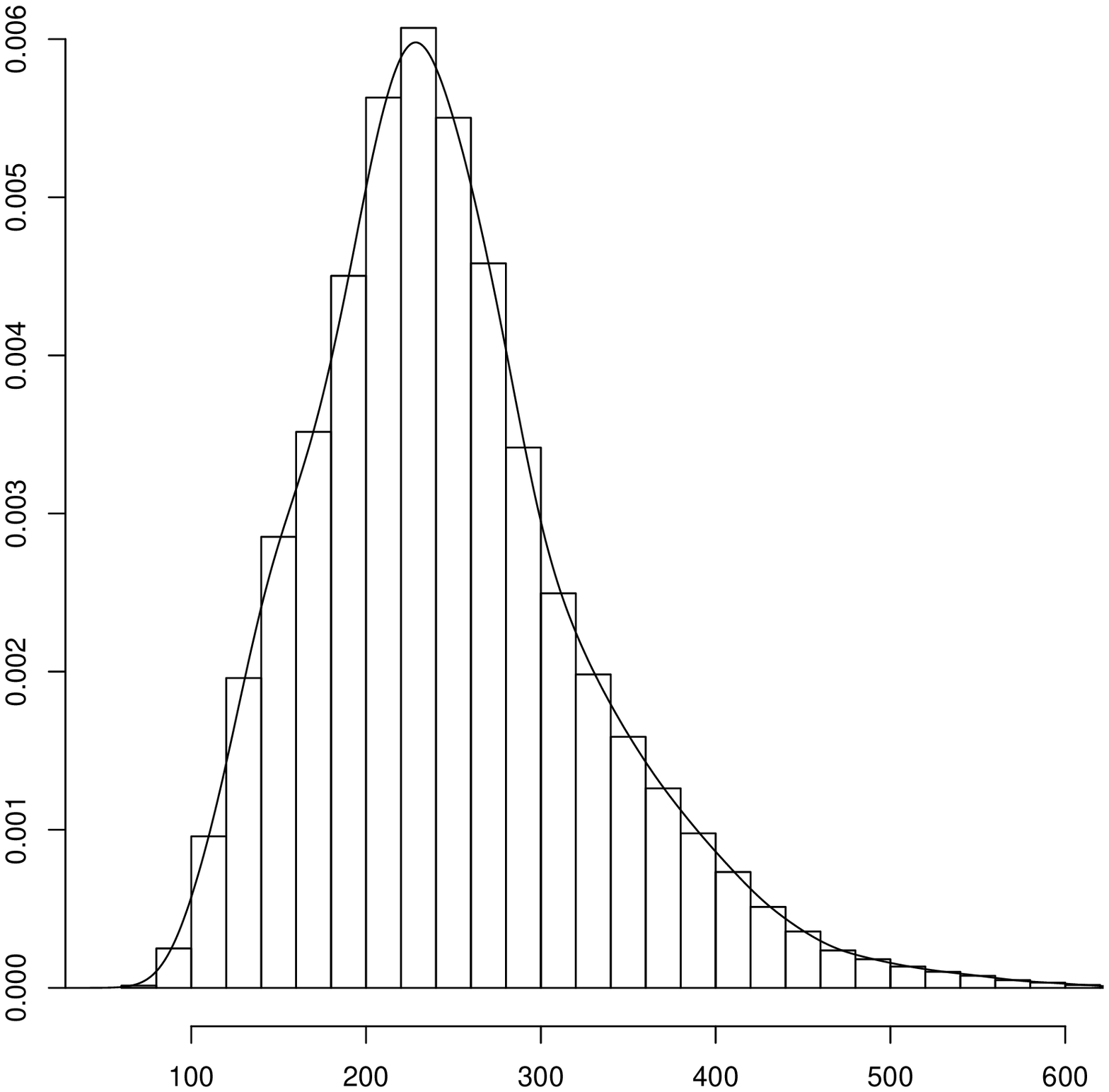}}
\caption{Approximate predictive posterior densities for $\mathsf{F_{16}}$, $\mathsf{M^R_{16}}$, $\mathsf{M^{R\to r}_{16}}$ and $\mathsf{M^{r\to r}_{16}}$, respectively, given $\overline{\mathcal{FM}}_{15}$ in Table \ref{E2}.} \label{comden2b}
\end{figure*}

Figure \ref{comden2b} shows the approximated predictive posterior distributions for $\mathsf{F_{16}}$, $\mathsf{M^R_{16}}$, $\mathsf{M^{R\to r}_{16}}$ and $\mathsf{M^{r\to r}_{16}}$, given $\overline{\mathcal{FM}}_{15}$ in Table \ref{E2}. Notice that these estimates are in accordance with the relation between the parameters and with the observed sample, where the $\R$-allele is the dominant one.

\begin{obs}The software environment \ for statistical \ \ computing and graphics $\mathbf{R}$ (``GNU  S'', see \cite{r}) has been used to perform the ABC methodology and the simulation study. To calculate the kernel density estimation the GenKern  package (see \cite{GenKern}) and sm package (see \cite{sm}) have been used. To implement the Bayesian t-test, the BayesFactor and BEST packages (see \cite{BayesFactor} and \cite{BEST}, respectively) have been applied.\end{obs}

\section{Concluding Remarks}

The aim of this work has been to develop Bayesian
inference theory for a Y-linked two-sex branching process with blind
choice which is useful to model the evolution of the number of carriers of two alleles (named as $\R$ and $\mathsf{r}$) of a Y-linked gene considering the possibility of mutations from $\R-$allele to the $\mathsf{r}-$allele.

We have focussed mainly on approximating the posterior distributions of the main parameters of such mo\-del considering for that, at the beginning, a realistic sampling scheme where the observation of the total number of females and males in each generation is assumed as well as the observation of the total number of each type of males (males with $\R-$allele and males with $\mathsf{r}-$allele) in the last generation. Then, we have described the development of a method based on the Approximate Bayesian Computation (ABC) methodology (Tolerance Rejection-ABC Algorithm) to approximate the posterior distributions of the model parameters based on such sample scheme.

We have shown throughout a simulated example that
the methodology presents difficulties to estimate the posterior
distribution of the probability of mutation, $\beta$, due to the fact that with the observed sample it is not possible to know how many of the observed $\mathsf{r}-$alleles stem from mutations. For that reason, we consider another sampling scheme where also is observed, in the last generation, the total number of $\mathsf{r}-$males stemming from $\R-$fathers as well as the total number of each type of males in the penultimate generation.

We have illustrated how the Tolerance Rejection-ABC Algorithm works based on this sampling scheme and considering different situations which can be observed in the sample in the case of coexistence of both alleles. In this sense, we have considered special situations which can be observed in the last generation of the sample: when there are the two types of males (i.e. $\MRrN>0$ and $\MrrN>0$), when there are not $\mathsf{r}-$males stemming from $\mathsf{r}-$fathers (i.e. $\MrrN=0$) and when there are not $\mathsf{r}-$males stemming from $\R-$fathers (i.e. $\MRrN=0$). In all cases, we have obtained accurate approximations to the posterior densities of all parameters with the $95\%$ HPD sets containing the true values of the parameters.

The case where $\MrrN=0$ is the special interest because it is no possible to know whether the mean number of individuals stemming from $\mathsf{r}-$couples, $\mathsf{m_r}$, is equal to 0 or strictly positive. Analogously, the case where $\MRrN=0$ is interesting because it is no possible to know whether the probability of mutation is equal to 0 or strictly positive. In both cases, after applying the ABC methodology, we have proposed a hypothesis test to decide the more plausible option (see (\ref{hypmr}) and (\ref{hypbeta})). In the two considered examples, the Bayes factor has lead us to conclude that the true situation was the supported one by the observed sample.

Notice that we have taken 15 generations in the sample schemes of all simulated examples considered in the paper. We considere that this is a balanced number in the sense that it is big enough to observe whether one of the alleles is the dominant and also it is a feasible number to be observed in many animal populations with sex reproduction. Moreover, in the examples, we have covered all possible situations between the parameters in the coexistence set taking into account the different magnitudes of the rates of growth.

We have also studied the robustness of the methodology by mean of a general simulated experiment where we have applied the methodology for different base distributions concluding that this is a robust methodology against the probability distribution used to simulate the processes.

Finally, we have been able to predict the future population size approximating the predictive distributions of the random variables related to the total number of females and the total number of each type of males in the following generation to the last one observed.

Note that the Approximate Bayesian Computation
is a proved statistical tool very useful for inference in
parameters of complex models in population genetics as is our case.
It is easy and fast to simulate from our model, and therefore,
in this case has been more convenient that the Gibbs sampler.

\begin{acknowledgements}
This research was supported by Grant MTM2015-70522-P (MINECO/FEDER, UE) and Grant IB16\-103 (Junta de Extremadura / Fondo Europeo de Desarrollo Regional, UE).
\end{acknowledgements}


\end{document}